%% file: paper_data.tex
\documentclass[aps,prd,twocolumn,superscriptaddress]{revtex4-2}
\usepackage{xcolor}
\usepackage{amsmath,amssymb}
\usepackage{appendix}
\usepackage{enumerate}
\usepackage{subfigure}
\usepackage{natbib}
\usepackage{url}
\usepackage{aas_macros}
\usepackage{graphicx}
\usepackage[colorlinks=true,allcolors=blue]{hyperref}
\usepackage{float}
\usepackage{footnote}
\usepackage{orcidlink}
\begin{document}
\setcounter{secnumdepth}{3}

\newcommand{\sm}[1]{\textcolor{red}{#1}}
\newcommand{\dg}[1]{\textcolor{orange}{#1}}
\newcommand{\add}[1]{\textcolor{blue}{#1}}
\newcommand{\ptcite}[1]{\textcolor{magenta}{#1}}

\title{How lonely are the Binary Compact Objects Detected by the LIGO-Virgo-KAGRA Collaboration?}

\author{Devesh Giri \orcidlink{0009-0006-3310-481X}}
\email{devesh.giri@students.iiserpune.ac.in}
\email{deveshgiri.edu@gmail.com}
\affiliation{Indian Institute of Science Education and Research, Pashan, Pune 411008, India}
\author{Suvodip Mukherjee \orcidlink{0000-0002-3373-5236}}
\email{suvodip@tifr.res.in}
\affiliation{Department of Astronomy and Astrophysics, Tata Institute of Fundamental Research, Mumbai 400005, India}

\begin{abstract}
Gravitational-wave (GW) observations of compact binary coalescences (CBCs) are traditionally interpreted under the assumption that the binary evolves in isolation. However, in realistic astrophysical environments, brief three-body encounters may perturb the binary's orbital evolution and imprint deviations on the emitted GWs. We develop a physically motivated model for such interactions, retaining Newtonian three-body dynamics supplemented by leading-order ($2.5$PN) radiation-reaction within the binary. We show that such encounters produce a distinctive morphology of dephasing and amplitude modulation in GWs. We search for this kind of distortion from the LIGO--Virgo--KAGRA (LVK) GW catalog GWTC-4 on three events: GW170817, GW190814, and GW230627\_015337, chosen based on high SNR and in-band duration $\gtrsim 10~\mathrm{s}$. We find no statistically significant deviation in the data, which translates into constraints on the absence of any intermediate-mass black hole in the mass range above $\sim 10^2$ M$_\odot$ in the vicinity of these binaries of radius approximately $10^{-1}~\mathrm{AU}$. This arises from robust exclusions arising from fly-by interactions that would dynamically disrupt the binary and are directly ruled out independent of waveform modelling, placing the first upper bound on intermediate-mass black holes near these GW events. In future, with the availability of long-duration GW signals, this new avenue can probe encounters of the binary GW sources with compact objects of lighter masses at distances farther away than 1 AU and hence opens a new window to probe the population of individual compact objects of both astrophysical and primordial origin in astrophysical systems of dense environments ranging from galactic centers to dense globular clusters.   
\end{abstract}

\maketitle

\section{\label{sec:intro}Introduction\protect}

Gravitational Waves (GWs) are a fundamental prediction of Einstein's General Theory of Relativity, arising from time-varying mass-quadrupole moments in gravitating systems~\cite{Einstein:1916cc}. These waves manifest themselves as distortions in the fabric of spacetime and induce a time-varying strain. Although any system with changing mass-quadrupole moments, in principle, emits gravitational radiation, the resulting strain amplitude is typically extremely weak, making most astrophysical sources infeasible to be observed through GWs with current detector sensitivites. Detectable signals from individual sources require compact objects (COs) undergoing rapid changes in their quadrupole moment. Compact Binary Coalescences (CBCs) which involve objects such as black holes and neutron stars, provide such conditions and constitute the primary class of individually resolvable GW sources~\cite{peters1964, CutlerFlanagan1994}.

Since the first direct detection of GWs in 2015~\cite{lvk2016feb}, the LIGO–Virgo–KAGRA (LVK) collaboration~\cite{LIGO, VIRGO, KAGRA} has observed $\sim200$ CBCs, including binary black hole (BBH), neutron star–black hole (NSBH), and binary neutron star (BNS) systems~\cite{gwtc4Catalog}. These detections have opened a new channel to hear deep into the Universe and learn about strong-field gravity, the formation and environments of GW sources, and CO populations across cosmic time~\cite{GW250114,gwtc4Tgr,gwtc4Rnp,gwtc4Cosmo}. At present sensitivities, individual CBC events represent the dominant resolvable GW sources, while the detection of single systems remains intrinsically challenging due to the weakness of gravitational radiation.

GW signals from CBCs are traditionally searched for using matched-filtering techniques, which rely on banks of modelled template waveforms to correlate against data from ground-based interferometers~\cite{signalextraction,Sathyaprakash:1991mt,Dhurandhar:1992mw}. These templates are constructed under the assumption that the binary evolves in vacuum, i.e., as an isolated two-body system governed solely by general relativity~\cite{Babak:templates}. This assumption enables precise waveform modelling and efficient searches, but it also implicitly neglects a range of plausible environmental effects that could perturb the dynamics of the binary and imprint measurable deviations on the observed gravitational waveforms.

Many studies have demonstrated that compact binaries embedded in nontrivial astrophysical environments can experience additional forces that modify their orbital evolution and, consequently, their GW phase and amplitude from effects due to three-body interactions or dynamical friction ~\cite{Gultekin:2005fd,Barausse:2014pra, Cardoso:2019rou, SamsingDephasing,Li:2022pnc,PhysRevD.107.023023,Hendriks:2024gpp}. Compact binaries may reside in dense stellar environments such as globular clusters~\cite{Maccarone:2007dd, 2010MNRAS.407.1946D, Tiwari:2023cpa}, young massive clusters~\cite{2022MNRAS.517.2953T}, and nuclear star clusters~\cite{Mapelli:2021gyv}, in galactic nuclei hosting supermassive black holes and active galactic nucleus (AGN) disks~\cite{Joshi:2025wcu, Vijaykumar:2023tjg}, or in hierarchical triple systems formed through stellar evolution~\cite{Britt:2021dtg}. In cosmological scenarios, they may also evolve within a population of Primordial Black Holes (PBHs) or other compact dark matter candidates that can act as tertiary perturbers and dynamically interact with the compact binaries~\cite{Bhalla:2024jbu, Biermann:2002ky}. A population of PBHs or other compact dark-matter candidates could similarly act as unbound perturbers~\cite{Bhalla:2025xce, DeLorenci:2025wbn, vanDie:2024htf, Gomez-Aguilar:2025wss, Garcia-Bellido:2021jlq, Afroz:2025urb}. Such encounters would be distinct from conventional stellar-dynamical channels on a population-level as they are not tied to stellar evolution as they may involve compact objects with masses below the minimum expected from stellar collapse~\cite{chandra1931, Tolman:1939jz, Oppenheimer:1939ne}. In particular, compact objects with masses $\lesssim 1\,M_\odot$ are not predicted by standard stellar evolution~\cite{Hawking:1971ei, Kouvaris:2018wnh}, so gravitational signatures of sub-solar-mass (SSM) perturbers would provide a qualitatively different probe of CO populations. In all such settings, interactions with additional compact objects, whether long-lived bound companions or transient fly-by encounters with stellar black holes, neutron stars, or PBHs, can modulate orbital elements including the semi-major axis, eccentricity, inclination, and orbital frequency, thereby altering both the phase and amplitude evolution of the gravitational waveform~\cite{Samsing:2024syt, Hendriks:2024zbu, Hendriks:2024gpp, Zwick:2025wkt}. These will be new signatures on top of the acceleration of the center of mass of the system due to a massive object present far away from the source \cite{Yunes:2010sm}.  

In this work, we focus exclusively on unbound fly-by interactions between a compact binary and a third compact object, and we address the following central question: \textit{What GW signatures do brief three-body encounters imprint on the inspiral of compact binaries, and can current GW observations be used to constrain such interactions?} Unlike long-lived hierarchical triples, fly-by encounters represent adiabatic and impulsive perturbations whose signatures depend sensitively on the mass, relative velocity, impact parameter, and compactness of the perturber. For reference, stellar-mass binaries enter the LIGO band at orbital separations of order $r_0 \sim 10^{-6}$--$10^{-5}\,\mathrm{AU}$, corresponding to orbital periods of $\lesssim \mathcal{O}(\mathrm{seconds})$, whereas inter-object separations in dense globular cluster cores are of order $10^3$--$10^4\,\mathrm{AU}$ (with number densities reaching $10^4$--$10^6\,\mathrm{pc^{-3}}$), decreasing to $\sim 10$--$200\,\mathrm{AU}$ in the innermost regions of nuclear star clusters and galactic nuclei~\cite{Bahramian:2013ihw, NSC2020, NSC2010, Gnedin:2013cda, Bahcall:1976aa, Campanelli:2005kr, MillerIMBHs, Gultekin:2005fd, Gultekin:2003xd}. Although these scales vastly exceed the binary separation in the LVK band, dynamical relaxation, mass segregation, and repeated scattering can bring compact objects onto intersecting trajectories, making close fly-by interactions during the late inspiral dynamically plausible. Encounter rates are further enhanced in galactic nuclei and AGN disks, where compact objects are confined to flattened, gas-rich environments~\cite{Trani:2023oqa, Laeuger:2023qyz}.

A key feature of fly-by interactions is that they can induce a characteristic and nontrivial modification of the GW signal, affecting both its phase and amplitude evolution. In particular, the perturbation can generate cumulative dephasing, transient frequency shifts due to Doppler motion of the binary center-of-mass, and radial perturbations corresponding to eccentricity excitation. Depending on the encounter parameters, the induced phase shift may accumulate gradually over many cycles or appear as a more localised feature in time. Importantly, even when such perturbations are present, the signal may still be detected by standard vacuum template banks provided that the waveform mismatch remains below the template-placement threshold. In such cases, detection proceeds conventionally, but the unmodelled deviations may be partially absorbed by intrinsic binary parameters, leading to systematic biases in parameter estimation. Conversely, sufficiently strong perturbations can drive the mismatch beyond acceptable levels, potentially causing signals to evade detection altogether.

The primary goal of this paper is twofold. First, we develop a physically motivated and computationally tractable model for GW waveforms from compact binaries undergoing a fly-by interaction with a third compact object, retaining only the leading-order (quadrupolar) radiation from the binary. While this approximation does not capture the full complexity of state-of-the-art waveform models, it allows us to isolate and characterise the qualitative and quantitative impact of three-body perturbations on the GW phase evolution. We explicitly assess the regime of validity of this approximation by identifying the portion of the inspiral over which the perturbed waveform remains in good agreement with vacuum inspiral models and restrict our analysis to this region. Second, we confront this model with data from the current ground-based GW detector network. In this work, we yield the first constraints on transient three-body encounters with compact binaries using GW data. We parameterise the perturber by its mass ratio relative to the secondary component of the binary and by its closest-approach distance in units of the binary’s initial separation. These constraints demonstrate that compact binary mergers can be used as indirect probes of their local dynamical environments, even in the absence of explicit waveform templates for environmental effects. While the constraints obtained with current LVK data are necessarily modest, they highlight the strong potential of longer-duration signals, such as those accessible to future ground-based detectors or multi-band observations combining space- and ground-based instruments, to accumulate larger dephasing and substantially improve sensitivity to three-body interactions. We show by a schematic diagram the physical parameter space which can be explored by GW observations. 

\begin{figure*}
    \centering
    \includegraphics[width=0.8\linewidth]{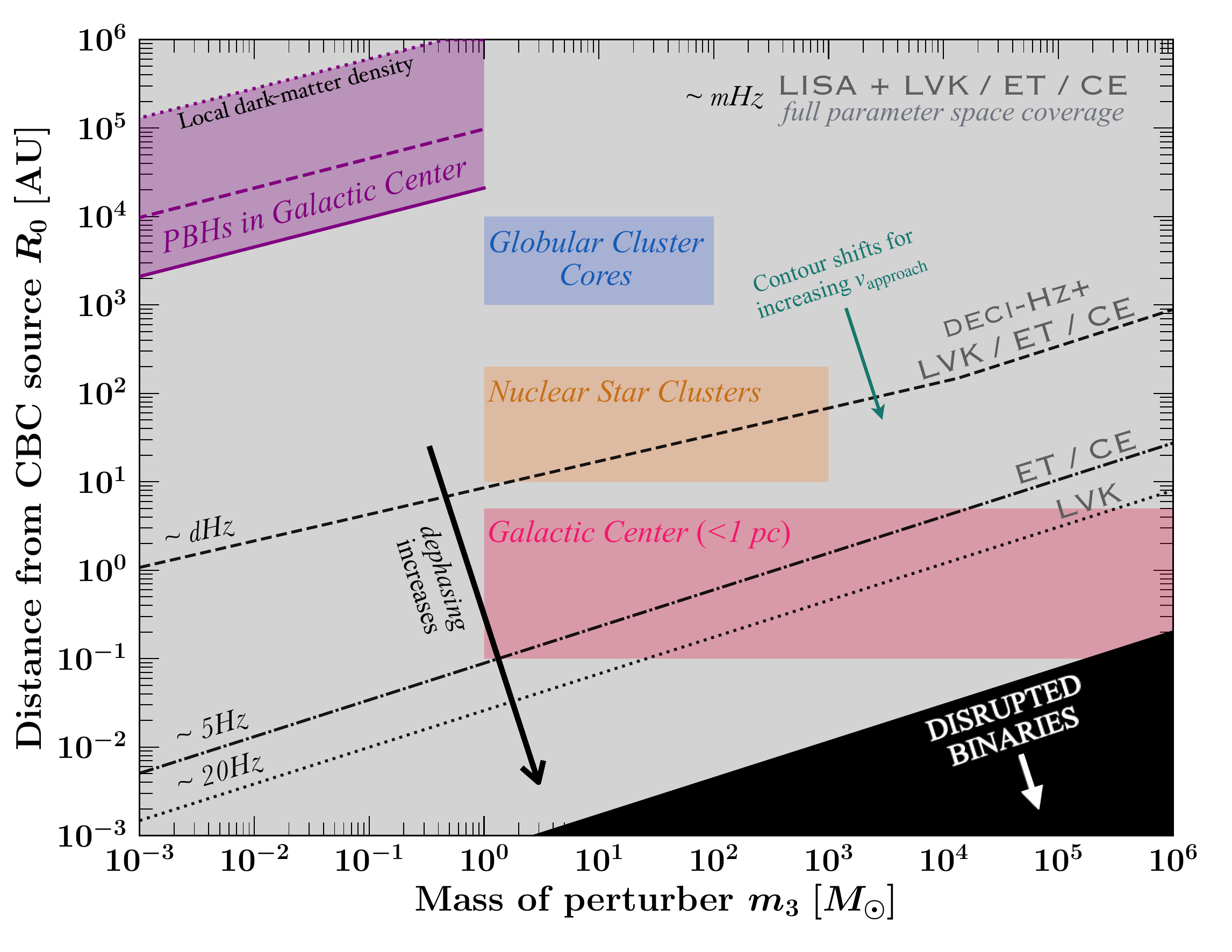}
    \caption{
    Schematic overview of the observable parameter space for three-body fly-by encounters as a function of the perturber mass $m_3$ and its initial separation $R_0$ from the binary center-of-mass. The solid black region marks configurations in which the fly-by interaction would dynamically disrupt the binary prior to merger; these are directly excluded by the very observation of a compact binary coalescence, independent of any waveform modelling (\textit{Disrupted Binaries}). The slanted lines are contours of fixed waveform mismatch between the perturbed and vacuum inspiral gravitational waveforms; configurations to the lower right of each contour produce a detectable imprint on the GW signal, while configurations to the upper left remain below the detection threshold. Each contour corresponds to a different detector configuration labelled by its lower frequency threshold $f_0$ from which the inspiral is tracked through to the innermost stable circular orbit: LVK at ${\sim}20\,\mathrm{Hz}$ (dotted), Einstein Telescope/Cosmic Explorer (ET/CE) at ${\sim}5\,\mathrm{Hz}$ (dash-dot), and proposed deci-hertz instruments at ${\sim}0.1\,\mathrm{Hz}$ (dashed). In the adiabatic regime, where the dynamical encounter timescale $t_\mathrm{enc} \sim R_0/v_3$ greatly exceeds the in-band inspiral duration $T$, the accumulated dephasing scales as $\delta\phi \propto G m_3 T / R_0^3$, so that the mismatch contour follows $R_0 \propto (m_3 T)^{1/3}$. Since the inspiral duration scales as $T \propto f_0^{-8/3}$, the contour anchor shifts as $f_0^{-8/9}$ between detectors; this accounts for the progressively larger accessible separations at lower frequencies. The LVK contour is anchored directly from numerical simulations; contours for other detectors in this regime are obtained by rescaling accordingly. For the deci-hertz configuration, a change in slope is visible at $R_0 \sim \mathcal{O}(10^2)\,\mathrm{AU}$: above this point the encounter is adiabatic, while below it the encounter timescale becomes comparable to the inspiral duration and the interaction begins to transition toward the impulsive regime, where the perturber physically traverses the binary orbit during the observation window. For LVK and ET/CE, this transition occurs at $R_0$ values below the plotted range and is therefore not shown. The exact contour slope in the impulsive regime depends sensitively on the encounter geometry and requires dedicated numerical simulation for each detector configuration; the slope shown below the kink for deci-hertz is therefore indicative only. The gray background denotes that the LISA multiband configuration (${\sim}\mathrm{mHz}$), combined with a ground-based detector tracking the same source to merger, spans an observational baseline of years; the corresponding dynamical encounter timescale $R_0^{\rm dyn} = v_3 T$ exceeds the entire plotted range, implying that the full visible parameter space constitutes an observationally probeable domain with LISA multiband studies. For increasing perturber approach velocity (varying in $\mathcal{O}(10\text{--}100)$ km/s), the contours shift downward (arrow), as a faster encounter requires a closer or more massive perturber to accumulate the same dephasing in a given inspiral time. Coloured bands indicate characteristic mean inter-object separations in representative astrophysical environments: galactic centre within $1\,\mathrm{pc}$ of Sgr~A$^*$ (pink, $\bar{d} \sim 0.1$--$5\,\mathrm{AU}$)~\cite{Amaro-Seoane:2012lgq, Bahcall:1976aa, Gnedin:2013cda}, nuclear star clusters (orange, $\bar{d} \sim 10$--$200\,\mathrm{AU}$)~\cite{Ozernoy:1997pa, Ott:2003gr, NSC2020, NSC2010}, and globular cluster cores (blue, $\bar{d} \sim 10^3$--$10^4\,\mathrm{AU}$)~\cite{Willems:2007xe, Ivanova:2003by, Miller:2002pg, Baumgardt2018, Bahramian:2013ihw, Baumgardt2018}. The purple band shows the mean inter-PBH separation computed from an NFW dark matter profile assuming all dark matter is in PBHs ($f_\mathrm{PBH}=1$), for galactocentric radii $r = 0.1\,\mathrm{pc}$ (solid purple boundary) to $r = 10\,\mathrm{pc}$ (dashed purple boundary); and the dotted purple line corresponds to the local dark matter density $\rho_s$ assuming $\rho_s=4\times10^{-3}~M_\odot~\mathrm{pc}^{-3}$~\cite{Navarro:1995iw, Bertone:2004pz, Cautun:2019eaf, Klypin2002, Sofue2012, Pato:2015dua}; the band spans the sub-solar mass range $m_3 \lesssim 1\,M_\odot$ for which PBH candidates are not predicted by standard stellar evolution. This figure is schematic; contour normalizations are illustrative and the exact slopes require numerical simulations.
    }
    \label{fig:paramspace_schematic}
\end{figure*}

To place our analysis in context, Fig.~\ref{fig:paramspace_schematic} provides a schematic map of the observable parameter space for three-body fly-by encounters as a function of the perturber mass $m_3$ and its initial separation $R_0$ from the merging binary. Three qualitatively distinct regions can be identified. At large $m_3$ and small $R_0$, the gravitational interaction is strong enough to dynamically disrupt the binary prior to merger; such configurations are directly excluded by the very observation of a coalescence, independent of any waveform model (\textit{Disrupted Binaries}, black region). At intermediate values, the binary survives but accumulates a measurable dephasing and amplitude deviation in the gravitational waveform; configurations to the lower-right of each detector contour line lie in this \textit{dephased} regime where the waveform mismatch exceeds a detectable threshold. Configurations well to the upper-left of the contours produce perturbations that remain below the noise floor of the respective detector.

The slanted mismatch contours depend on both the perturber parameters and the detector configuration through the in-band inspiral duration $T$. In the adiabatic regime, where the dynamical encounter timescale $t_\mathrm{enc} \sim R_0/v_3$ greatly exceeds $T$, the perturber acts as a quasi-static external tidal field. The accumulated dephasing scales as $\delta\phi \propto G m_3 T / R_0^3$, so at fixed mismatch threshold the contour follows $R_0 \propto (m_3\,T)^{1/3}$. Since the inspiral duration scales as $T \propto f_0^{-8/3}$ with the lower frequency threshold $f_0$, the contour anchor shifts between detectors as $f_0^{-8/9}$, producing the progressively higher contours at lower frequencies seen in Fig.~\ref{fig:paramspace_schematic}. The LVK~\cite{LIGO, VIRGO, KAGRA} contour is anchored directly from the numerical simulations of this work; contours for Einstein Telescope (ET)~\cite{EinsteinTelescope} / Cosmic Explorer (CE)~\cite{CosmicExplorer} and deci-hertz configurations~\cite{DECIGO, LGWA, LILA, deciHz_status} are obtained by rescaling the anchor accordingly.

At sufficiently small $R_0$, or equivalently for detectors with sufficiently long in-band durations, the encounter timescale becomes comparable to $T$ and the interaction transitions from adiabatic to impulsive: the perturber physically traverses the binary orbit during the observation window, delivering a velocity kick rather than a slow tidal deformation. For the deci-hertz configuration, this transition occurs at $R_0 \sim \mathcal{O}(10^2)\,\mathrm{AU}$, producing the visible change in slope in Fig.~\ref{fig:paramspace_schematic}. For LVK and ET/CE, the transition falls below the plotted range and the contours are purely adiabatic throughout. The exact contour slope in the impulsive regime is sensitive to the encounter geometry and requires dedicated numerical simulation; the slope shown for deci-hertz below the kink is therefore approximate and indicative only.

For LISA multiband observations~\cite{LISA:2024hlh}, in which the same source is tracked from the millihertz band through to merger in a ground-based detector, the observational baseline spans years. The corresponding dynamical scale $v_3 T$ exceeds the entire plotted range of $R_0$, implying that the full visible parameter space constitutes an impulsive regime for LISA: any perturber within $10^6\,\mathrm{AU}$ of the binary would interact dynamically during the multi-year baseline. This is represented by the uniform gray background in Fig.~\ref{fig:paramspace_schematic}, indicating full parameter space coverage by the LISA multiband configuration.

We note that $v_3$ does not alter the qualitative structure of the diagram but shifts the contours: a higher approach speed requires a closer or more massive perturber to accumulate the same dephasing (downward shift of the contours, indicated by the arrow in Fig.~\ref{fig:paramspace_schematic}).

The coloured bands in Fig.~\ref{fig:paramspace_schematic} indicate the characteristic mean inter-object separations in representative astrophysical environments. Current LVK observations are sensitive only to the disruption boundary at $R_0 \lesssim \mathcal{O}(10^{-1})\,\mathrm{AU}$, many orders of magnitude below the characteristic separations in any known stellar environment. Next-generation detectors operating at lower frequencies will progressively
extend the accessible separations: ET and CE at ${\sim}5\,\mathrm{Hz}$ will reach galactic-centre scales, deci-hertz instruments will probe nuclear star cluster separations, and LISA multiband observations will in principle cover the entire parameter space shown, including the mean inter-object separations
of globular cluster cores. Of particular interest is the primordial black hole (PBH) population: the purple band in Fig.~\ref{fig:paramspace_schematic} shows the mean inter-PBH separation computed from an NFW dark matter profile assuming $f_\mathrm{PBH} = 1$, evaluated for galactocentric radii $r = 0.1$--$10\,\mathrm{pc}$~\cite{Navarro:1995iw, Bertone:2004pz, Cautun:2019eaf, Klypin2002, Sofue2012, Pato:2015dua}. For different dark matter density profiles, the particle separation will vary. Moreover, a core like density profile at the galactic center can also change the mean particle separation \cite{2010AdAst2010E...5D}.   Since compact objects with masses $m_3 \lesssim 1\,M_\odot$ are not predicted by standard stellar evolution~\cite{chandra1931,Tolman:1939jz,Oppenheimer:1939ne}, a gravitational detection of sub-solar-mass perturbers through this channel would constitute unambiguous evidence for a population of primordial or exotic compact objects.

This paper is organised as follows. In Sec. \ref{sec:physical_setup}, we describe the physical setup of the three-body fly-by interaction and outline the waveform model employed in this work, including a limited validation against state-of-the-art waveform families and the definition of the analysis window. In Sec. \ref{sec:residual_cc}, we present our residual cross-correlation methodology and its application to real GW data. Section \ref{sec:constraints} presents statistical constraints on the perturber parameter space derived from noise realisations and injected signals. We discuss the astrophysical implications of our results and prospects for future detectors in Sec. \ref{sec:implications}, and we conclude in Sec. \ref{sec:conclusion}.

\section{\label{sec:physical_setup}Three-Body Fly-by Model and Waveform Construction}

In this section, we describe the physical setup, dynamical evolution, and waveform construction for compact binaries undergoing transient fly-by interactions with a third compact object. We first outline the three-body configuration and the parametrisation of the encounter (Sec.~\ref{sec:flyby_setup}), then present the equations of motion and the procedure used to construct GW signals from the perturbed binary evolution (Sec.~\ref{sec:waveform_generation}). Finally, we define an event-specific analysis window by performing a limited validation of the approximate waveform model against state-of-the-art inspiral waveforms used in the LVK analyses (Sec.~\ref{sec:validation}).

\begin{figure*}
    \centering
    \includegraphics[width=\linewidth]{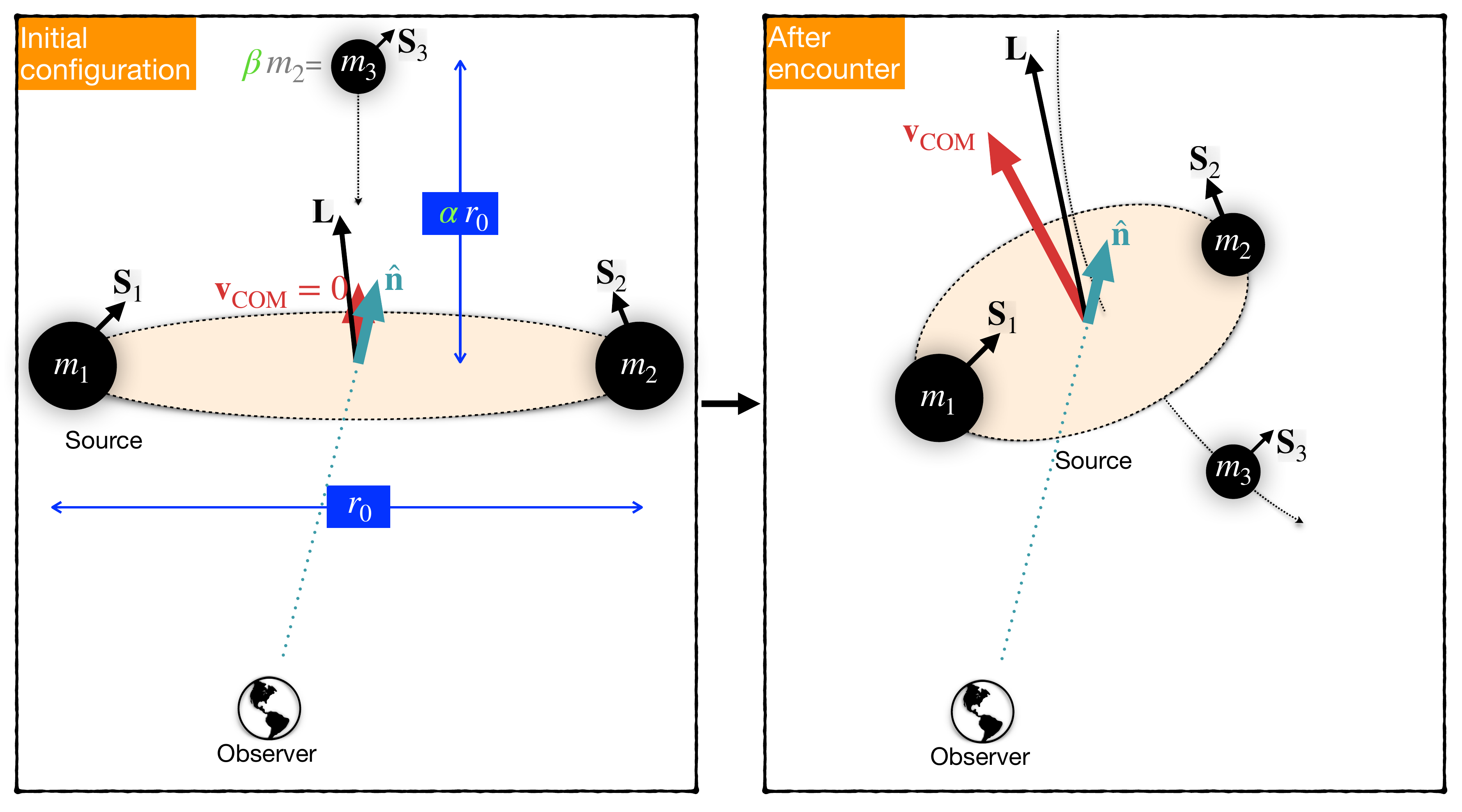}
    \caption{
    Schematic illustration of the three-body fly-by configuration considered in this work. \textit{Left:} Initial configuration, showing a circular compact binary with component masses $m_1$ and $m_2$, initial separation $r_0$, and vanishingly small velocity of the center-of-mass of the binary, approached by a third black hole of mass $m_3=\beta m_2$ located at a distance $R_0=\alpha r_0$. \textit{Right:} Configuration at a later time (after a plausible encounter), where the fly-by interaction induces a nonzero center-of-mass velocity $\mathbf{v}_{\rm COM}$ of the binary and alters its orbital geometry. GWs are observed along the line-of-sight direction $\hat{\mathbf{n}}$, leading to a Doppler modulation of the observed signal along with perturbations in the amplitude and phase of the GW signal induced by perturbation in the orbit of the binary. $\mathbf{S}_1, \mathbf{S}_2$, and $\mathbf{S}_3$ denote the spins of the three compact objects and $\mathbf{L}$ denotes the net angular momentum of the binary system. In this work, we have considered zero-spin systems in our model.
    }
    \label{fig:schematic}
\end{figure*}

A schematic overview of the three-body fly-by configuration at the initial time and during the subsequent dynamical interaction is shown in Fig.~\ref{fig:schematic}. The left panel illustrates the initial configuration at the time when the GW signal enters the detector band: a circular compact binary with component masses $m_1$ and $m_2$, separated by $r_0$, and with vanishing center-of-mass velocity, is approached by a third compact object of mass $m_3=\beta m_2$ located at a distance $R_0=\alpha r_0$ along the direction perpendicular to the binary’s orbital plane. The binary orbital angular momentum $\mathbf{L}$ is initially aligned with this axis, and the observer direction $\hat{\mathbf{n}}$ defines the line of sight. The right panel depicts a later stage of the evolution after a representative encounter. The fly-by interaction modifies the orbital geometry of the binary and induces a nonzero center-of-mass velocity $\mathbf{v}_{\rm COM}$, leading to Doppler modulation of the observed GW frequency. The perturbation also alters the instantaneous orbital separation and phase evolution, thereby imprinting amplitude and phase deviations on the GW signal. The spin vectors $\mathbf{S}_1$, $\mathbf{S}_2$, and $\mathbf{S}_3$ are shown for completeness; however, in the present analysis we restrict attention to non-spinning configurations. The perturbations also generically excite orbital eccentricity in the binary; however, the subsequent eccentricity evolution and the associated radial GW frequency modes are beyond the scope of the present analysis and are deferred to future work.

\subsection{\label{sec:flyby_setup}Physical Setup of the Fly-by Interaction}

We consider a compact binary system composed of two Schwarzschild black holes of masses $m_1$ and $m_2$ (with $m_2 \leq m_1$), initially on a circular orbit, together with a third compact object of mass $m_3$. The system is specified at an initial reference time corresponding to a GW frequency $f_0 = 20~\mathrm{Hz}$ for the binary. At this frequency, the binary separation is uniquely determined (under the assumption of circular motion) by
\begin{equation}
    r_0 = \left( \frac{G (m_1 + m_2)}{\pi^2 f_0^2} \right)^{1/3}.
\end{equation}

The initial configuration of the three-body system is defined as follows:

\begin{enumerate}
    
    \item \textbf{Binary initialization:}
    The binary components are placed on a circular orbit in the $x$--$y$ plane with separation $r_0$ and corresponding orbit velocity
    \begin{equation}
        v_0 = \sqrt{\frac{G (m_1 + m_2)}{r_0}}.
    \end{equation}
    The initial orbital phase is arbitrary.

    \item \textbf{Perturber position:}
    The third body is initialised at an initial distance
    \begin{equation}
        R_0 = \alpha \, r_0
    \end{equation}
    from the center-of-mass of the binary along $\hat{\mathbf{k}}$ such that the perturber is at $(R_0 \hat{\mathbf{k}})$, where $\alpha$ is a dimensionless parameter characterising the initial separation in units of the binary separation.

    \item \textbf{Perturber mass:}
    The perturber mass is parameterised relative to the secondary component of the binary as
    \begin{equation}
        m_3 = \beta \, m_2,
    \end{equation}
    where $\beta$ defines the perturber-to-secondary mass ratio.

    \item \textbf{Perturber velocity:}
    The perturber is assigned an initial velocity $v_3$  directed toward the binary center-of-mass along a specified axis. The approach direction is taken to be perpendicular to the binary orbital plane unless otherwise stated.

    \item \textbf{Reference frame:}  
    The system is evolved in the center-of-mass frame of the full three-body system. Initial positions and velocities are assigned such that the total linear momentum vanishes.

\end{enumerate}

In this setup, the term \textit{encounter} refers to the dynamical interaction due to gravitational force that arises from evolving this initial configuration forward in time under gravitational forces. The perturber need not pass through or closely penetrate the binary; even moderate gravitational interactions can transfer energy and angular momentum and alter the binary evolution.

All bodies are treated as point masses interacting purely through gravity. We include leading-order ($2.5$PN) radiation-reaction within the binary~\cite{Burke:1970dnm, Blanchet:2002av, Blanchet:2013haa} but neglect conservative post-Newtonian corrections, spin effects, tidal interactions, mass transfer, and hydrodynamical forces. The binary is initialised as circular, and no eccentricity is introduced by hand. In the future, a more accurate waveform for making parameter estimation of the third body can be constructed.

\subsection{\label{sec:waveform_generation}Equations of Motion and Waveform Generation}

\begin{figure*}
    \centering
    \includegraphics[width=\textwidth]{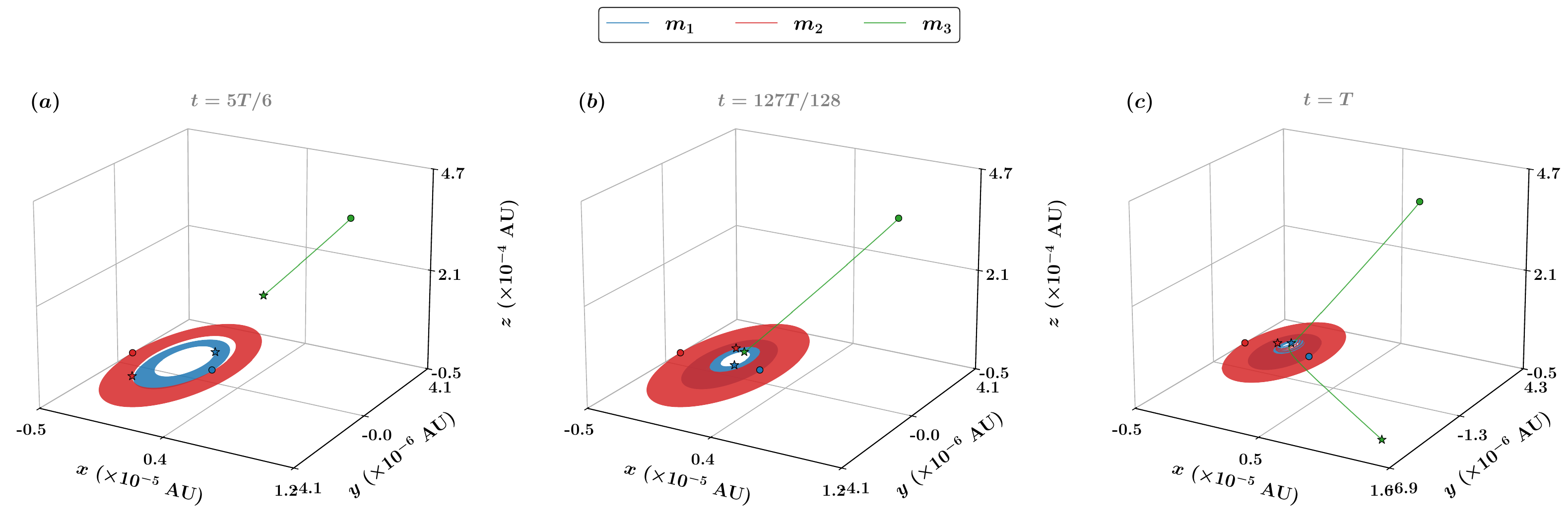}
    \caption{
    Three-dimensional snapshots of the three-body evolution for a representative configuration with $\alpha = 78.04$, $\beta = 0.1$, and $v_3 = 100~\mathrm{km\,s^{-1}}$, with the perturber initially approaching along $\hat{\mathbf{k}} = (2,0,78)/\sqrt{6088}$. The blue and red curves denote the binary components $m_1$ and $m_2$, respectively, while the green trajectory corresponds to the perturber $m_3$. The panels $(a)$, $(b)$, and $(c)$ show the configuration at times $t = 5T/6$, $t = 127T/128$, and $t = T$, respectively, where $T$ is the time at which the binary reaches ISCO. The solid dots denote the initial configuration, and the solid stars denote the configuration at time $t$ in each panel. The fly-by interaction induces a transient distortion of the binary orbit and a displacement of the center-of-mass, while the system remains bound.
    }
    \label{fig:trajectory_snapshots}
\end{figure*}

\begin{figure*}
    \centering
    \includegraphics[width=\textwidth]{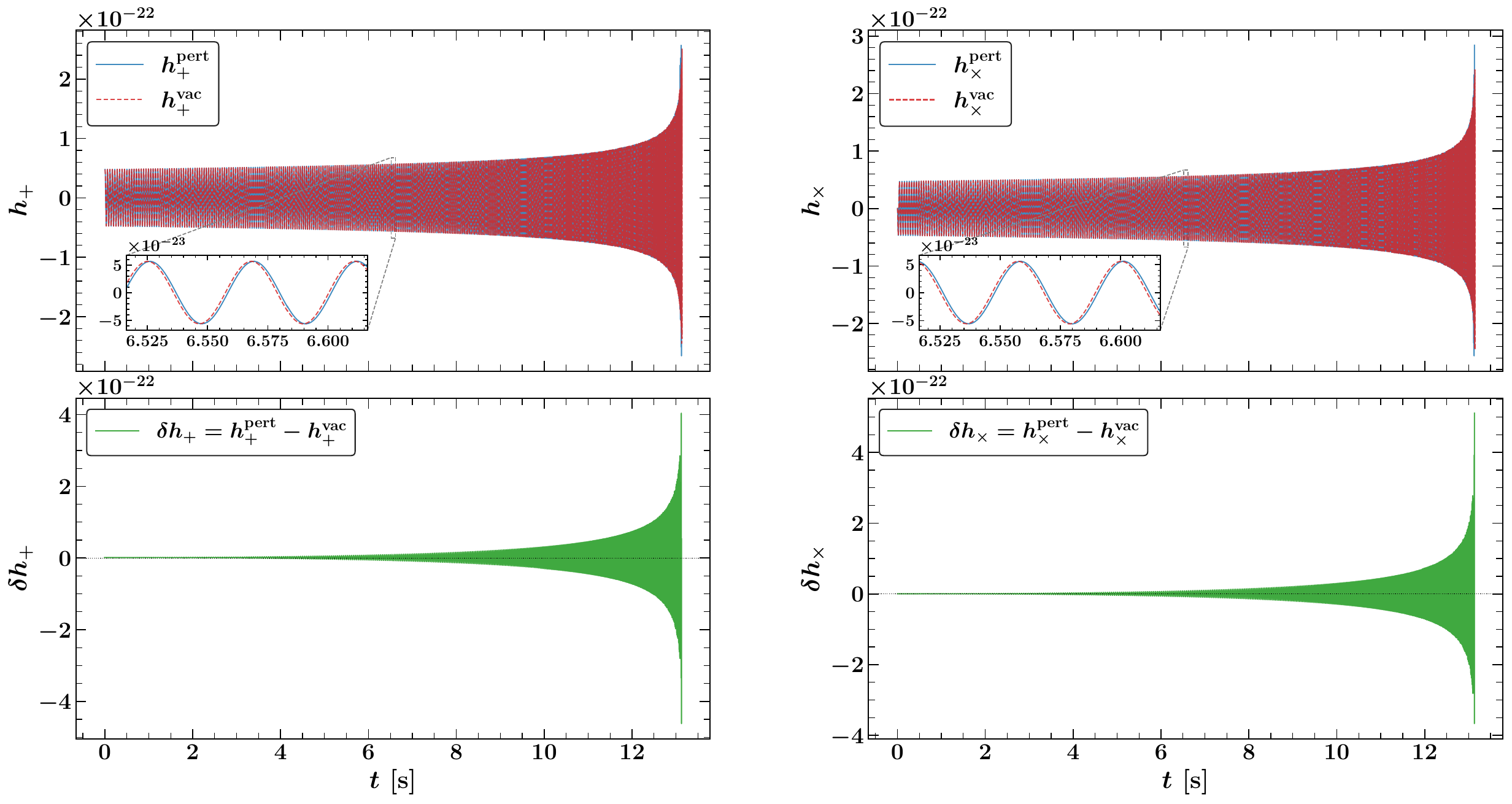}
    \caption{
    Comparison between the perturbed ($h^{\mathrm{pert}}$) and vacuum ($h^{\mathrm{vac}}$) GW polarisations for the same configuration shown in Fig.~\ref{fig:trajectory_snapshots}. The upper panels show the plus and cross polarisations evolved up to ISCO, while the lower panels display the residual $\delta h = h^{\mathrm{pert}} - h^{\mathrm{vac}}$. The fly-by interaction produces a cumulative phase deformation that becomes more pronounced toward late times.
    }
    \label{fig:waveform_example}
\end{figure*}

The dynamical evolution of the system is obtained by numerically integrating the equations of motion for the three bodies under their mutual gravitational interaction. The binary components are evolved, including leading-order radiation-reaction effects, while the interaction with the third body is treated purely at the Newtonian level.

Let $\mathbf{r}_i$ and $\mathbf{v}_i$ denote the position and velocity of body $i$ in the center-of-mass frame of the three-body system. We define the relative separation and velocity of the binary as
\begin{equation}
    \mathbf{r} = \mathbf{r}_1 - \mathbf{r}_2, \qquad
    \mathbf{v} = \mathbf{v}_1 - \mathbf{v}_2,
\end{equation}
with $r = |\mathbf{r}|$. The equations of motion for the three bodies are given by
\begin{align}
    \ddot{\mathbf{r}}_1 &= - G \frac{m_2}{r^3} \mathbf{r}
    - G \frac{m_3}{|\mathbf{r}_1 - \mathbf{r}_3|^3} (\mathbf{r}_1 - \mathbf{r}_3)
    + \mathbf{a}^{\mathrm{2.5PN}}_1, \label{eq:a1_full} \\
    \ddot{\mathbf{r}}_2 &= + G \frac{m_1}{r^3} \mathbf{r}
    - G \frac{m_3}{|\mathbf{r}_2 - \mathbf{r}_3|^3} (\mathbf{r}_2 - \mathbf{r}_3)
    + \mathbf{a}^{\mathrm{2.5PN}}_2, \label{eq:a2_full} \\
    \ddot{\mathbf{r}}_3 &= 
    G \frac{m_1}{|\mathbf{r}_3 - \mathbf{r}_1|^3} (\mathbf{r}_1 - \mathbf{r}_3)
    + G \frac{m_2}{|\mathbf{r}_3 - \mathbf{r}_2|^3} (\mathbf{r}_2 - \mathbf{r}_3),
    \label{eq:a3_full}
\end{align}
where $\mathbf{a}^{\mathrm{2.5PN}}_{1,2}$ denote the leading-order radiation-reaction accelerations acting on the binary components.

The explicit form of the 2.5PN radiation-reaction acceleration is given by~\cite{Blanchet:2002av, Blanchet:2013haa}
\begin{align}
    \mathbf{a}^{\mathrm{2.5PN}}_1 &=
    \frac{4}{5c^5} \frac{G^2 m_1 m_2}{r^3}
    \Bigg[
        \left(
            \frac{2G}{r} (m_1 - 4 m_2) - v^2
        \right) \mathbf{v}
        \nonumber \\
        &\hspace{1.2cm}
        + \frac{\mathbf{r}\cdot\mathbf{v}}{r}
        \left(
            \frac{52 G m_2}{3r}
            - \frac{6 G m_1}{r}
            + 3 v^2
        \right)
        \frac{\mathbf{r}}{r}
    \Bigg], \label{eq:a25_1}
\end{align}
and
\begin{align}
    \mathbf{a}^{\mathrm{2.5PN}}_2 &=
    -\frac{4}{5c^5} \frac{G^2 m_1 m_2}{r^3}
    \Bigg[
        \left(
            \frac{2G}{r} (m_2 - 4 m_1) - v^2
        \right) \mathbf{v}
        \nonumber \\
        &\hspace{1.2cm}
        + \frac{\mathbf{r}\cdot\mathbf{v}}{r}
        \left(
            \frac{52 G m_1}{3r}
            - \frac{6 G m_2}{r}
            + 3 v^2
        \right)
        \frac{\mathbf{r}}{r}
    \Bigg], \label{eq:a25_2}
\end{align}
where $v = |\mathbf{v}|$ and $c$ is the speed of light. No conservative post-Newtonian corrections are included; radiation-reaction acts only within the compact binary, and no dissipative terms are applied to the perturber.

The equations of motion are integrated numerically using the \texttt{solve\_ivp} routine from the \texttt{SciPy} library, employing adaptive, high-order ODE solvers. Specifically, we use 
the Dormand--Prince method \texttt{DOP853} which is an Explicit Runge-Kutta method of order 8~\cite{HairerNorsettWanner1993, hairer_software, 2020SciPy-NMeth}
. The integration is performed with relative and absolute tolerances $\mathtt{rtol}=10^{-12}$ and $\mathtt{atol}=10^{-14}$, respectively. The system is evolved forward in time until one of the following termination conditions is satisfied:

\begin{enumerate}

    \item \textbf{Binary reaches ISCO:}  
    The binary separation decreases to the innermost stable circular orbit (ISCO). These configurations are classified as \textit{surviving binaries}.

    \item \textbf{Binary disruption:}  
    The total mechanical energy of the binary becomes positive indicating that the binary has been broken due to energy transfer from the interaction. These cases are classified as \textit{disrupted} or \textit{broken binaries}.

    \item \textbf{Direct plunge:}  
    Any compact object approaches within a few Schwarzschild radii of another object.

    \item \textbf{Breakdown of dynamical model:}  
    The velocity of any object becomes comparable to the speed of light.
    In this regime, the Newtonian three-body treatment supplemented by leading-order radiation-reaction is no longer self-consistent, as higher-order relativistic corrections to both conservative and dissipative dynamics become comparable to the retained terms which no more remain the only leading-order terms. Such configurations are therefore excluded as lying outside the regime of validity of our model.

\end{enumerate}
This classification makes a clear separation between physically disrupted systems, binaries that survive during their full inspiral, and configurations in which the adopted dynamical approximation breaks down.
The evolution is terminated when the binary separation reaches the innermost stable circular orbit (ISCO), implemented via an event condition corresponding to $r = r_{\mathrm{ISCO}}$. In addition, the integration is terminated if the binary becomes unbound, if any body in the system attains velocities approaching the speed of light, or if a direct plunge or other physical inconsistency occurs, as mentioned above. The solution is then interpolated onto a uniform time grid using cubic spline interpolation for subsequent waveform construction~\cite{deBoor1978PracticalGuideSplines}.

At each time step, the instantaneous GW frequency of the binary is computed from the relative separation as 
\begin{equation}
    f(t) = \frac{1}{\pi}
    \sqrt{\frac{G (m_1 + m_2)}{|\mathbf{r}(t)|^3}} .
\end{equation}
Here, we have taken the GW frequency to be twice the orbital frequency as we do not consider any higher harmonics or excitations in different fundamental frequency modes in this work. To account for the motion of the binary center-of-mass induced by the fly-by interaction, we apply a Doppler correction to the frequency due to the line-of sight velocity of the binary. Let $\mathbf{v}_{\mathrm{COM}}(t)$ denote the velocity of the binary center-of-mass and $\hat{\mathbf{n}}$ the unit vector pointing from the observer to the source. The observed frequency is taken to be
\begin{equation}
    f_{\mathrm{obs}}(t) =
    f(t) \sqrt{\frac{1 - \mathbf{v}_{\mathrm{COM}}(t)\cdot\hat{\mathbf{n}}/c}
                     {1 + \mathbf{v}_{\mathrm{COM}}(t)\cdot\hat{\mathbf{n}}/c}} .
\end{equation}

The GW phase is obtained by numerically integrating
\begin{equation}
    \frac{d\Phi}{dt} = 2\pi f_{\mathrm{obs}}(t),
\end{equation}
with an arbitrary reference phase $\Phi_0$. Finally, the GW strain is constructed using the leading-order quadrupole approximation. The plus and cross polarisations are given by~\cite{Maggiore:2007ulw}
\begin{align}
    h_+(t) &= \frac{1}{D} \, A(t) \, \frac{1 + \cos^2\iota}{2} \cos\Phi(t), \\
    h_\times(t) &= \frac{1}{D} \, A(t) \, \cos\iota \, \sin\Phi(t),
\end{align}
where $D$ is the luminosity distance to the source, $\iota$ is the inclination angle defined as the angle between the line-of-sight from the observer to the source and the orbital angular momentum of the binary. A(t) is the amplitude factor given by
\begin{equation}
    A(t) =
    4 \left( \frac{G \mathcal{M}}{c^2} \right)^{5/3}
    \left( \frac{\pi f_{\mathrm{obs}}(t)}{c} \right)^{2/3},
\end{equation}
with $\mathcal{M}$ denoting the chirp mass of the binary. This construction yields a time-domain waveform that consistently incorporates perturbative three-body effects through the modified orbital evolution.


To illustrate the dynamical behaviour of the system and the corresponding waveform deformation, we present a representative example corresponding to parameters motivated by GW230627\_015337~\cite{gwtc4Catalog}. In this example, we set $\beta = 0.1$, $\alpha = 78.04$, and an initial perturber velocity $v_3 = 100~\mathrm{km\,s^{-1}}$. The perturber is initialised along the direction $\hat{\mathbf{k}} = \frac{1}{\sqrt{6088}} ( 2\,\hat{\mathbf{x}} + 78\,\hat{\mathbf{z}} ),$ and moves towards the center-of-mass if the binary along this direction.

Figure~\ref{fig:trajectory_snapshots} shows three snapshots of the three-body trajectories during the evolution. The panels correspond to times $t = 5T/6$, $t = 127T/128$, and $t = T$, where $T$ denotes the total evolution time until the binary reaches ISCO. The binary components remain bound throughout the interaction, while the perturber approaches from a distance of $4.18\times10^{-4}$ AU, induces a distortion in the binary orbit near closest approach, and subsequently recedes. The perturbation produces both a displacement of the binary center-of-mass and a transient modification of the orbital geometry, without breaking the system.

The resulting GW polarisations are shown in Fig.~\ref{fig:waveform_example}, where we compare the full perturbed waveform (evolved up to ISCO) denoted by $h^{\rm pert}$ with the corresponding vacuum waveform denoted by $h^{\rm vac}$ for both the polarization plus ($h_+$) and cross ($h_\times$). The upper panels display the full waveforms, while the lower panels show the residuals
\begin{equation}
    \delta h_{+,\times}(t) = h^{\mathrm{pert}}_{+,\times}(t) - h^{\mathrm{vac}}_{+,\times}(t).
\end{equation}

Although the overall morphology of the perturbed and vacuum signals remains similar, the fly-by interaction induces a cumulative phase shift that grows toward late times. The residual amplitude increases monotonically as the binary approaches ISCO, reflecting the integrated dynamical effect of the three-body interaction.

As the dynamical evolution of the binary system includes only Newtonian conservative interactions supplemented by leading-order ($2.5$PN) radiation-reaction for the compact binary, conservative post-Newtonian corrections (1PN, 2PN, etc.) and higher-order dissipative terms are neglected. So the late inspiral and strongly relativistic regimes are not modelled with full accuracy. To control the impact of these omissions, we do not use the entire simulated inspiral in the data analysis. Instead, as described in Sec.~\ref{sec:validation}, we perform a mismatch-based validation against state-of-the-art waveform models and restrict all residual cross-correlation analyses to the portion of the signal over which our approximate model waveform remains consistent with the reference models given a particular mismatch threshold. This procedure ensures that constraints are derived only from the regime where higher-order post-Newtonian effects do not dominate the dynamics and the adopted approximation remains self-consistent.

\subsection{\label{sec:validation}Waveform Validation and Analysis Window}

The waveform model described above is approximate by construction, as it retains only Newtonian conservative dynamics supplemented by leading-order ($2.5$PN) radiation-reaction and quadrupolar GW emission. Higher-order post-Newtonian corrections, higher harmonics, spin effects, and merger-ringdown dynamics are not included. It is therefore necessary to determine the portion of the inspiral over which the model remains internally consistent with state-of-the-art inspiral descriptions which are deployed in the LVK search and inference pipelines.

To assess this, we perform a systematic waveform-consistency study for a small set of representative mass configurations. The mass choices are motivated by the specific events selected from GWTC-4.0 that are later used in our residual cross-correlation analysis. These configurations provide concrete reference systems for determining the portion of the inspiral over which the approximate waveform model remains consistent with state-of-the-art inspiral descriptions.

\begin{figure*}
    \centering
    \begin{minipage}{0.32\textwidth}
        \centering
        \includegraphics[width=\linewidth]{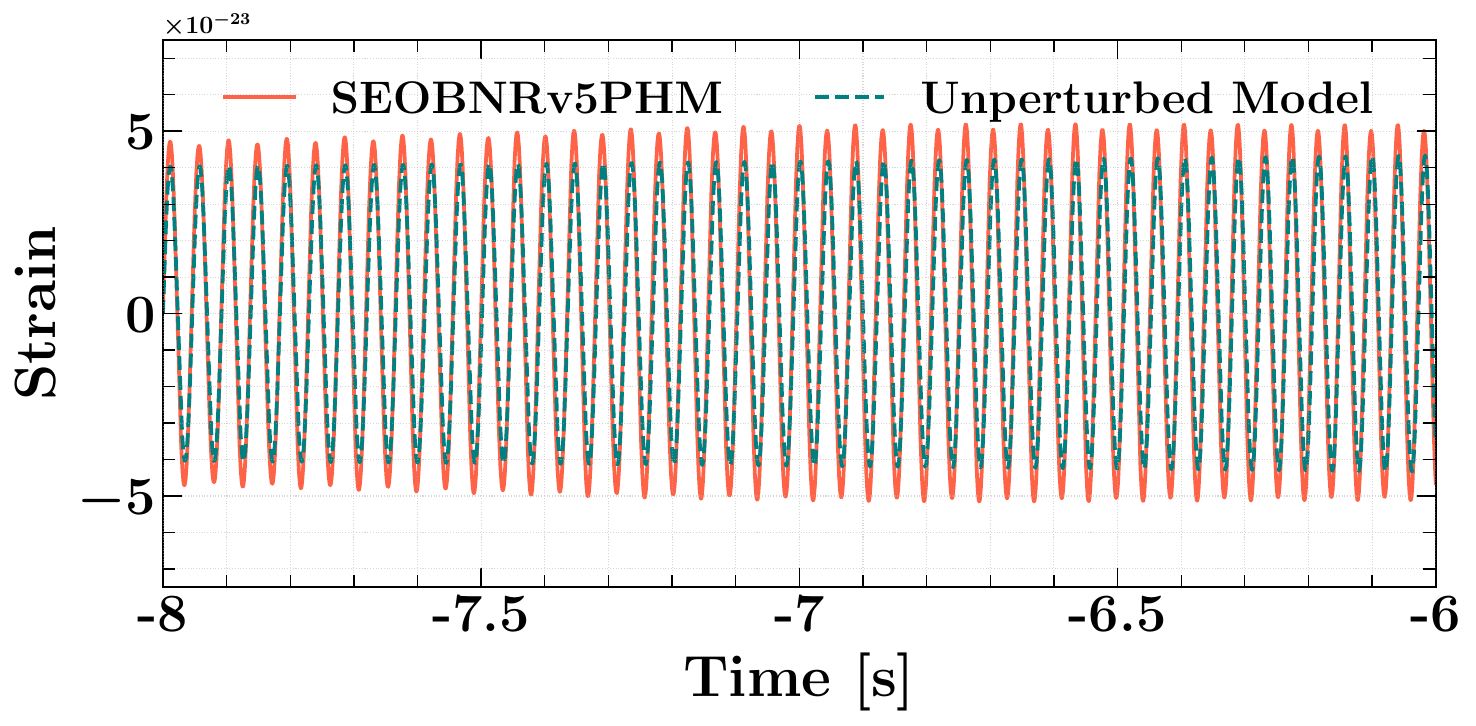}
        \vspace{0.3em}
        {\small (a) GW230627\_015337-like}
    \end{minipage}
    \hfill
    \begin{minipage}{0.32\textwidth}
        \centering
        \includegraphics[width=\linewidth]{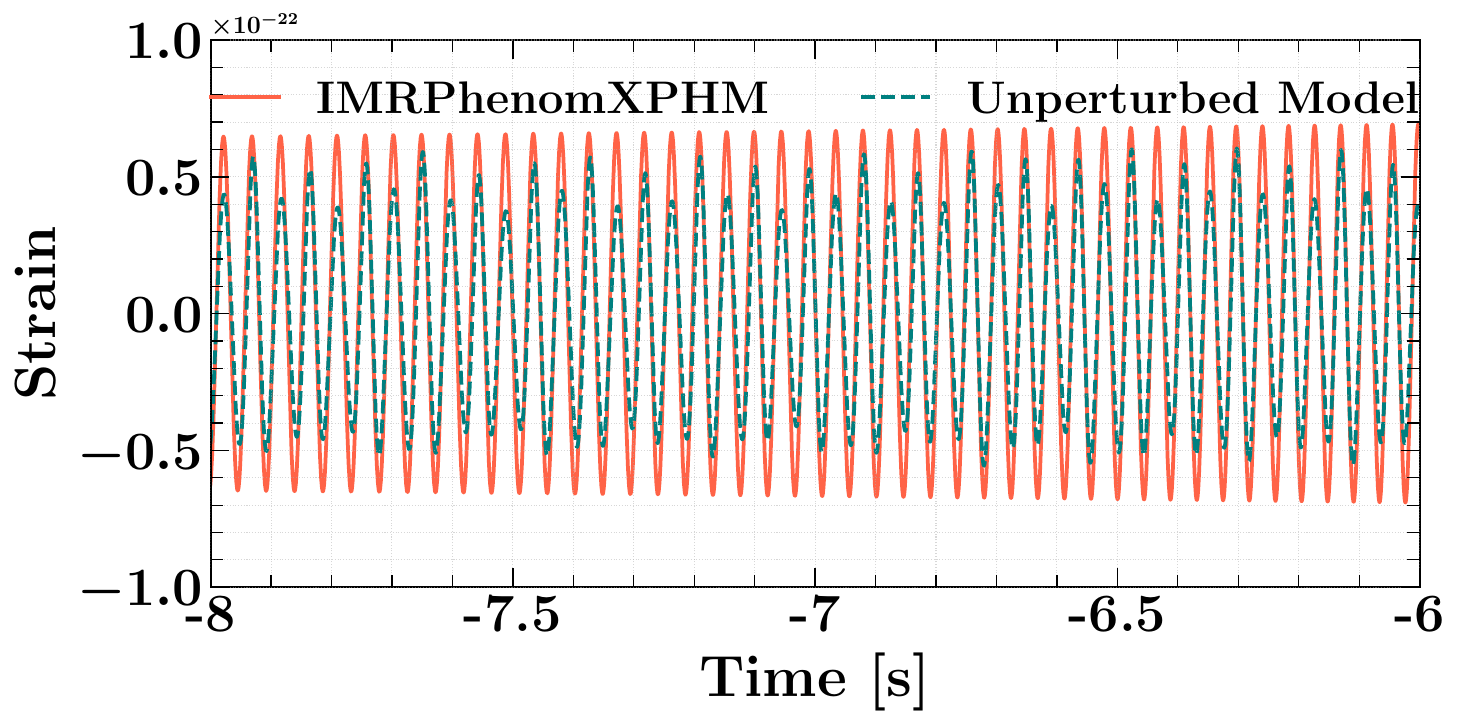}
        \vspace{0.3em}
        {\small (b) GW190814-like}
    \end{minipage}
    \hfill
    \begin{minipage}{0.32\textwidth}
        \centering
        \includegraphics[width=\linewidth]{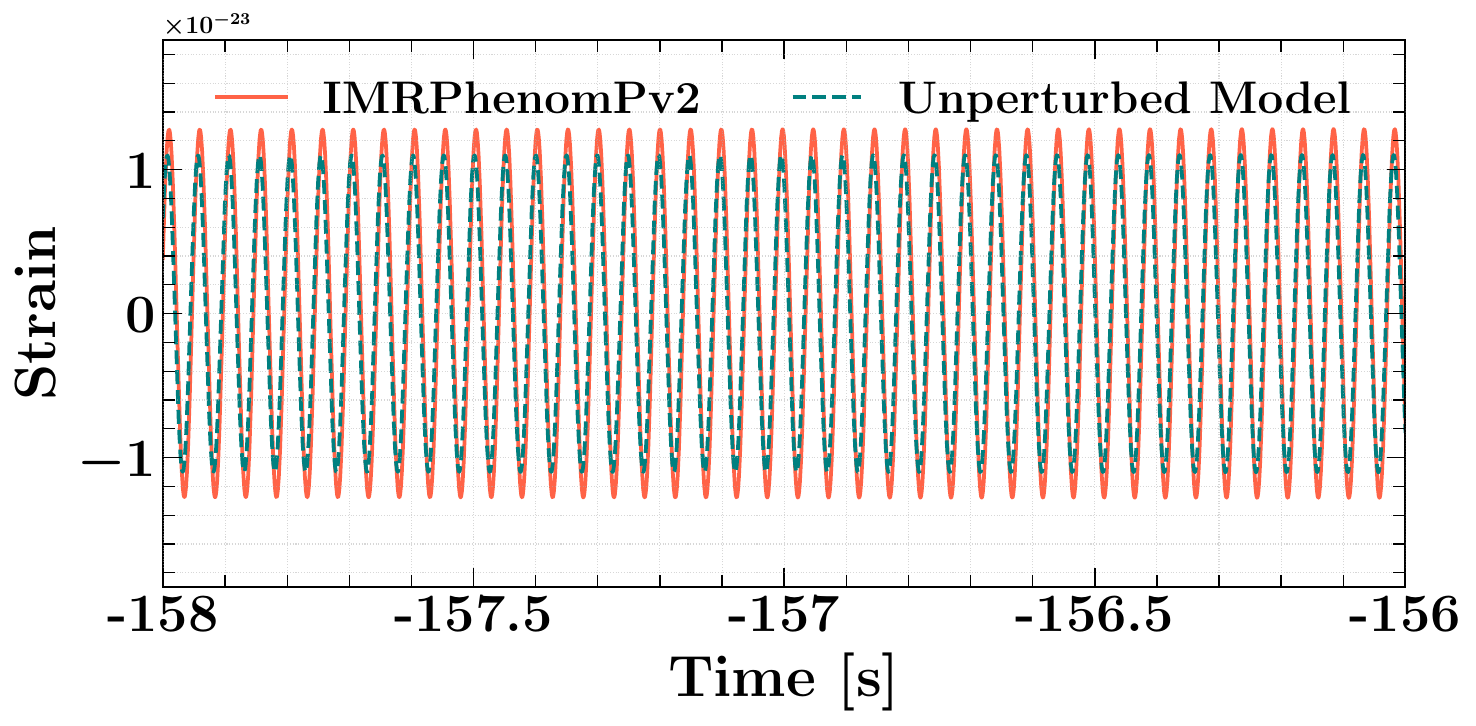}
        \vspace{0.3em}
        {\small (c) GW170817-like}
    \end{minipage}
    \caption{
    Time-domain comparison between the reference waveform used in the LVK parameter-estimation analysis and the corresponding unperturbed waveform constructed in this work for three representative mass configurations motivated by the events: GW170817 (right): $m_1=1.48M_\odot,\,m_2=1.27M_\odot$, GW190814 (center): $m_1=24.44M_\odot,\,m_2=2.73M_\odot$, and GW230627\_015337 (left): $m_1=9.65M_\odot,\,m_2=5.70M_\odot$. In all cases, the phase evolution remains in close agreement within the selected analysis window, while a small but systematic amplitude mismatch is visible due to the use of a leading-order quadrupole approximation in our model. This residual amplitude difference reflects the absence of higher-order post-Newtonian amplitude corrections and higher harmonics in the present construction, being most significant for the most asymmetric-system GW190814. The leading-order gravitational waveform shows an oscillating nature in amplitude for GW190814 because of the asymmetry in the masses and the non-monotonic decay of the orbital separation on small time scales within an orbital period. This figure is illustrative and is included to visualise the level of agreement motivating the choice of the analysis window described in Sec.~\ref{sec:validation}.
    }
    \label{fig:waveform_comparison}
\end{figure*}

For each representative mass configuration, we generate:
(i) an unperturbed waveform using the dynamical model developed in Sec.~\ref{sec:physical_setup}, and 
(ii) a reference waveform from a state-of-the-art inspiral-merger-ringdown model appropriate for that mass scale (e.g., \texttt{IMRPhenomPv2}, \texttt{IMRPhenomXPHM}, or \texttt{SEOBNRv5PHM}~\cite{IMRPhenomPv, IMRPhenomXPHM, SEOBNRv5PHM}). Figure~\ref{fig:waveform_comparison} shows a time-domain visual comparison between the reference waveform used in the LVK analysis and the corresponding unperturbed waveform constructed in this work for three representative systems.

The comparison is carried out in the time domain. To determine the region of agreement, we systematically truncate both waveforms from the end of the signal and compute their mutual match, maximised over relative time and phase shifts. The match $\mathcal{M}$ between two waveforms $h_1$ and $h_2$ is defined as
\begin{equation}
    \mathcal{M} = \max_{\Delta t,\,\Delta\phi}\,\mathcal{O}\left(h_1, h_2\right), \label{eq:match_def}
\end{equation}
where $\mathcal{O}(h_1,h_2)$ denotes the normalised overlap,
\begin{equation}
    \mathcal{O}(h_1,h_2) =
    \frac{\langle h_1, h_2 \rangle}
         {\sqrt{\langle h_1, h_1 \rangle \langle h_2, h_2 \rangle}},
\end{equation}
with the inner product
\begin{equation}
    \langle h_1, h_2 \rangle = 4\,\mathrm{Re}\int_{f_\mathrm{low}}^{f_\mathrm{high}} \frac{\tilde{h}_1(f)\,\tilde{h}_2^*(f)}{S_n(f)}\,df \label{eq:inner_prod_def}.
\end{equation}
where $S_n(f)$ is the one-sided power spectral density (PSD) of the noise. In practice, the waveforms are bandpassed and the integration limits are taken to be $f_\mathrm{low}$ and $f_\mathrm{high}$, which are taken to be $20~\mathrm{Hz}$ and twice the ISCO frequency, respectively. For the purpose of waveform validation, the inner product is not weighted by any detector noise PSD. This choice isolates the intrinsic agreement between waveform models and avoids mixing modelling discrepancies with detector sensitivity. The truncation-based procedure identifies the earliest portion of the inspiral over which the approximate waveform remains consistent with the reference description.

We adopt a conservative mismatch threshold of $1 - \mathcal{M} < 0.03$ to define the validated region of the inspiral. The time interval over which this condition is satisfied defines the analysis window used in subsequent sections. All constraint calculations are restricted to this window to ensure that conclusions are derived only from the regime in which the approximate waveform model remains self-consistent. This validation procedure does not constitute a full agreement across the inspiral–merger–ringdown regime. Instead, it provides a controlled and conservative restriction of the analysis to the portion of the signal where higher-order post-Newtonian effects do not dominate and the adopted approximation remains reliable. This validation procedure is sufficient for the present study for two reasons. First, our analysis is based on a residual cross-correlation statistic rather than direct matched-filtering based \textit{search} with the approximate waveform model. The purpose of the validation is therefore not to ensure parameter-estimation accuracy, but to guarantee that the waveform used for obtaining the residual does not introduce spurious residual structure at a level comparable to or larger than the perturbations being probed. By restricting the analysis to the region where the intrinsic vacuum mismatch remains below $3\%$, we ensure that any residual power is dominated by genuine three-body perturbations rather than modelling systematics of the baseline inspiral description used in the LVK analyses. Second, within the validated window, the phase evolution of the approximate model tracks the reference waveform sufficiently closely that cumulative dephasing from neglected higher-order post-Newtonian terms remains subdominant compared to the perturbations induced by the fly-by interaction over the same interval. Since the residual cross-correlation statistic is sensitive primarily to coherent phase deviations accumulated over many cycles, controlling the intrinsic vacuum mismatch to this level ensures that the statistic responds to dynamical perturbations rather than systematic modelling errors. In this sense, the validation threshold defines a regime in which the hierarchy of effects is preserved: higher-order vacuum corrections remain smaller than, or at most comparable to, the environmental perturbations being constrained.

\section{\label{sec:residual_cc}Search for the signal from GWTC-4}

In this section, we describe the residual cross-correlation statistic used to search for coherent excess power between the LIGO Hanford (H1) and Livingston (L1) detectors after subtracting the best-fit vacuum waveform from the data. At this stage, no three-body waveforms are involved; the goal is solely to assess whether any statistically significant residual correlation between different detectors remains in the data beyond what is expected from instrumental noise.

\subsection{Data conditioning and template subtraction}

We first select those events from GWTC-4.0 which have individual detector SNR $\gtrsim10$ and an in-band signal duration of atleast 10 s given a reference frequency of 20Hz. The events that qualify these conditions are: GW170817, GW190814, and GW230627\_015337, and we restrict our analyses to only these events. For each event, we analyse strain time series from the H1 and L1 detectors centered on the maximum-likelihood geocentric time reported by the LVK Collaboration~\cite{GWOSC_GWTC4_EventAPI}. The data are bandpassed between 20 Hz and 512 Hz for the two events: GW190814 and GW230627\_015337, and between 20Hz and 1592 Hz for GW170817. The segment length is chosen to be substantially longer than the signal duration to avoid edge effects. Since the event GW190425 \cite{LIGOScientific:2020aai}, with an in-band duration greater than 10 seconds, was a single detector detection, we have not considered this event in this work.

The best-fit vacuum waveform is generated using the maximum-likelihood source parameters and the waveform family employed in the corresponding LVK parameter-estimation analysis~\cite{LVK_GWTC4_PE_2025}. The waveform is projected onto each detector using the appropriate antenna response functions~\cite{Finn:1992xs}, accounting for the sky location and polarisation angle.

Both the data and the template are whitened using the one-sided noise PSD estimated from data surrounding the event. The whitened time series are then filtered using high-pass and low-pass finite impulse response filters consistent with the frequency band of the analysis. After whitening and filtering, the template is subtracted from the data to obtain the residual strain time series,
\begin{equation}\label{eq:res}
    r_I(t) = d_I(t) - h_I(t),
\end{equation}
where $d_I(t)$ and $h_I(t)$ denote the whitened strain data and best-fit template in detector
$I \in \{\mathrm{H1}, \mathrm{L1}\}$, respectively.

\subsection{Cross-correlation estimator}

Using Eq. \eqref{eq:res}, we search for coherent excess power between the Hanford (H1) and Livingston (L1) residual time series. Cross-correlation based statistics have previously been used in GW data analysis to identify coherent signals between spatially separated detectors, particularly in the context of CBC events, stochastic background searches and unmodelled burst analyses~\cite{Allen:1997ad, LIGOScientific:2007gwp, LIGOScientific:2016jlg, Klimenko:2008fu, Klimenko:2015ypf, LIGOScientific:2019gaw, Dideron:2022tap, Chakraborty:2024mbr, Chakraborty:2024net, Dideron:2024xwm, Chakraborty:2025maj}. Here we adapt this framework to search for correlated residual power following template subtraction.

Let $r_{\mathrm{H1}}(t)$ and $r_{\mathrm{L1}}(t)$ denote the whitened and bandpassed residual strain time series in the two detectors. The detectors are first aligned by applying the relative time delay $\Delta t_{\mathrm{H1L1}}$ between the detectors inferred from the maximum-likelihood sky location of the event. A relative sign flip is applied to the data from one of the detectors to account for the differing detector orientations.

We define a short-time cross-correlation estimator,
\begin{equation}
    Y(t) = \frac{1}{\tau}
    \int_{t-\tau/2}^{t+\tau/2}
    r_{\mathrm{H1}}(t')
    \, r_{\mathrm{L1}}(t' + \Delta t_{\mathrm{H1L1}})
    \, dt',
    \label{eq:Yt}
\end{equation}
where $\tau$ is the integration window~\cite{Dideron:2022tap, Dideron:2024xwm}. For stationary, uncorrelated Gaussian noise in the two detectors, the expectation value satisfies of $Y(t)$ should be consistent with zero in the absence of any correlated physical signal.To quantify the accumulation of coherent residual power over the inspiral, we define the cumulative cross-correlation statistic,
\begin{equation}
    C(t) = \int_{t_0}^{t} Y(t') \, dt',
    \label{eq:Ct}
\end{equation}
where $t_0$ denotes the start of the analysed data segment. In discretized form, $C(t)$ is computed as a cumulative sum over time samples.

Under the null hypothesis of independent detector-noise, $C(t)$ behaves as a random walk with zero mean. The presence of a coherent residual signal $\delta h(t)$ common to both detectors introduces a nonzero mean contribution,
\begin{equation}
    \langle C(t) \rangle
    \sim \int_{t_0}^{t}
    \delta h_{\mathrm{H1}}(t') \,
    \delta h_{\mathrm{L1}}(t') \,
    dt',
\end{equation}
leading to systematic growth of $C(t)$ relative to the noise expectation.

We therefore define a normalised residual cross-correlation signal-to-noise ratio (SNR),
\begin{equation}
    \mathcal{S}_{\mathrm{res}}(T)
    = \frac{C(T)}{\sigma_{C}(T)},
\end{equation}
where $\sigma_C(T)$ denotes the standard deviation of the cumulative cross-correlation under the noise-only hypothesis estimated from segments of real data released by the LVK collaboration. In the limit of stationary Gaussian noise, $\mathcal{S}_{\mathrm{res}}$ follows an approximately standard normal distribution in the absence of correlated residual power.

This approach offers a key advantage over model-dependent searches: it does not assume any specific morphology for the three-body perturbation. Instead, it tests for the generic presence of coherent residual structure between detectors after subtraction of the best-fit vacuum waveform. As a result, the method remains sensitive to a broad class of deviations from the isolated two-body inspiral model, including perturbations that may not be captured by a finite template bank of three-body waveforms.

To assess statistical significance, we estimate the variance of the cumulative cross-correlation statistic directly from segments of real detector data that do not contain astrophysical signals. These off-source segments are processed identically (whitening and filtering) to construct an empirical distribution of $C(t)$ under the noise-only hypothesis.

\begin{figure*}
    \centering
    \begin{minipage}[t]{0.32\textwidth}
        \centering
        \includegraphics[width=\linewidth]{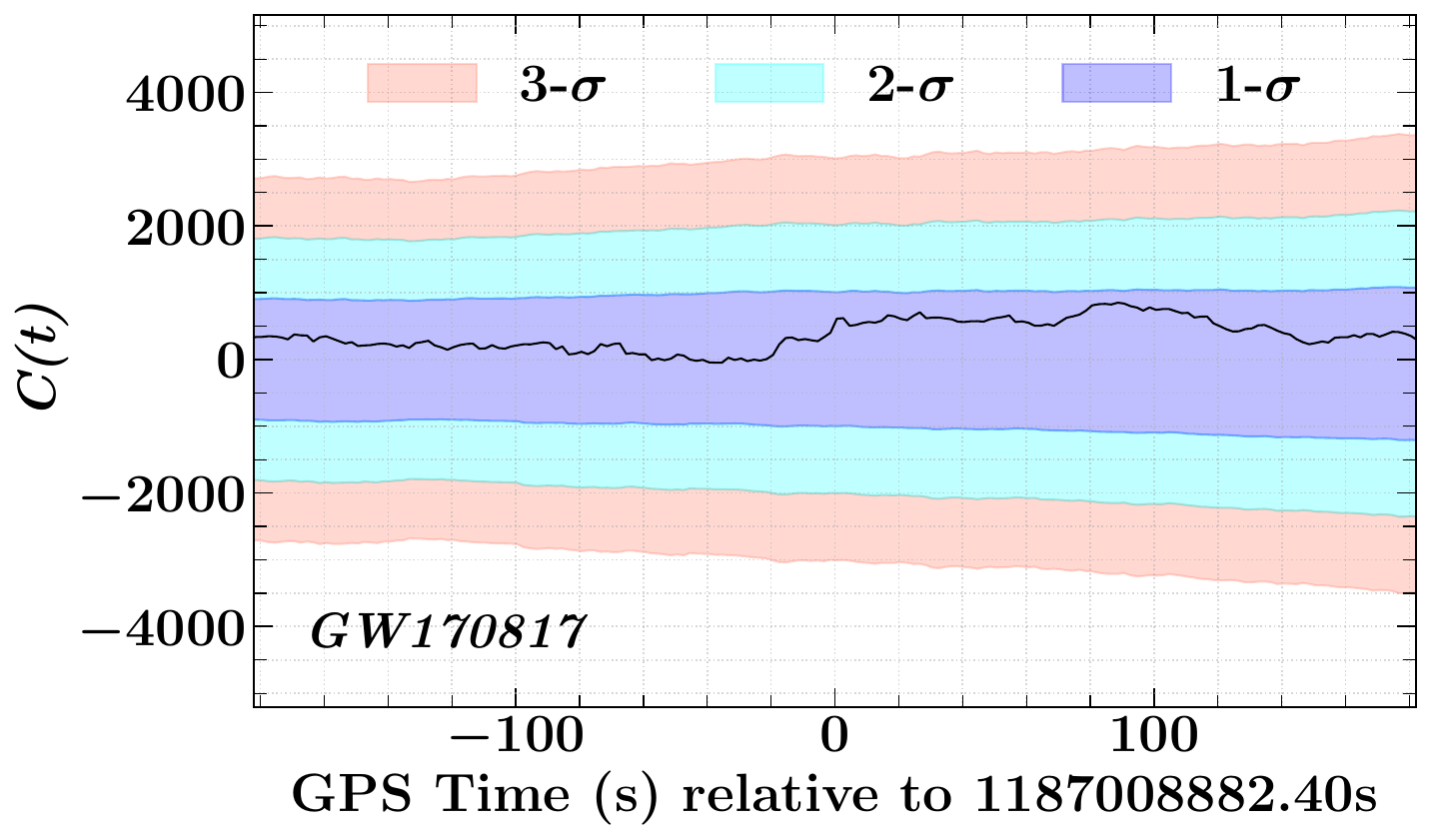}
    \end{minipage}
    \hfill
    \begin{minipage}[t]{0.32\textwidth}
        \centering
        \includegraphics[width=\linewidth]{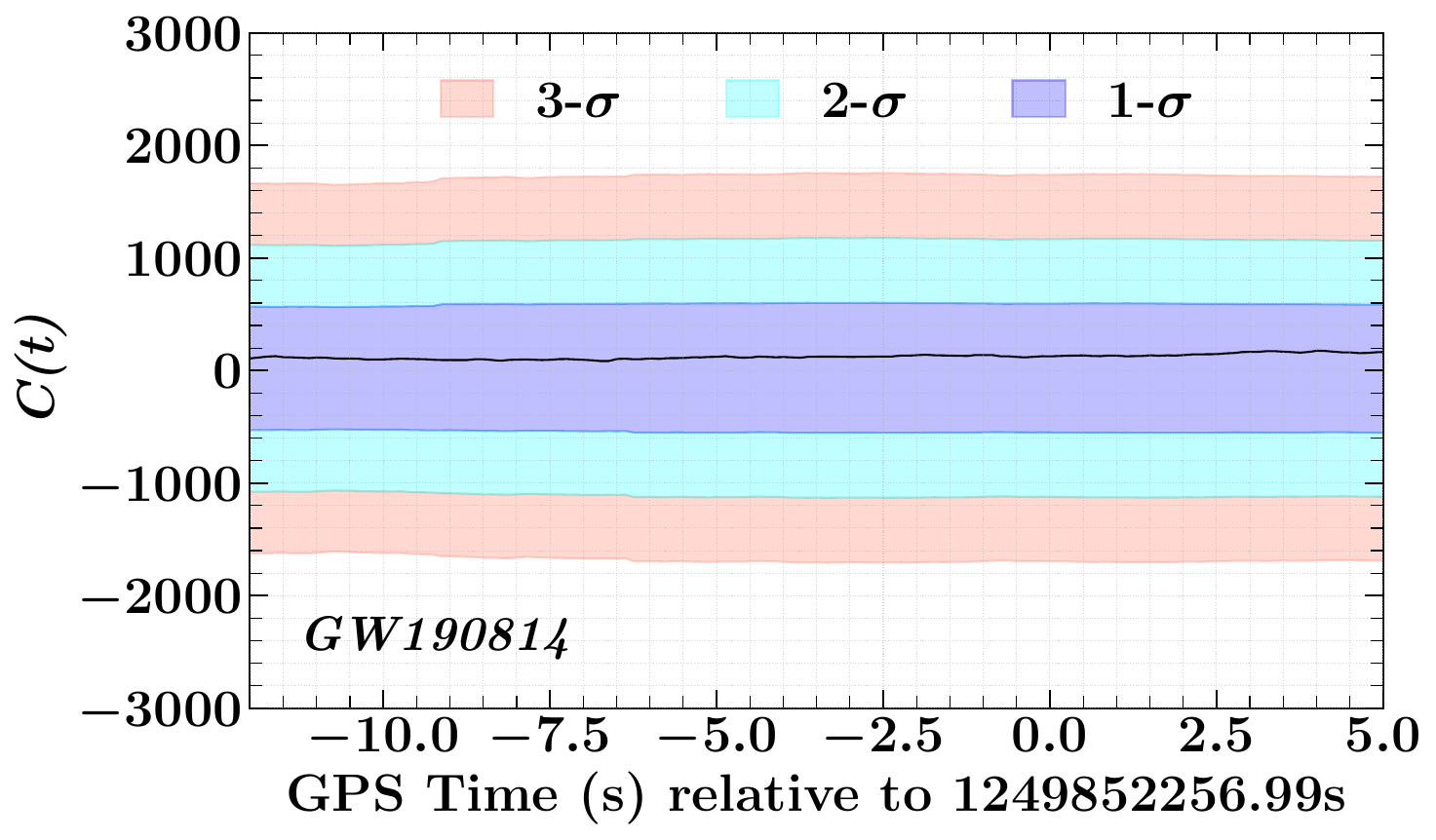}
    \end{minipage}
    \hfill
    \begin{minipage}[t]{0.32\textwidth}
        \centering
        \includegraphics[width=\linewidth]{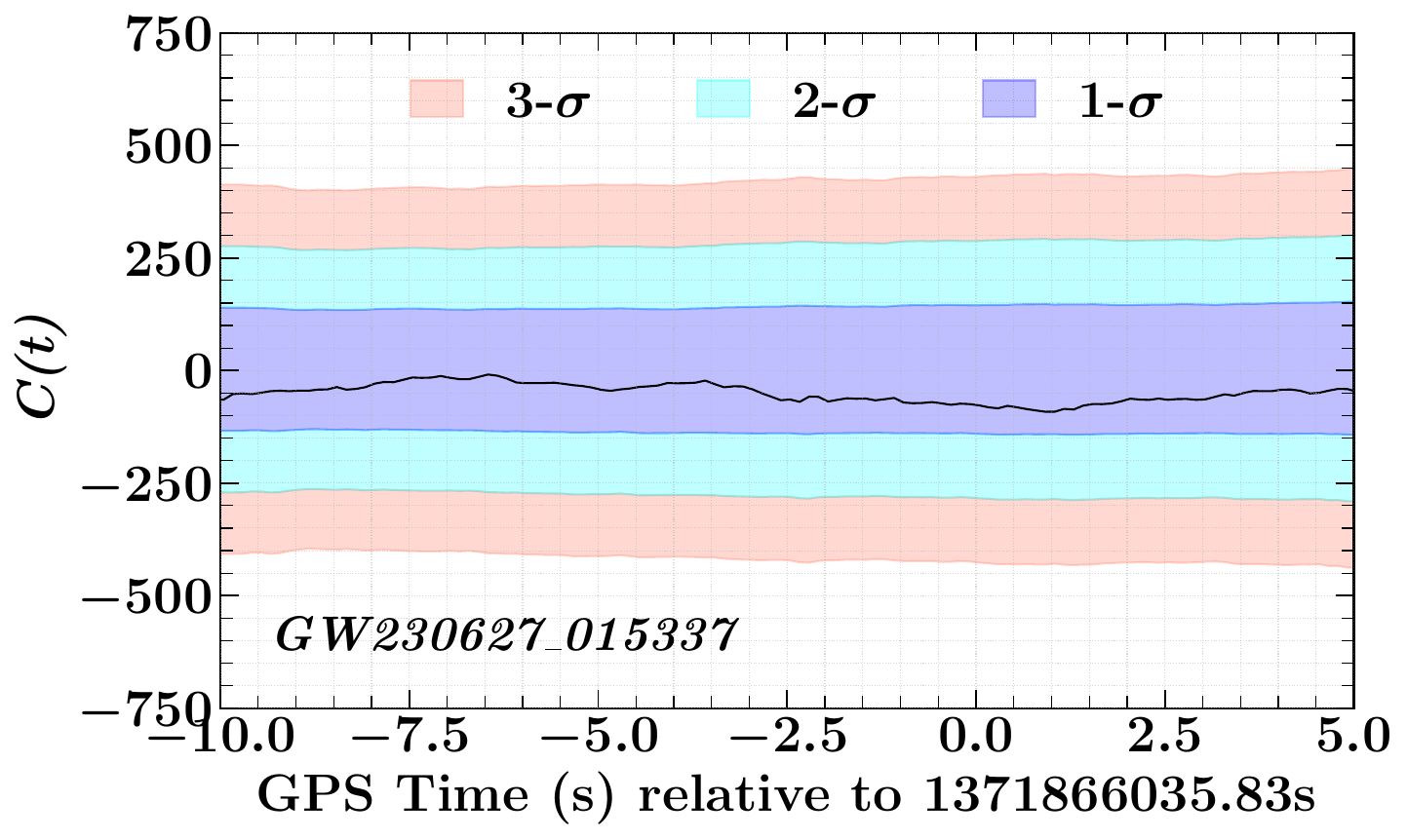}
    \end{minipage}
    \caption{
    Cumulative residual cross-correlation between the LIGO Hanford (H1) and Livingston (L1) detectors for the events analysed in this work. The black solid curve in each panel shows the cumulative cross-correlation computed from the residual strain after subtracting the best-fit vacuum waveform. The shaded regions indicate the $1\sigma$, $2\sigma$, and $3\sigma$ confidence bands under the noise-only hypothesis. In all three events, the residual cross-correlation remains consistent with the noise expectation. 
    }
    \label{fig:residual_cc}
\end{figure*}

Figure~\ref{fig:residual_cc} shows the cumulative residual cross-correlation for the analysed events. In each case, the residual remains well within the confidence bands of the obtained noise statistics, indicating no statistically significant correlated excess beyond the best-fit vacuum waveform. This null result motivates the constraint-based analysis presented in the following section, where simulated three-body perturbations are injected and tested against the empirically measured noise fluctuations. This null result is not impacted by systematic uncertainty in our model waveform as compared to the state-of-the-art waveform models, as the template subtracted from the data is the maximum-likelihood waveform obtained using the full LVK parameter-estimation framework, in which the waveforms include higher-order post-Newtonian corrections, higher harmonics, and merger--ringdown effects. Any residual arising purely from imperfect modelling within that framework would therefore reflect the intrinsic accuracy of state-of-the-art waveform families rather than the simplified dynamical model developed in this work.

\begin{figure}
    \centering
    \includegraphics[width=\linewidth]{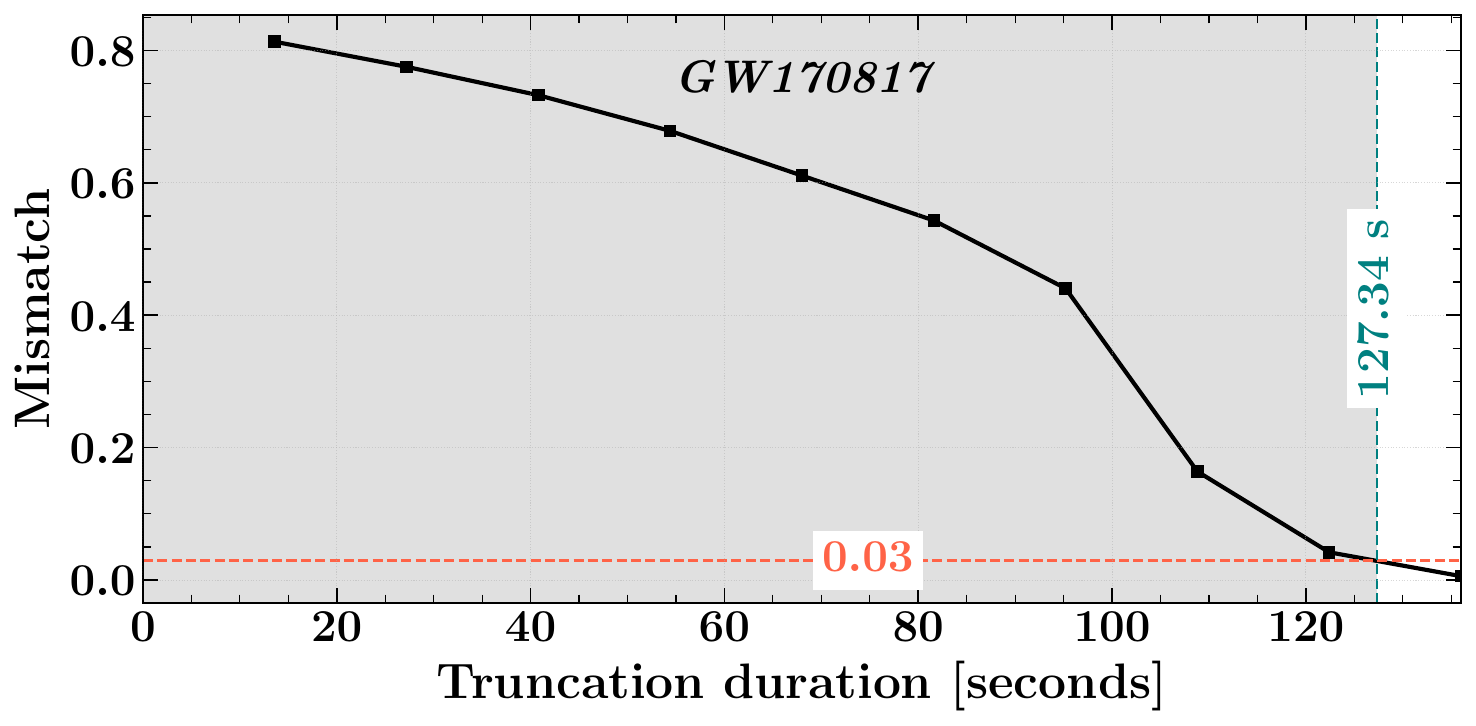}
    \includegraphics[width=\linewidth]{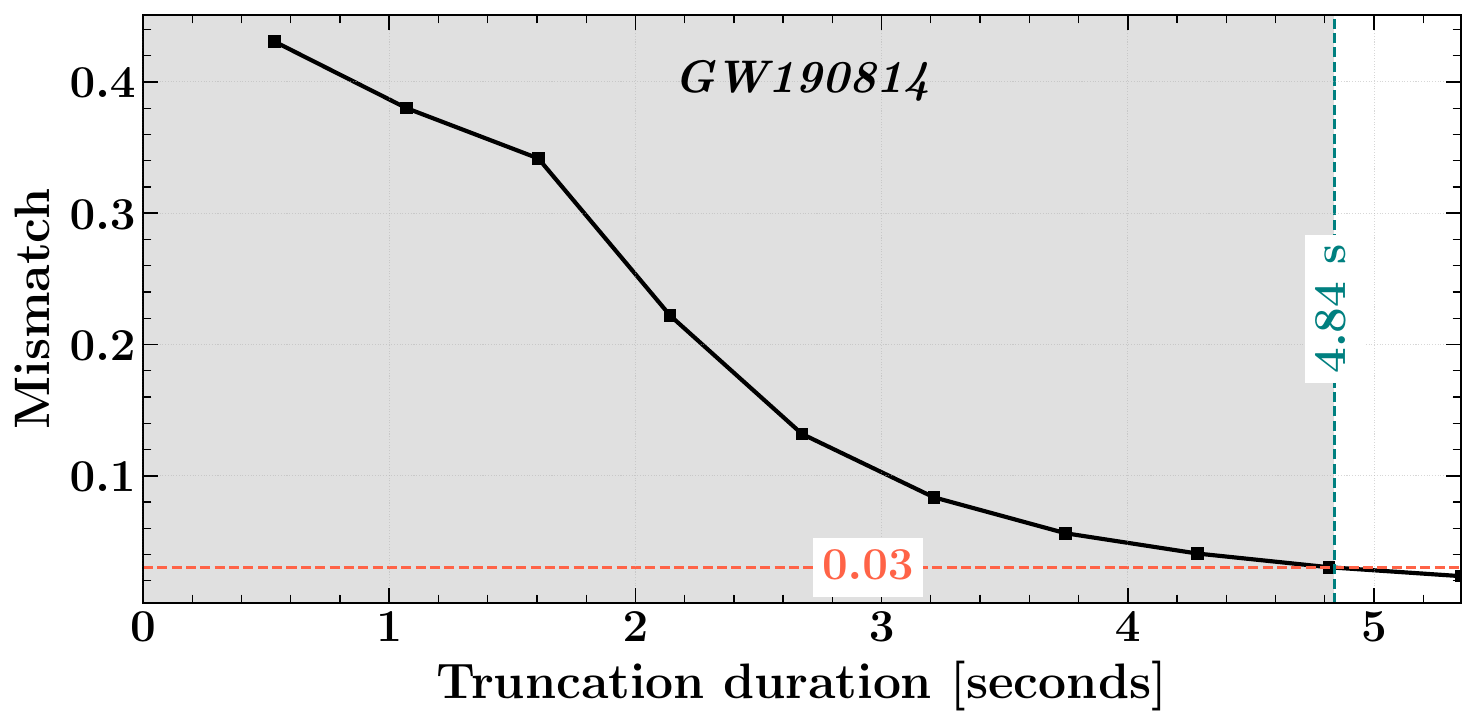}
    \includegraphics[width=\linewidth]{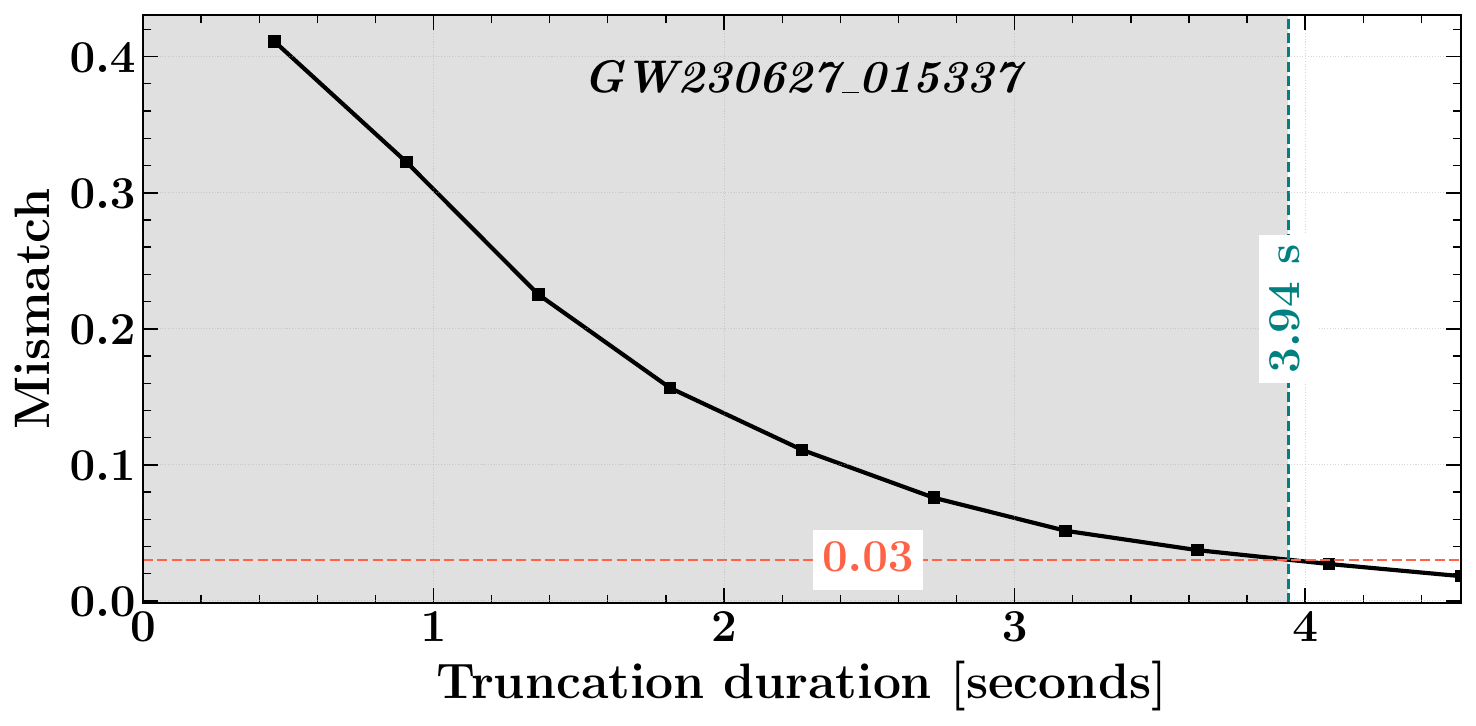}
    \caption{
    Mismatch between the unperturbed waveform constructed in this work and the corresponding reference waveform used in the LVK analysis, shown as a function of the truncation duration from the end of the signal. Each panel corresponds to one of the three events considered in this work. The dashed horizontal line indicates a mismatch threshold of $0.03$, while the vertical dashed line marks the truncation point beyond which the mismatch falls below this threshold. This procedure is used to define the event-specific analysis window used in subsequent residual cross-correlation analyses. The gray region corresponds to the truncation duration adopted in this work given the fiducial mismatch threshold of 0.03.
    }
    \label{fig:mismatch_validation}
\end{figure}

\begin{figure}
    \centering
    \includegraphics[width=\linewidth]{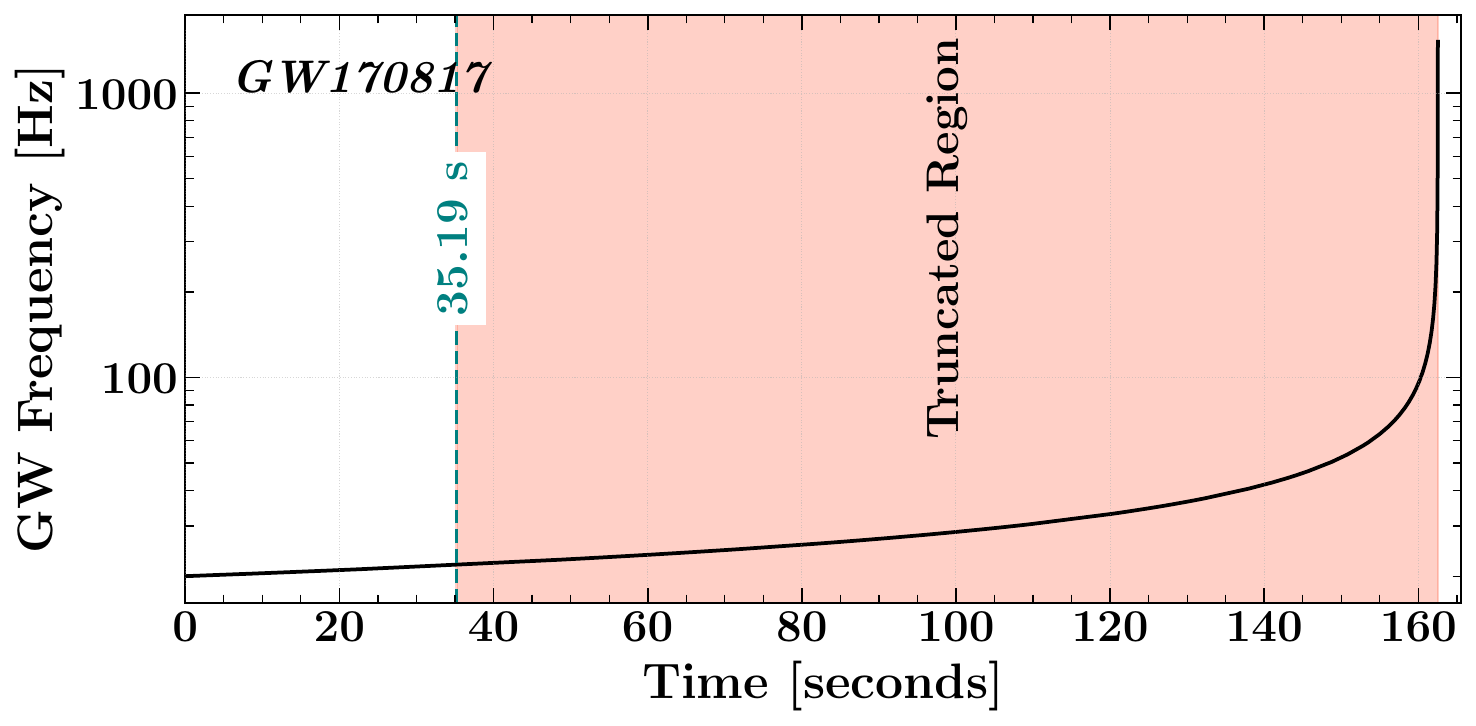}
    \includegraphics[width=\linewidth]{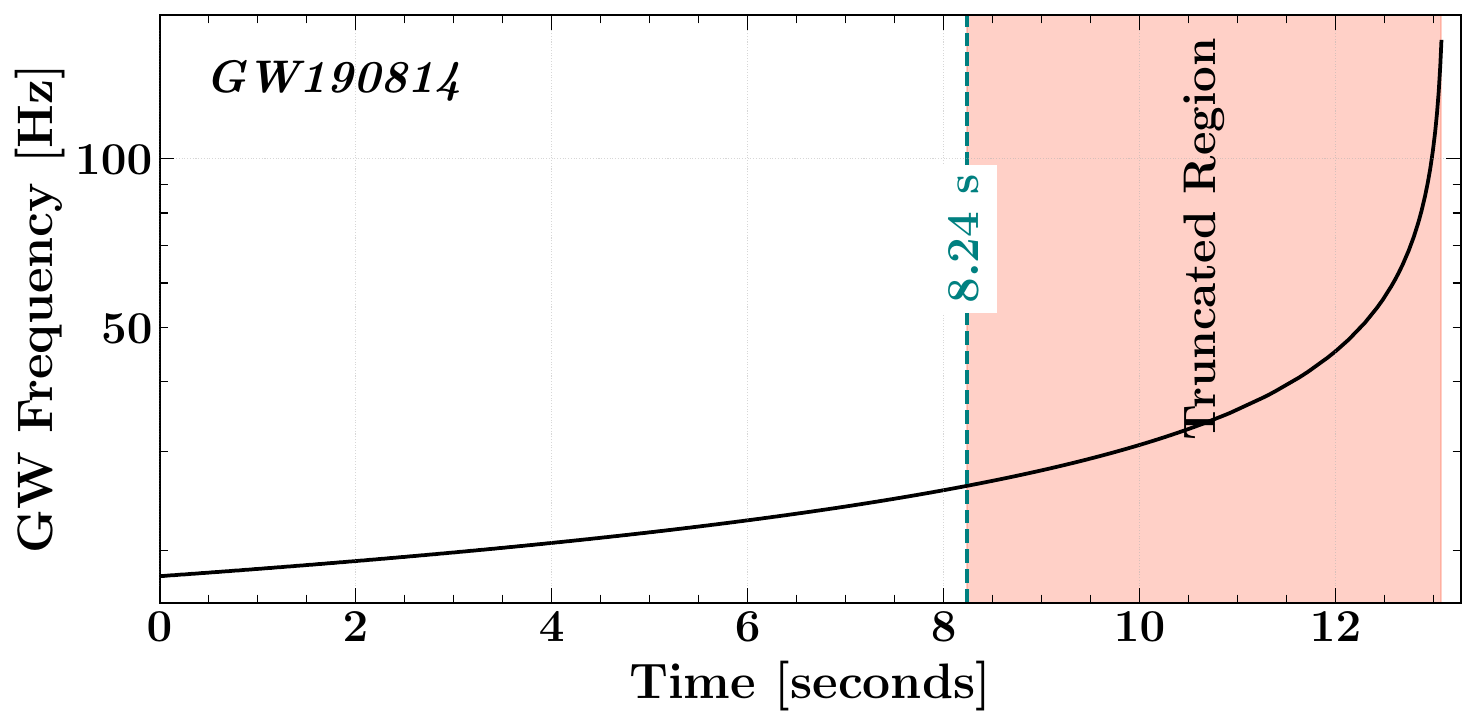}
    \includegraphics[width=\linewidth]{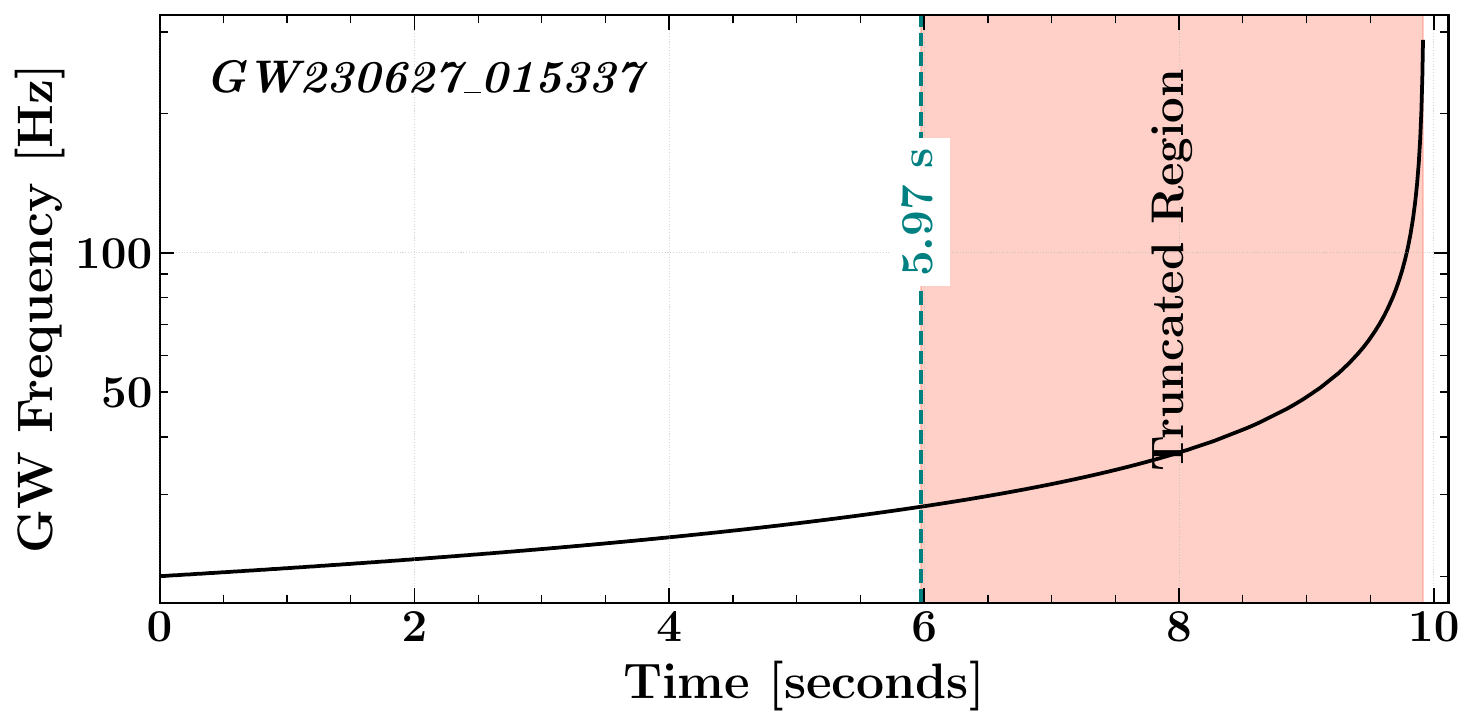}
    \caption{
    Determination of the validated analysis window for the three events considered in this work, assuming a conservative mismatch threshold of $1-\mathcal{M}<0.03$. Each panel shows the time evolution of the GW frequency inferred from the quadrupole radiation of the unperturbed waveform, together with a marker indicating the latest time up to which the mismatch with the corresponding LVK reference waveform remains below this threshold. The resulting event-specific time intervals define the validated portion of the signal used in all subsequent residual cross-correlation analyses.
    }
    \label{fig:validation_window}
\end{figure}

\section{\label{sec:constraints}Constraints from Three-Body Fly-by Perturbations}

\begin{table}
\centering
\caption{
Initial binary separation $r_0$ at $f_0 = 20~\mathrm{Hz}$ for the events considered in this work. These values provide a physical scale for the dimensionless separation parameter $\alpha = R_0 / r_0$.
}
\label{tab:r0_values}
\begin{tabular}{lc}
\hline\hline
Event & $r_0$ [AU] \\
\hline
GW170817              & $3.03\times10^{-6}$ \\
GW190814              & $6.49\times10^{-6}$ \\
GW230627\_015337      & $5.36\times10^{-6}$ \\
\hline\hline
\end{tabular}
\end{table}

\begin{table}
\centering
\caption{
Event-specific durations of the validated analysis window $T$ obtained from the
waveform-validation procedure described in Sec.~\ref{sec:validation}, for two different
mismatch thresholds. The threshold $1-\mathcal{M}<0.03$ is adopted for the constraint
analysis in this work, while $1-\mathcal{M}<0.1$ is shown for comparison.
}
\label{tab:analysis_windows}
\begin{tabular}{lcc}
\hline\hline
Event & $T\ [1-\mathcal{M}<0.03]$ [s] & $T\ [1-\mathcal{M}<0.1]$ [s] \\
\hline
GW170817              & 35.19 & 48.22 \\
GW190814              &  8.24 & 10.09 \\
GW230627\_015337      &  5.97 &  7.51 \\
\hline\hline
\end{tabular}
\end{table}

In this section, we translate the null residual cross-correlation results of Sec.~\ref{sec:residual_cc} into constraints on transient three-body encounters. As shown in Sec.~\ref{sec:residual_cc}, no statistically significant coherent residual power is observed in any of the analysed events after subtraction of the best-fit vacuum waveform. This implies that any additional physical effect beyond the isolated two-body inspiral must either (i) be sufficiently weak to remain below the empirical noise fluctuations within the validated analysis window, or (ii) lead to a configuration incompatible with the observed CBC.

These two possibilities naturally define distinct constraints in the parameter space related to the initial separation of the perturber from the binary $R_0$ and the mass of the perturber $m_3$. First, sufficiently strong fly-by interactions can dynamically disrupt the binary prior to merger or drive the system into a regime where our dynamical approximation breaks down; such configurations are directly excluded independent of waveform-based residual analysis. Second, for binaries that survive the interaction within the regime of validity of the model, the absence of a statistically significant residual cross-correlation constrains the strength of the induced perturbation. The resulting exclusions thus arise from a combination of dynamical survival criteria and residual-based statistical limits.

\subsection{The valid physical parameter space used in the search from data}

For each event, we simulate an ensemble of three-body fly-by encounters on a grid in the dimensionless parameters
\begin{equation}
    \alpha \equiv \frac{R_0}{r_0}, \qquad \beta \equiv \frac{m_3}{m_2},
\end{equation}
where $R_0$ is the initial perturber separation from the binary center-of-mass, $r_0$ is the initial binary separation at $f_0 = 20~\mathrm{Hz}$, and $m_3$ and $m_2$ are the perturber mass and secondary component mass, respectively. The approach velocity is fixed to $v_3 = 100~\mathrm{km\,s^{-1}}$ for the fiducial analysis, directed perpendicular to the binary orbital plane and toward the binary center-of-mass. Table~\ref{tab:r0_values} lists the physical scale $r_0$ for each event, which converts the dimensionless parameter $\alpha$ into a physical initial separation $R_0$. We additionally explore alternative configurations, for completeness, including higher approach speeds (e.g.\ $v_3 = 10^4~\mathrm{km\,s^{-1}}$) and initially receding trajectories, and the corresponding results are shown separately in Appendix~\ref{sec:appendix_alternative}.

For all configurations shown in this section for the fiducial velocity $v_3 = 100~\mathrm{km\,s^{-1}}$, the perturber does not cross the plane of the binary orbit before the binary reaches the innermost stable circular orbit (ISCO). In these cases, the binary completes its in-band inspiral while the third body is still approaching. The perturbation, therefore, arises from a gradually increasing tidal influence rather than from an impulsive kick at closest approach. This places the system in a regime where the cumulative phase deformation builds up monotonically over the validated analysis window.

For each $(\alpha, \beta)$ grid point, we generate a perturbed waveform and its vacuum counterpart and then obtain the residual $\delta h(t) \equiv h_{\mathrm{pert}}(t) - h_{\mathrm{vac}}(t)$. Antenna response functions appropriate to the sky position and polarisation of each event are applied to project $\delta h(t)$ onto the H1 and L1 detectors. All computations of the residual cross-correlation statistic $\mathcal{S}_\mathrm{res}(T)$ are restricted to the validated analysis window of duration $T$ as prescribed in Sec.~\ref{sec:validation} and tabulated in Table~\ref{tab:analysis_windows}. The restriction to a validated analysis window is essential because our dynamical model retains only Newtonian conservative dynamics supplemented by leading-order ($2.5$PN) radiation-reaction. As the binary approaches merger, higher-order post-Newtonian effects and strong-field corrections become increasingly important. To prevent spurious residual power from being generated by intrinsic modelling inaccuracies of the vacuum waveform, we truncate the late inspiral until the mismatch between our unperturbed model and the corresponding LVK reference waveform falls below a conservative threshold.

Figure~\ref{fig:mismatch_validation} shows the mismatch between the unperturbed waveform constructed in this work and the corresponding reference waveform used in the LVK analysis as a function of truncation duration from the end of the signal. The truncation point at which the mismatch drops below $1-\mathcal{M}=0.03$ defines the latest time up to which the model waveform is considered self-consistent. Figure~\ref{fig:validation_window} illustrates the corresponding validated time intervals in the frequency domain. The durations of these validated windows are summarised in Table~\ref{tab:analysis_windows} for two representative mismatch thresholds, $1-\mathcal{M}<0.03$ (used for the main analysis) and $1-\mathcal{M}<0.1$ (shown for comparison).

This procedure ensures that the residual cross-correlation analysis is sensitive only to physically meaningful deviations induced by the three-body interaction, rather than to inaccuracies of the approximate vacuum waveform model in the strongly relativistic regime.

\begin{figure*}
    \centering
    \begin{minipage}[t]{0.32\textwidth}
        \centering
        \includegraphics[width=\linewidth]{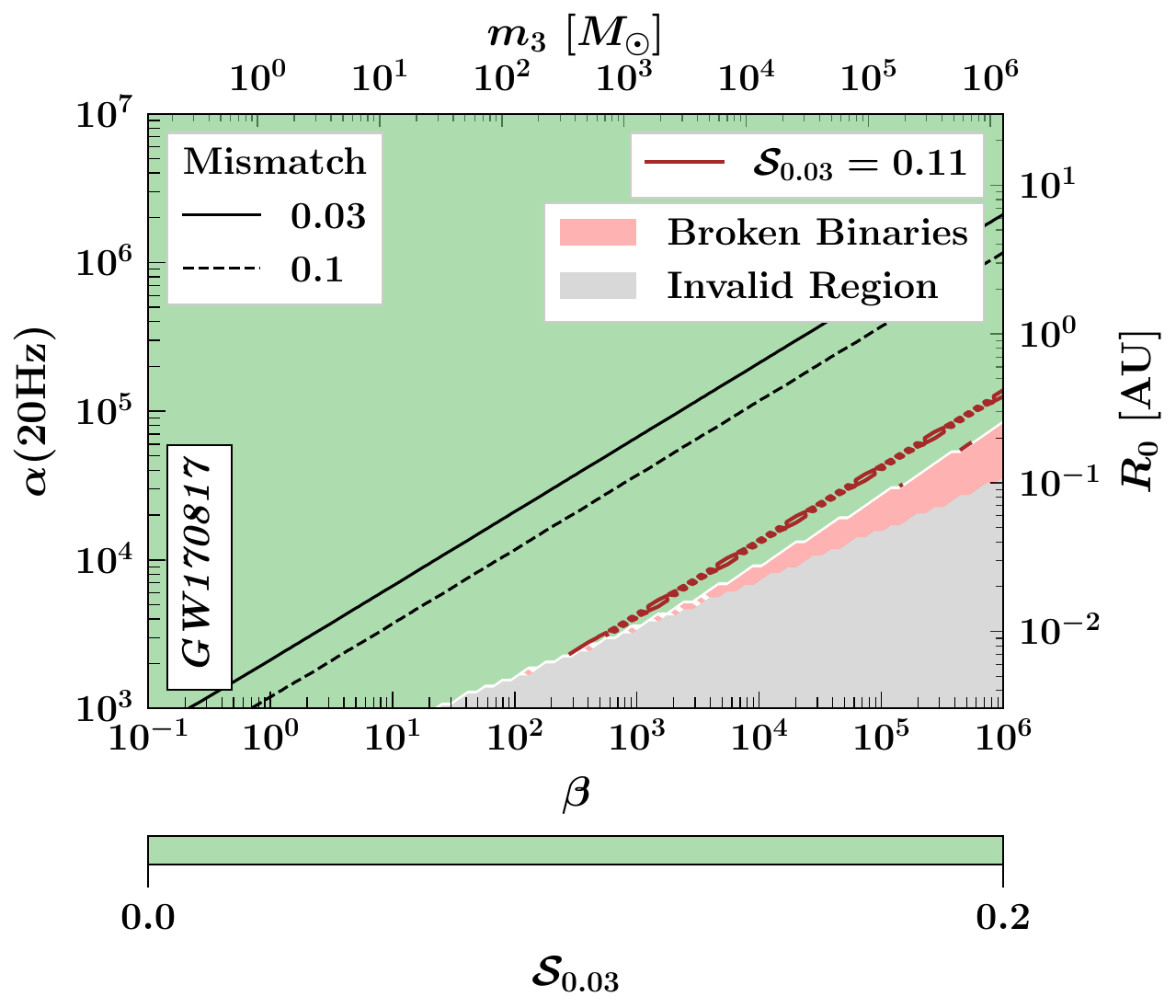}
    \end{minipage}
    \hfill
    \begin{minipage}[t]{0.32\textwidth}
        \centering
        \includegraphics[width=\linewidth]{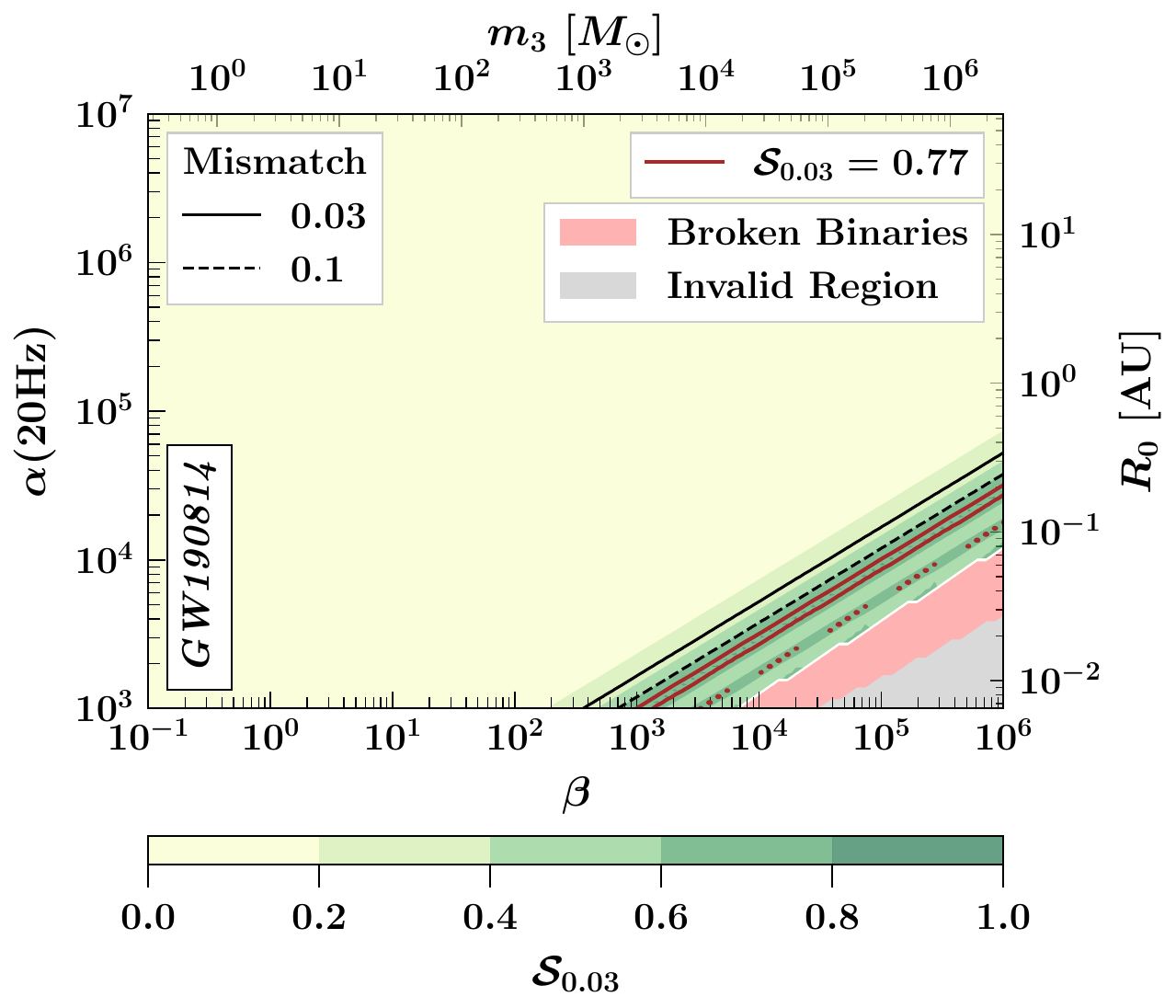}
    \end{minipage}
    \hfill
    \begin{minipage}[t]{0.32\textwidth}
        \centering
        \includegraphics[width=\linewidth]{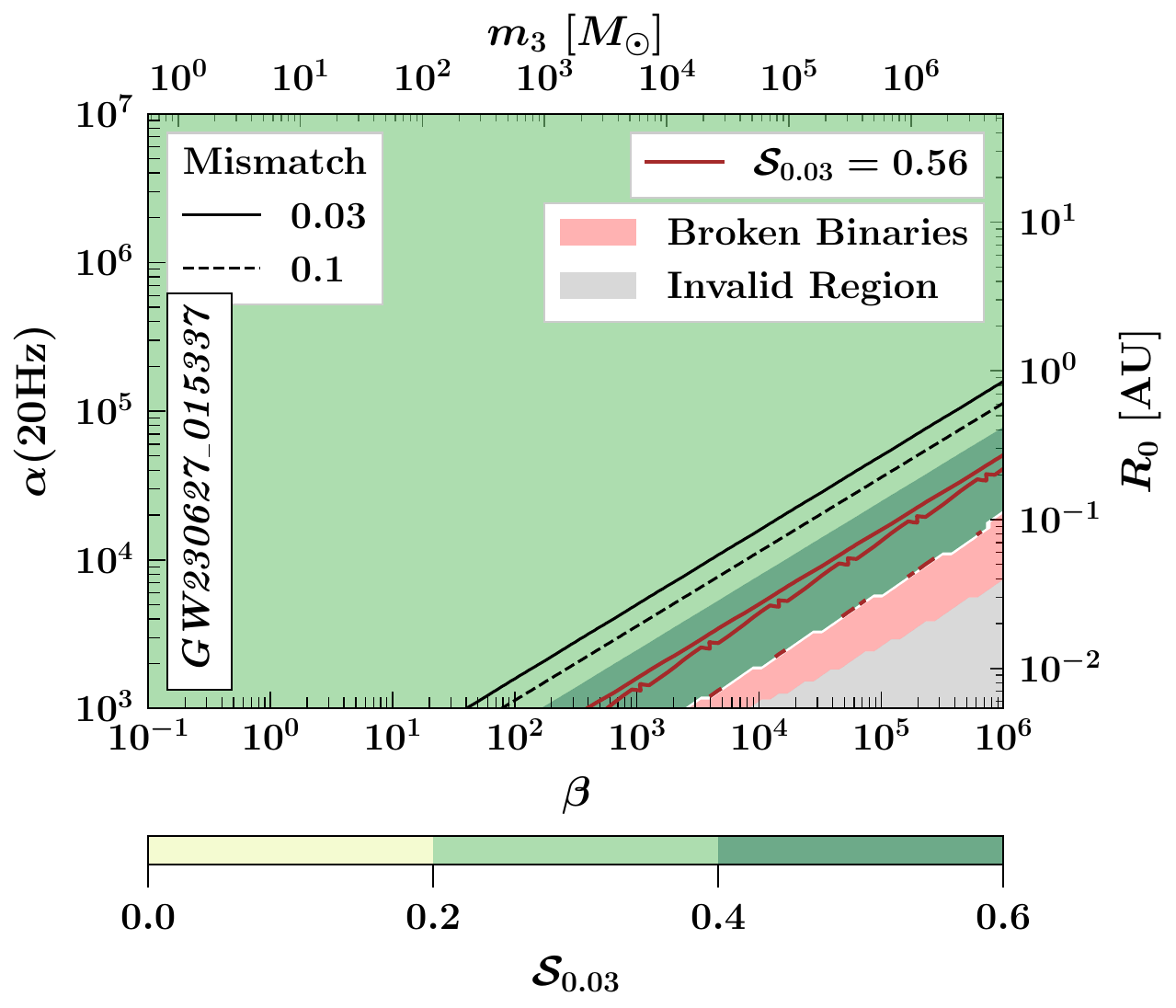}
    \end{minipage}
    \caption{
    Constraints on the perturber mass $m_3$ and initial separation $R_0$ from three-body fly-by interactions for GW170817 (left): $m_1=1.48M_\odot,\,m_2=1.27M_\odot$, GW190814 (center): $m_1=24.44M_\odot,\,m_2=2.73M_\odot$, and GW230627\_015337 (right): $m_1=9.65M_\odot,\,m_2=5.70M_\odot$. The top and right axes show the corresponding perturber mass $m_3$ in solar masses and separation $R_0$ in astronomical units, respectively, with the bottom and left axes showing the dimensionless parameters $\beta = m_3/m_2$ and $\alpha = R_0/r_0$ at $f_0 = 20~\mathrm{Hz}$. The gray region marks configurations where the dynamical model is invalid (velocities approach $c$ or a direct plunge occurs). The red region identifies fly-by configurations that would disrupt the binary prior to merger and are directly excluded by the observation of a coalescence, independent of waveform modelling. The solid and dashed black contours denote noise-weighted mismatch thresholds $1-\mathcal{M} = 0.03$ and $0.1$, computed using the full perturbed waveform without truncation; these indicate the sensitivity achievable with improved waveform models extending into the late inspiral. The brown solid contour marks the boundary of the near-maximum $\mathcal{S}_\mathrm{res}(T)$ achieved within the validated analysis window (defined by the $1-\mathcal{M}<0.03$ truncation criterion). The contour map shows the residual SNR $\mathcal{S}_{0.03}$ values across the parameter space in units of $0.2\sigma$. In all three events, the maximum residual SNR remains below unity, indicating no statistically significant residual cross-correlation, and owing to this, no uniform trend is observed.
    }
    \label{fig:constraint_maps_astro}
\end{figure*}

\subsection{Results}

The maximum residual cross-correlation SNR across all simulated configurations for the three events is
\begin{equation}
    \mathcal{S}_\mathrm{res}^{\mathrm{max}} =
    \begin{cases}
        0.11 & \text{GW170817,} \\
        0.81 & \text{GW190814,} \\
        0.59 & \text{GW230627\_015337.}
    \end{cases}\label{eq:res_SNR}
\end{equation}
In all cases, $\mathcal{S}_\mathrm{res} < 1$, meaning the induced residual cross-correlation does not rise above the $1\sigma$ noise level within the validated analysis window. Consequently, no statistically significant residual-based constraint is obtained from any of the three events using the truncated waveform. The color map in Fig.~\ref{fig:constraint_maps_astro} shows the residual SNR $\mathcal{S}_{0.03}$ values across the parameter space and does not yield a meaningful exclusion beyond what is already excluded by binary disruption. The constraint maps in the physical $(\alpha, \beta)$ plane, equivalently parameterised as $(m_3, R_0)$, are shown in Fig.~\ref{fig:constraint_maps_astro} for all three events. Three distinct regions can be clearly identified in each panel:
\begin{itemize}
    \item \textit{Invalid region.} The gray region at large $\beta$ and small $\alpha$ marks configurations in which the adopted dynamical model breaks down: the velocities of one or more bodies become comparable to the speed of light, or a direct plunge to within a few Schwarzschild radii occurs. These configurations are excluded from the analysis entirely, as the Newtonian-plus-leading-order-radiation-reaction treatment is not self-consistent in this regime.
    \item \textit{Disrupted binaries.} The red region immediately adjacent to the invalid region identifies configurations in which the fly-by interaction transfers sufficient energy to unbind the binary. Since a CBC is observed, configurations that would disrupt the binary prior to merger are directly excluded, independent of any waveform modelling. This constitutes a robust, model-independent constraint.
    \item \textit{Surviving binaries: residual cross-correlation.} For configurations outside these two regions, the binary survives the encounter and produces a perturbed GW signal. These are constrained through the residual cross-correlation statistic $\mathcal{S}_\mathrm{res}(T)$.
\end{itemize}

Two families of contours are shown in each panel. The solid and dashed black contours correspond to noise-weighted mismatch thresholds of $1-\mathcal{M} = 0.03$ and $1-\mathcal{M} = 0.1$, respectively, computed using the full perturbed waveform without any truncation. These contours indicate the parameter-space boundary at which the full perturbed signal would be distinguishable from the vacuum waveform at the corresponding mismatch level, and serve as an approximate reference for the improvement achievable with more accurate waveform models that extend further into the late inspiral. The brown solid contour shows the boundary of the region where $\mathcal{S}_\mathrm{res}(T)$ attains its maximum value (with a small margin) within the truncated analysis window defined by the $1-\mathcal{M} < 0.03$ threshold.

A notable feature of Fig.~\ref{fig:constraint_maps_astro} is the absence of a smooth transition in $\mathcal{S}_\mathrm{res}$ as one approaches the disruption boundary. Naively, one would expect the residual SNR to grow monotonically as the perturber becomes more massive or closer, transitioning continuously into the disruption region. That this trend is not seen is a direct consequence of the waveform truncation imposed by the analysis window. For configurations in the vicinity of the disruption boundary, where the perturber is either very massive ($\beta \gg 1$) or relatively nearby (moderate $\alpha$), the dominant phase perturbation accumulates predominantly during the late inspiral, after the validated analysis window of duration $T$ (Table~\ref{tab:analysis_windows}) has ended. Since the residual cross-correlation is evaluated only within this window, the most dynamically significant portion of the waveform is discarded, and the statistic $\mathcal{S}_\mathrm{res}(T)$ remains low even for configurations that generate a large mismatch when evaluated over the full waveform. In contrast, Fig.~\ref{fig:constraint_maps_astro} clearly shows that the full-waveform mismatch contours (solid and dashed black curves) do exhibit the expected monotonic growth toward the disruption boundary, confirming that the underlying physical signal is present and growing; it is simply being cut off by the truncation. Waveform models that extend consistently through the late inspiral and merger-ringdown would recover this sensitivity and exhibit the expected smooth transition from weak residual perturbation through strong dephasing to binary disruption.

It is also important to be noted that configurations at very small $\alpha$ (i.e., $R_0 \lesssim 10^{-4}~\mathrm{AU}$) do in principle place the encounter within the analysis window, since the perturber is already very close at the start of the inspiral and its dynamical effect manifests early in the waveform. However, these configurations are not astrophysically meaningful: they require a third compact object to be located within a fraction of a solar radius of the merging binary at the moment the signal enters the LIGO band. These configurations are therefore not discussed further.

The disruption-based exclusions, however, provide robust, waveform-independent limits. For GW230627\_015337 and GW190814, the disruption boundary directly excludes perturbers with $m_3 \gtrsim 10^4$--$10^6\,M_\odot$ at initial separations $R_0 \lesssim 10^{-1}$--$10^{-2}~\mathrm{AU}$. For GW170817 the excluded region is at $R_0 \lesssim 0.2~\mathrm{AU}$ for perturbers with $m_3 \gtrsim 10^3$--$10^7\,M_\odot$. We discuss the astrophysical significance of these constraints in the following section.

\subsection{Limitations of the current results}

Although the truncated-window residual analysis yields no significant constraints, the full-waveform mismatch contours (black solid and dashed in Fig.~\ref{fig:constraint_maps_astro}) illustrate the parameter space that \textit{would} be constrained if waveform models of sufficient accuracy were available throughout the late inspiral. These contours display the expected physical behaviour: the mismatch grows monotonically as one moves toward larger $\beta$ or smaller $\alpha$. For GW230627\_015337, the $1 - \mathcal{M} = 0.03$ contour reaches perturber masses of order $m_3 \sim 100\,M_\odot$ at separations $R_0 \lesssim 10^{-2}~\mathrm{AU}$. For GW190814, the corresponding contours probe qualitatively similar separations. For GW170817, the corresponding contours show that it might be possible to provide constraints on sub-solar mass compact objects across a very wide range of $R_0$, benefiting from its longer in-band signal duration, in principle, by considering more accurate waveforms.

The key limiting factor in the current analysis is the waveform-model truncation imposed by the $1-\mathcal{M}<0.03$ consistency criterion, which restricts the usable signal to $T \lesssim 5-35~\mathrm{s}$ (Table~\ref{tab:analysis_windows}) and discards the most sensitive late-inspiral portion of the signal where even small dephasing can correspond to a large mismatch between the waveforms. For configurations at large $\alpha$, the fly-by perturbation grows gradually throughout the inspiral but reaches its largest amplitude only in the final seconds before merger and in the portion that falls outside the analysis window. A compounding limitation is the short in-band duration of CBC signals in the LVK band starting from $f_0 \sim 20~\mathrm{Hz}$: even without truncation, the total inspiral duration is only of order one to tens of seconds for the events considered here, leaving little signal over which perturbation-induced dephasing can accumulate. Longer-duration signals, such as those from lower-mass or high-mass-ratio systems that enter the detector band earlier and spend more cycles in band, would allow the tidal perturbation from a fly-by to build up over a substantially larger number of orbital cycles, probing both lower perturber masses and larger initial separations. Advanced waveform models that consistently incorporate third-body perturbations throughout the full inspiral-merger-ringdown would recover the sensitivity currently lost to truncation and would naturally reproduce the smooth transition from weak dephasing to disruption visible in the full-waveform residual correlation maps. These considerations together point to next-generation ground-based detectors such as Einstein Telescope~\cite{EinsteinTelescope} and Cosmic Explorer~\cite{CosmicExplorer} as prime instruments for such studies: their lower frequency thresholds ($f_0 \sim 5~\mathrm{Hz}$) will extend the in-band duration by orders of magnitude, their improved sensitivities will lower the noise floor against which residual signals must compete, and their broader frequency coverage will enable accurate waveform modelling further into the inspiral. We discuss the astrophysical implications of the current constraints and the prospects for next-generation detectors in Sec.~\ref{sec:implications}.

\section{\label{sec:implications}Astrophysical implications and Future Prospects}
The constraints derived in Sec.~\ref{sec:constraints} represent the first application of a three-body fly-by residual search to real GW data. In this section we discuss the astrophysical implications of the findings from the current LVK data. 

The three events analysed here have full in-band inspiral durations (from $f_0 = 20~\mathrm{Hz}$ to ISCO) of $T \approx 160~\mathrm{s}$ (GW170817), $13~\mathrm{s}$ (GW190814), and $10~\mathrm{s}$ (GW230627\_015337). The validated analysis windows used in this work are shorter still: $35.19$, $8.24$, and $5.97~\mathrm{s}$ respectively (Table~\ref{tab:analysis_windows}), due to the waveform truncation imposed by the consistency criterion of Sec.~\ref{sec:validation}. At the fiducial approach speed $v_3 = 100~\mathrm{km\,s^{-1}}$, even the full (untruncated) inspiral durations translate to dynamically accessible separations of:
\begin{equation}
    R_0^{\rm dyn} \;\lesssim\; v_3\,T \;\approx\;
    \begin{cases}
        1.07\times10^{-4}~\mathrm{AU} & \text{GW170817},\\
        8.69\times10^{-6}~\mathrm{AU} & \text{GW190814},\\
        6.68\times10^{-6}~\mathrm{AU} & \text{GW230627\_015337.}
    \end{cases}
    \label{eq:R0_access}
\end{equation}
This quantity $R_0^{\rm dyn}$ sets an approximate length scale that is probed using these events. The sensitivity of a fly-by search is fundamentally governed by the in-band signal duration $T$. A perturber approaching at velocity $v_3$ from initial separation $R_0$ reaches closest approach on a timescale approximately $t_\mathrm{enc} \sim R_0/v_3$. For the perturbation to accumulate coherently during the inspiral, one requires $t_\mathrm{enc} \lesssim T$, i.e.\ $R_0 \lesssim v_3 T$. This condition sets the maximum perturber separation that is dynamically accessible to any detector configuration.

These separations are many orders of magnitude smaller than the characteristic inter-object distances in any known astrophysical environment (Sec.~\ref{ssec:environments}). The disruption-based constraints, which do not require the perturbation to accumulate over the inspiral and are therefore not subject to the timescale condition above, remain the primary astrophysical output of the current analysis.

\subsection{\label{ssec:IMBH}Novel constraints on intermediate-mass black holes}

The observation of a CBC provides a direct, model-independent constraint on the presence of a massive perturber in the immediate vicinity of the merging binary. Any compact object that would have disrupted the binary prior to merger must have been absent: this is the physical content of the disruption-excluded region in Fig.~\ref{fig:constraint_maps_astro}. Reading the disruption boundary directly from Fig.~\ref{fig:constraint_maps_astro} for each event:
\begin{itemize}
    \item \textit{GW170817}: disruption excludes $m_3 \gtrsim 10^2~M_\odot$ at $R_0 \lesssim 10^{-2}~\mathrm{AU}$, rising along the tidal disruption boundary to $m_3 \gtrsim 10^6~M_\odot$ at $R_0 \lesssim 0.3~\mathrm{AU}$.
    \item \textit{GW190814}: disruption excludes $m_3 \gtrsim 10^4~M_\odot$ at $R_0 \lesssim 10^{-2}~\mathrm{AU}$, rising to $m_3 \gtrsim 10^6~M_\odot$ at $R_0 \lesssim 0.1~\mathrm{AU}$.
    \item \textit{GW230627\_015337}: essentially identical to GW190814, with disruption excluding $m_3 \gtrsim 10^4~M_\odot$ at $R_0 \lesssim 10^{-2}~\mathrm{AU}$, rising to $m_3 \gtrsim 10^6~M_\odot$ at $R_0 \lesssim 0.1~\mathrm{AU}$, consistent with the comparable chirp masses of the two systems.
\end{itemize}
These constraints are robust to the perturber trajectory and approach speed, as discussed in Appendix~\ref{sec:appendix_alternative}: both, a receding trajectory and a high-velocity approach ($v_3 = 10^4~\mathrm{km\,s^{-1}}$), produce essentially identical disruption boundaries for such systems.

\paragraph{Volume exclusion.}
Since $v_3 T \ll R_0$ for all disruption-excluded configurations (the perturber barely moves during the inspiral; Sec.~\ref{sec:constraints}), the relevant question is simply whether a massive compact object was present within the exclusion volume $V_\mathrm{excl} = \frac{4}{3}\pi R_0^3$ at the start of the inspiral. The observation of a non-disrupted merger requires the expected number of such objects $\mathcal{N} = n_\mathrm{IMBH}\,V_\mathrm{excl} < 1$, where $n_\mathrm{IMBH}$ denotes the number density of the IMBHs, giving:
\begin{equation}
    n_\mathrm{IMBH} \;\lesssim\; \frac{3}{4\pi R_0^3}.\label{eq:n_IMBH}
\end{equation}
Evaluating this at representative points along the disruption boundary for each event:
\begin{widetext}
\begin{equation}
    n_\mathrm{IMBH}(m_3) \;\lesssim\;
    \begin{cases}
        0.24~\mathrm{centiAU}^{-3} & 
        m_3 \gtrsim 10^2\,M_\odot,\ R_0 \lesssim 10^{-2}~\mathrm{AU}
        \ [\text{GW170817}],\\[4pt]
        0.87\times10^{-2}~\mathrm{deciAU}^{-3} & 
        m_3 \gtrsim 10^6\,M_\odot,\ R_0 \lesssim 0.3~\mathrm{AU}
        \ [\text{GW170817}],\\[4pt]
        0.24~\mathrm{centiAU}^{-3} & 
        m_3 \gtrsim 10^4\,M_\odot,\ R_0 \lesssim 10^{-2}~\mathrm{AU}
        \ [\text{GW190814, GW230627\_015337}],\\[4pt]
        0.24~\mathrm{deciAU}^{-3} & 
        m_3 \gtrsim 10^6\,M_\odot,\ R_0 \lesssim 0.1~\mathrm{AU}
        \ [\text{GW190814, GW230627\_015337}].
    \end{cases}
    \label{eq:n_IMBH_values}
\end{equation}
\end{widetext}

The current limits, while numerically modest, are the first direct event-by-event GW-based exclusions on the presence of massive compact objects in the immediate dynamical vicinity of merging binaries. Unlike population-level statistical arguments, each non-disrupted merger provides a specific statement: no compact object of the excluded mass was within $R_0$ of this binary during the final $T_\mathrm{insp}$ seconds of its inspiral.

From the residual analysis (Fig.~\ref{fig:constraint_maps_astro}), we can provide further constraints from the data using the residual cross-correlation SNR, at however, very low statistical significance of less than $1\sigma$ of the noise statistics (Eq.~\ref{eq:res_SNR}). At a statistical significance of $0.1\sigma$, $0.8\sigma$, and $0.6\sigma$ for the events GW170817, GW190814, GW230627\_015337, respectively, we obtain:
\begin{widetext}
\begin{equation}
    n_\mathrm{IMBH}(m_3) \;\lesssim\;
    \begin{cases}
        0.24~\mathrm{centiAU}^{-3} & 
        m_3 \gtrsim 10^2\,M_\odot,\ R_0 \lesssim 10^{-2}~\mathrm{AU}
        \ [\text{GW170817}],\\[4pt]
        0.37\times10^{-2}~\mathrm{deciAU}^{-3} & 
        m_3 \gtrsim 10^6\,M_\odot,\ R_0 \lesssim 0.4~\mathrm{AU}
        \ [\text{GW170817}],\\[4pt]
        0.24~\mathrm{centiAU}^{-3} & 
        m_3 \gtrsim 10^3\,M_\odot,\ R_0 \lesssim 10^{-2}~\mathrm{AU}
        \ [\text{GW190814, GW230627\_015337}],\\[4pt]
        0.30\times10^{-1}~\mathrm{deciAU}^{-3} & 
        m_3 \gtrsim 10^6\,M_\odot,\ R_0 \lesssim 0.2~\mathrm{AU}
        \ [\text{GW190814, GW230627\_015337}].
    \end{cases}
    \label{eq:n_IMBH_values_res}
\end{equation}
\end{widetext}

As the GW event catalog grows, waveform models improve, and detector networks push to lower frequencies, this channel will develop into a complementary waveform-based probe of the IMBH occupation fraction in dense stellar environments, independent of electromagnetic detection.

\subsection{Future Prospects to Explore Other Astrophysical Systems}
\subsubsection{Astrophysical systems}\label{ssec:environments}
The other astrophysical systems where three-body fly-by interactions can be important are the dense environments such as globular clusters, nuclear star clusters, and galactic center. We briefly describe below the expected length scale over which we can expect three-body interactions on these systems. 

\textit{Globular clusters and Nuclear star clusters:} Globular cluster cores reach stellar number densities of $n \sim 10^4$--$10^6~\mathrm{pc}^{-3}$ \cite{Willems:2007xe, Ivanova:2003by, Miller:2002pg, Baumgardt2018}, with mean inter-object separations $\bar{d} \sim n^{-1/3} \sim 10^3$--$10^4~\mathrm{AU}$ and one-dimensional velocity dispersions $\sigma \sim 10$--$30~\mathrm{km\,s^{-1}}$~\cite{Bahramian:2013ihw, Baumgardt2018}. Nuclear star clusters in the centres of galaxies reach $n \sim 10^6$--$10^8~\mathrm{pc}^{-3}$ \cite{Ozernoy:1997pa, Ott:2003gr}, with mean separations $\bar{d} \sim 10$--$200~\mathrm{AU}$ and velocity dispersions $\sigma \sim 50$--$150~\mathrm{km\,s^{-1}}$~\cite{NSC2020, NSC2010}.  These environments are among the most productive known sites for compact binary formation through dynamical exchange interactions. The mean separations exceed the separations probed by current disruption-based constraints by four to five orders of magnitude, and remain a factor of $\sim10$--$100$ beyond the reach of next-generation ground-based detectors. Multi-band observations with milli-hertz detectors (as discussed in the following section) will first begin to probe characteristic globular cluster separations.

\textit{Galactic center.} The innermost regions of the Milky Way represent the most extreme accessible environment. Within $\lesssim 0.1~\mathrm{pc}$ of Sgr~A$^*$, stellar and compact object number densities are estimated at $n \sim 10^7$--$10^8~\mathrm{pc}^{-3}$, with mean separations $\bar{d} \sim 0.1$--$5~\mathrm{AU}$ and velocity dispersions $\sigma \sim 100$--$1000~\mathrm{km\,s^{-1}}$~\cite{Amaro-Seoane:2012lgq, Bahcall:1976aa, Gnedin:2013cda}. These separations sit within the range accessible to deci-hertz detectors and partially within the reach of ET/CE at $1~\mathrm{Hz}$ (Table~\ref{tab:detector_reach}). If compact binary mergers occur in such environments, as expected for a fraction of 
events hosted in galactic nuclei and AGN disks, they represent the most promising targets for next-generation fly-by searches. Even current disruption constraints, while weak in absolute terms, provide the first direct GW-based limits on the presence of massive compact objects at AU-scale separations from merging binaries in any environment. 

Moreover, dark matter candidates such as PBHs can also be constrained by this window \cite{Carr:2020xqk}. The expected number of PBH encounters assuming the PBH number density follows a Milky Way NFW dark matter profile~\cite{Navarro:1995iw, Weber_2010, Iocco:2015xga,Salucci:2018hqu}
\begin{equation}
    \rho_\mathrm{NFW}(r) \;=\; \frac{\rho_s}{(r/r_s)(1+r/r_s)^2},\label{eq:NFW}
\end{equation}
with $\rho_s = 4\times10^{-3}~M_\odot\,\mathrm{pc}^{-3}$ and $r_s = 23.8~\mathrm{kpc}$~\cite{Cautun:2019eaf, Klypin2002, Sofue2012, Pato:2015dua}. The PBH number density at galactocentric radius $r$ is:
\begin{equation}
    n_\mathrm{PBH}(r) \;=\; \frac{f_\mathrm{PBH}\,\rho_\mathrm{NFW}(r)}{m_\mathrm{PBH}},\label{eq:nPBH}
\end{equation}
giving representative values:
\begin{widetext}
\begin{equation}
    n_\mathrm{PBH}(r) \;\approx\;\frac{f_\mathrm{PBH}}{m_\mathrm{PBH}/M_\odot}\times
    \begin{cases}
        6.67\times10^{-3}~\mathrm{pc}^{-3} & r = 8~\mathrm{kpc}\ (\text{solar neighbourhood}),\\
        9.44\times10^{-1}~\mathrm{pc}^{-3}            & r = 100~\mathrm{pc},\\
        95.2~\mathrm{pc}^{-3}            & r = 1~\mathrm{pc}.
    \end{cases}
    \label{eq:nPBH_values}
\end{equation}
\end{widetext}
As a result, sub-solar PBHs can have a large number density near the galactic center and will enhance the chance of three-body fly-by interaction with coalescing GW events which can be observed for a long duration. 

\subsubsection{Future prospects from upcoming detectors}

The way to astrophysically relevant dynamical separations requires substantially longer in-band durations, which future detectors will provide through lower frequency thresholds. Since the inspiral duration scales as $T \propto f_0^{-8/3}$, the accessible separation scales correspondingly. We discuss below briefly the accessibility of this signal from upcoming detectors. 

\textit{Above 1 Hz GW band: } The Einstein Telescope~\cite{EinsteinTelescope} and Cosmic 
Explorer~\cite{CosmicExplorer} are designed to operate from $f_0 \sim 5~\mathrm{Hz}$, extending the in-band duration by a factor of $(20/5)^{8/3} \approx 40$ over LIGO. This gives $T \approx 1.8~\mathrm{hr}$ for a GW170817-like BNS and $T \approx 500~\mathrm{s}$ for the BBH systems analysed here, with dynamically accessible separations $R_0^{\rm dyn} \sim 4\times10^{-3}~\mathrm{AU}$ and $\sim 3\times10^{-4}~\mathrm{AU}$ respectively. Lowering the threshold further to $1~\mathrm{Hz}$ extends this by an additional factor of $(5)^{8/3} \approx 73$, reaching $T \sim 5.5~\mathrm{days}$ for GW170817-like systems ($R_0^{\rm dyn} \sim 0.3~\mathrm{AU}$) and $T \sim 10~\mathrm{hrs}$ for the BBH systems ($R_0^{\rm dyn} \sim 0.02~\mathrm{AU}$).

\textit{Deci-hertz GW band: } Proposed deci-hertz instruments~\cite{Mandel:2017pzd, Ajith:2024mie} operating around $f_0 \sim 0.1~\mathrm{Hz}$ extend inspiral durations by a factor of $(20/0.1)^{8/3} \sim 10^6$ over LVK, giving $T \sim \mathcal{O}(\mathrm{years})$ for BNS systems and $T \sim \mathcal{O}(\mathrm{months})$ for the BBH systems considered here. The dynamically accessible separation reaches $R_0^{\rm dyn} \sim 1$--$10~\mathrm{AU}$, beginning to overlap with the mean inter-object separations in the densest known environments.

\begin{table*}
\centering
\caption{
Dynamically accessible perturber separation $R_0^{\rm dyn} \sim v_3 T$ for different detector configurations at $v_3 = 100~\mathrm{km\,s^{-1}}$, for a GW170817-like BNS and a GW230627-like BBH systems. Characteristic mean inter-object separations $\bar{d}$ in relevant environments are shown for comparison.
}
\label{tab:detector_reach}
\begin{tabular}{lccc}
\hline\hline
Configuration              & $f_\mathrm{low}$ & $R_0^{\rm dyn}$ (BNS) & $R_0^{\rm dyn}$ (BBH) \\
\hline
Current LIGO/Virgo/KAGRA   & 20~Hz   & $\sim10^{-4}$~AU & $\sim10^{-6}$~AU \\
ET/CE                      & 5~Hz    & $\sim10^{-3}$~AU   & $\sim10^{-4}$~AU \\
ET/CE                      & 1~Hz    & $\sim10^{-1}$~AU              & $\sim10^{-2}$~AU \\
Deci-Hz detectors           & 0.1~Hz  & $\sim100$~AU          & $\sim10$~AU         \\
LISA multiband             & cHz--mHz  & $\sim10^4$~AU--$\mathcal{O}(\mathrm{pc})$  & $\sim10^3$~AU--$\mathcal{O}(\mathrm{pc})$ \\
\hline
\multicolumn{4}{l}{\textit{Environment reference scales}:}\\
Globular cluster cores      & ---     & \multicolumn{2}{c}{$\bar{d} \sim 10^3$--$10^4$~AU}  \\
Nuclear star clusters       & ---     & \multicolumn{2}{c}{$\bar{d} \sim 10$--$200$~AU}     \\
Galactic center ($<0.1$~pc) & ---     & \multicolumn{2}{c}{$\bar{d} \sim 0.1$--$5$~AU}      \\
\hline\hline
\end{tabular}
\end{table*}

\textit{Multiband observations with LISA.} Stellar-mass compact binary systems detectable in the LVK band will, for a range of masses and mass ratios, also be visible to LISA~\cite{LISA:2024hlh} years to decades before merger, as their GW frequency passes through the mHz band at orbital separations of $r \sim 10^{-3}$--$10^{-1}~\mathrm{AU}$. Continuous tracking of the same source from the LISA band through merger in a ground-based 
detector~\cite{Sesana:2017vsj} provides an observational baseline spanning years, corresponding to $R_0^{\rm dyn} \lesssim 10$--$100~\mathrm{AU}$ at $v_3 \sim 100~\mathrm{km\,s^{-1}}$. At these separations the fly-by search begins to probe the mean inter-object separations in nuclear star clusters and the galactic center, making multi-band observations a qualitatively new channel for three-body environmental probes for both astrophysical black holes and PBHs. We summarise the full hierarchy in Table~\ref{tab:detector_reach}.

\section{\label{sec:conclusion}Conclusions}

In this work, we have developed a physically motivated model for fly-by three-body interactions on compact binary inspirals and applied it for the first time to GW data from the LVK detector network. The model evolves the full three-body system under Newtonian gravity supplemented by leading-order radiation-reaction within the binary, and constructs perturbed GW waveforms that consistently incorporate the resulting orbital modifications: cumulative phase deformation, amplitude modulations, and Doppler modulation from binary center-of-mass motion. The regime of validity of this approximate model was assessed through a mismatch-based comparison against state-of-the-art LVK waveform families, and all subsequent analyses were restricted to the validated portion of the inspiral.

Applying a residual cross-correlation analysis to GW170817, GW190814, and GW230627\_015337, we find no statistically significant correlated residual power beyond the best-fit vacuum waveform in any of the three events, with residual SNRs all below the $1\sigma$ noise level. This null result is translated into constraints on the perturber mass $m_3$ and initial separation $R_0$ through two complementary exclusion mechanisms. The disruption-based constraints, which follow solely from the observation of a non-disrupted coalescence and require no waveform modelling, exclude perturbers with $m_3 \gtrsim 10^2$--$10^6\,M_\odot$ at $R_0 \lesssim 10^{-2}$--$0.3~\mathrm{AU}$ for GW170817, and $m_3 \gtrsim 10^4$--$10^6\,M_\odot$ at $R_0 \lesssim 10^{-2}$--$0.1~\mathrm{AU}$ for GW190814 and GW230627\_015337. These limits are robust to the perturber trajectory and approach speed, as demonstrated in Appendix~\ref{sec:appendix_alternative}. The residual cross-correlation analysis does not yet yield constraints beyond the disruption boundary, primarily due to the short in-band signal durations available to the LIGO network and the waveform truncation required by the model validity criterion.

The primary limitations of the current analysis are the short in-band signal durations available to LVK ($T \lesssim 160~\mathrm{s}$ for the most favorable event) and the need to truncate the waveform to the regime of model validity, which together restrict the dynamically accessible perturber separations to $R_0^{\rm dyn} \lesssim 10^{-4}~\mathrm{AU}$ which is many orders of magnitude below characteristic inter-object separations in any known astrophysical environment. Both limitations will be substantially alleviated by next-generation detectors. The Einstein Telescope and Cosmic Explorer, operating from $f_0 \sim 1$--$5~\mathrm{Hz}$, will extend in-band durations by factors of tens to thousands, pushing $R_0^{\rm dyn}$ to $0.01$--$0.3~\mathrm{AU}$ and improving the $T^2$-amplified sensitivity correspondingly. Deci-hertz instruments and LISA-based multiband observations will extend the accessible baseline to years, reaching $R_0^{\rm dyn} \sim 10$--$100~\mathrm{AU}$ and beginning to probe the characteristic inter-object separations of several astrophysical systems. At these scales, the methodology developed here will transform individual CBC detections into sensitive, event-by-event dynamical probes of their local astrophysical and cosmological environments which will be complementary to, and independent of, all existing observational channels for compact dark objects and can shed light on their formation scenarios.

\section*{Acknowledgements}
The authors are very thankful to Mohit Raj Sah for reviewing the manuscript as part of the LIGO Publication \& Presentation policy and providing useful comments on the draft.
This work is a part of the $\boldsymbol{\langle}\texttt{data|theory}\boldsymbol{\rangle}$ \texttt{Universe-Lab} which is supported by the TIFR and the Department of Atomic Energy, Government of India. We acknowledge the support of the Department of Atomic Energy, Government of India, under Project Identification No. RTI 4012. This research is supported by the Prime Minister Early Career Research Award, Anusandhan National Research Foundation, Government of India.  The authors would like to thank the LIGO/Virgo scientific collaboration for providing the GW strain data. LIGO is funded by the U.S. National Science Foundation. Virgo is funded by the French Centre National de Recherche Scientifique (CNRS), the Italian Istituto Nazionale della Fisica Nucleare (INFN), and the Dutch Nikhef, with contributions by Polish and Hungarian institutes. This material is based upon work supported by NSF’s LIGO Laboratory, which is a major facility fully funded by the National Science Foundation. We acknowledge the use of the following software packages in this work: \textsc{NumPy}~\cite{numpy}, \textsc{SciPy}~\cite{2020SciPy-NMeth}, \textsc{PyCBC}~\cite{pycbc}, \textsc{Numba}~\cite{numba}, \textsc{Joblib}~\cite{joblib}, \textsc{PESummary}~\cite{pesummary}, \textsc{LAL}~\cite{lalsuite}, \textsc{Pandas}~\cite{pandas}, \textsc{h5py}~\cite{h5py}, and \textsc{matplotlib}~\cite{matplotlib}.

\appendix

\section{\label{sec:appendix_alternative}Alternative Perturber Configurations}

\begin{figure*}
    \centering
    \includegraphics[width=0.45\linewidth]{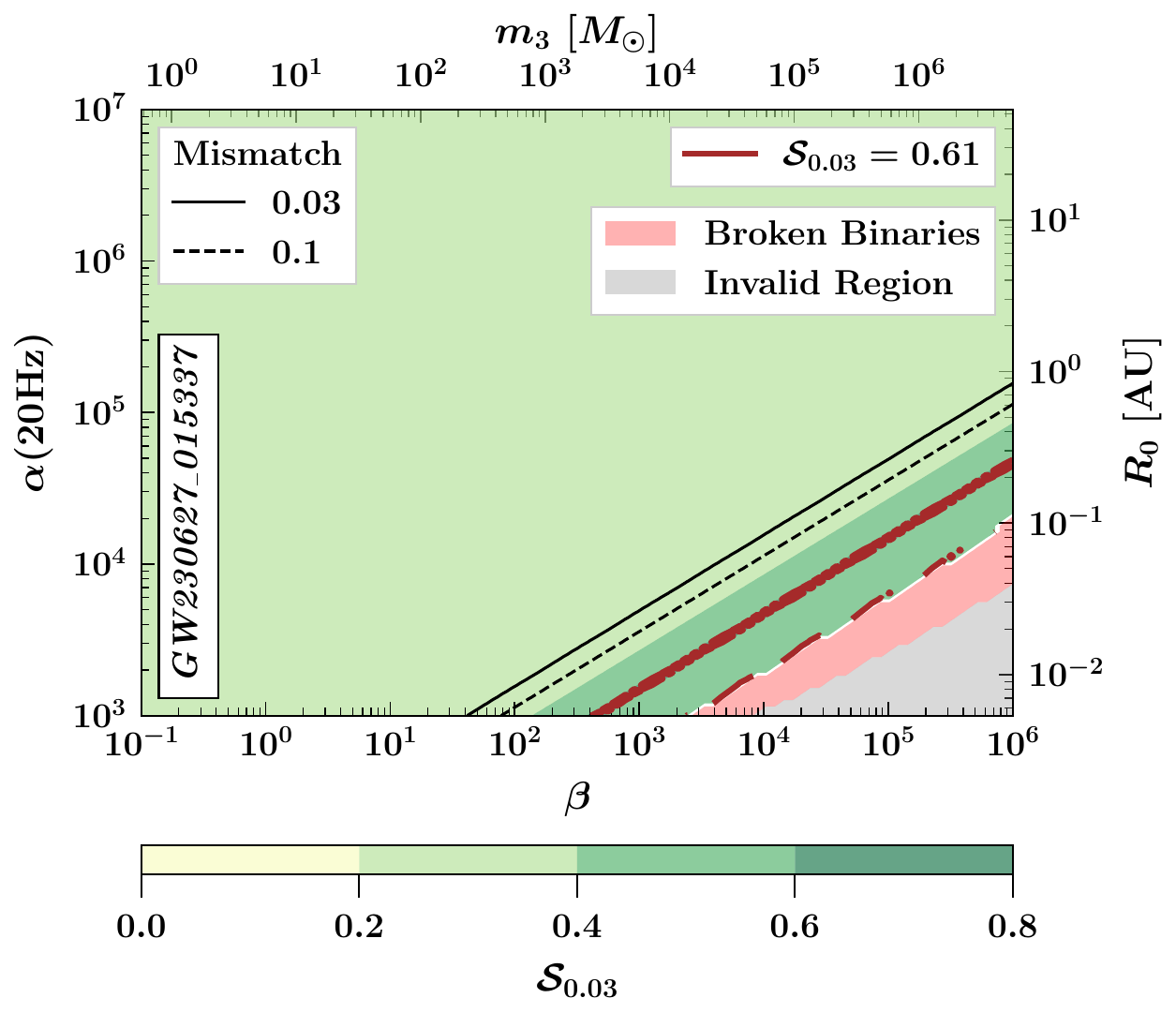}
    \includegraphics[width=0.45\linewidth]{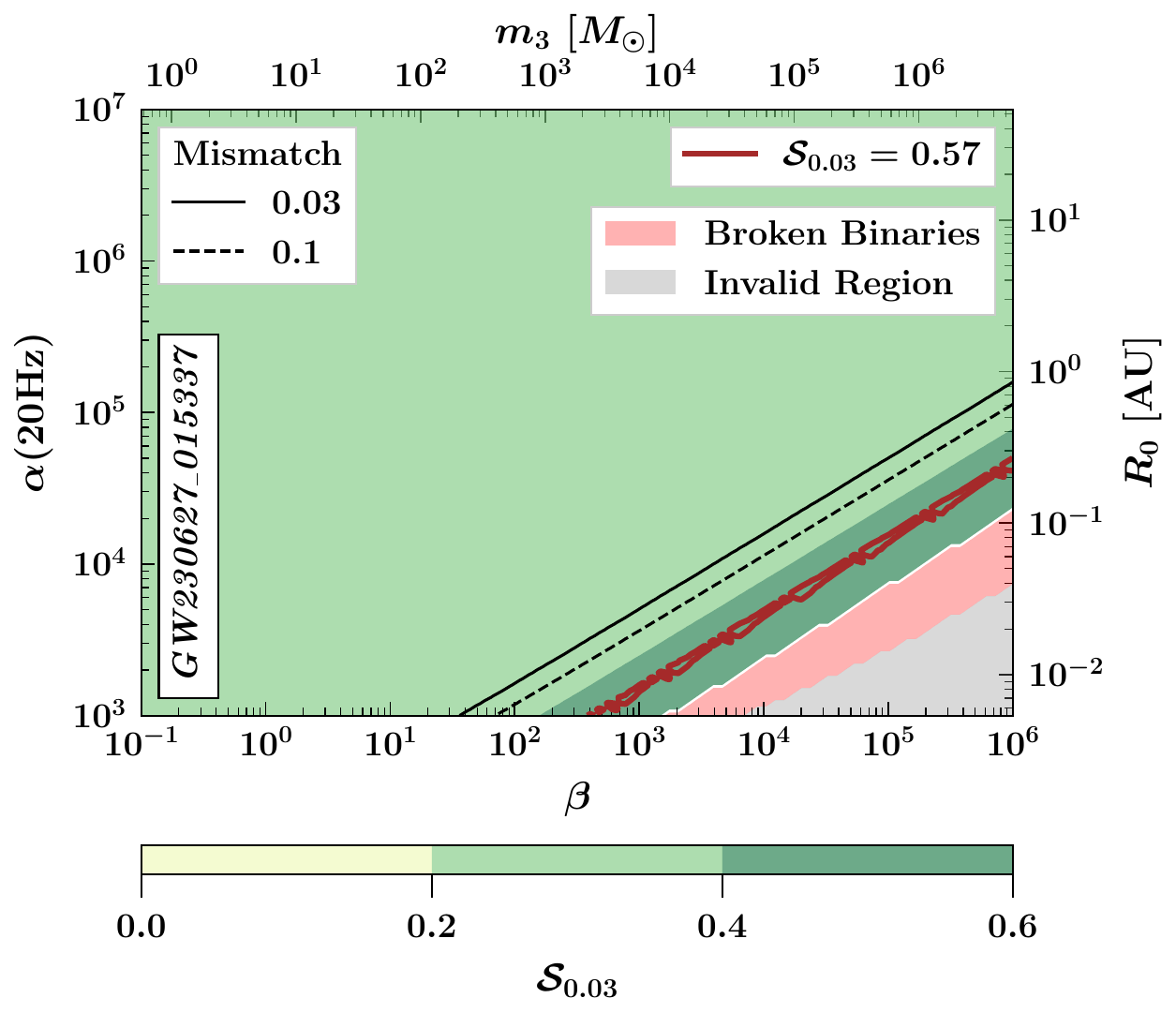}
    \caption{
    Constraint maps for GW230627\_015337 under two alternative perturber configurations: a receding trajectory at $v_3 = 100~\mathrm{km\,s^{-1}}$ (left) and an approaching trajectory at $v_3 = 10^4~\mathrm{km\,s^{-1}}$ (right). The layout and color scheme follow Fig.~\ref{fig:constraint_maps_astro}. In both cases, the structure of the contour maps and the disruption boundary are essentially unchanged, the residual SNR is always less than unity, and no statistically significant residual cross-correlation is found. The robustness of these results to a factor of $10^2$ variation in $v_3$ and to reversal of the approach direction confirms that, within the validated analysis window, the dominant perturbation is quasi-static and set by the instantaneous tidal field at separation $R_0$, rather than by the dynamical details of the encounter trajectory.
    }
    \label{fig:constraint_maps_appendix}
\end{figure*}

In the main analysis of Sec.~\ref{sec:constraints}, the fiducial perturber trajectory is directed \textit{toward} the binary center-of-mass at an approach speed of $v_3 = 100~\mathrm{km\,s^{-1}}$, perpendicular to the binary orbital plane. Here we examine two alternative configurations for the representative event GW230627\_015337: (i) an approaching perturber at a substantially higher speed $v_3 = 10^4~\mathrm{km\,s^{-1}}$, two orders of magnitude above the fiducial value and (ii) a perturber with the same initial conditions but moving \textit{away} from the binary center-of-mass (receding trajectory). The constraint maps for both cases are shown in Fig.~\ref{fig:constraint_maps_appendix}. We present the results from the numerical simulation for a representative event but as we will discuss below: the results remain valid for other two events considered here as well if the velocity of the perturber satisfies the condition $v_3 \cdot T \ll R_0$ where $T$ denotes the time of the binary inspiral.

In both alternative configurations, the maximum residual cross-correlation SNR remains well below the $1\sigma$ noise thresholds. These values are essentially identical to the fiducial result (shown in Sec.~\ref{sec:constraints}), and the overall structure of the constraint maps: the disruption boundary, the invalid region, and the distribution of residual SNR across the surviving parameter space, is qualitatively unchanged. We discuss the physical reasons for this robustness below.

\subsection*{Impact of higher velocity perturber}

The near-identical residual SNR between the fiducial ($v_3 = 100~\mathrm{km\,s^{-1}}$) and high-velocity ($v_3 = 10^4~\mathrm{km\,s^{-1}}$) cases can be understood from the following argument: for an approaching perturber initially at distance $R_0$ with velocity $v_3$ directed toward the binary, the time for the perturber to reach closest approach scales as $t_{\mathrm{enc}} \sim R_0 / v_3$. In the fiducial case, as noted in Sec.~\ref{sec:constraints}, the perturber does not cross the binary orbital plane before the binary reaches ISCO: for $v_3 = 100~\mathrm{km\,s^{-1}}$ and the range of $\alpha$ values probed, we have $t_{\mathrm{enc}} \gg T$, where $T \lesssim 6~\mathrm{s}$ is the validated analysis window for GW230627\_015337. The perturber therefore barely moves during the entire inspiral, and the binary evolves in a quasi-static, slowly varying external tidal field.

For $v_3 = 10^4~\mathrm{km\,s^{-1}}$, the encounter timescale is reduced by a factor of $10^2$, yet the linear displacement of the perturber during the inspiral is
\begin{equation}
    \Delta R \approx v_3 \cdot T \approx 10^4~\mathrm{km\,s^{-1}} \times 6~\mathrm{s} \approx 6 \times 10^4~\mathrm{km} \approx 4 \times 10^{-4}~\mathrm{AU}.
\end{equation}
For all but the smallest values of $\alpha$ in the surviving parameter space, $R_0 = \alpha r_0 \gg 4 \times 10^{-4}~\mathrm{AU}$, so $\Delta R / R_0 \ll 1$: the perturber remains effectively stationary relative to its initial distance even at $v_3 = 10^4~\mathrm{km\,s^{-1}}$. In this regime, both velocity choices place the system in the \textit{same} quasi-static tidal perturbation regime. The dominant perturbation to the orbit of the binary is determined by the instantaneous tidal effects from the stationary perturber at distance $R_0$, which depends on $m_3$ and $R_0$ but is insensitive to $v_3$ so long as $v_3 \cdot T \ll R_0$.

The velocity $v_3$ would become a discriminating parameter only when the perturber approaches to within $R_0 \lesssim v_3 \cdot T$, i.e.\ when the encounter is genuinely impulsive within the analysis window. For stellar-mass binaries in the LVK band with $T \lesssim \mathcal{O}(10~\mathrm{s})$, this requires $R_0 \lesssim 4 \times 10^{-4}~\mathrm{AU}$ even at the higher speed, corresponding to configurations that are either within the disruption boundary or not astrophysically meaningful, as discussed in Sec.~\ref{sec:constraints}. Consequently, no significant sensitivity gain or loss is expected from varying $v_3$ over two orders of magnitude in the LVK band, and the present analysis is robust to this choice within the valid surviving parameter space. The effects of the perturber velocities, however, will play crucial role in multiband studies of three-body encounters, where the binary inspiral duration is expected to be much longer: $\mathcal{O}(\text{seconds--minutes})$ in ground-based detectors to $\mathcal{O}(\text{days--months})$ in space-based detectors.

\subsection*{Impact of a receding perturber}

For the receding-trajectory configuration, the perturber begins at $R_0$ and moves away from the binary center-of-mass. The tidal perturbation experienced by the binary therefore starts at its maximum value (at $t = t_0$) and then decreases during the subsequent evolutionary stages. This is the time-reverse of the approaching case, in which the tidal field grows throughout the window (since the perturber is still approaching).

Although the temporal structure of the perturbation differs between the two cases: more tidal perturbations in the start for the receding trajectory versus more tidal perturbations during the later stages for the approaching case, neither produces a significantly larger cumulative phase deformation over the short validated analysis window. In both cases, the perturber is at a distance $R(t) \approx R_0$ throughout the window, and the integrated perturbation over $T \lesssim 6~\mathrm{s}$ is comparably small. The slight increase in $\mathcal{S}_\mathrm{res}^{\mathrm{max}}$ for the receding trajectory reflects the marginally stronger initial tidal impulse at $t = t_0$ in the receding case, though this difference is statistically negligible at current sensitivity and nothing speaks above the $1\sigma$ noise background.

The disruption boundary is also qualitatively unchanged between the two direction choices, as binary disruption is primarily governed by the total energy exchanged during closest approach which is determined largely by $m_3$, $R_0$, and $v_3$ rather than by the sign of the initial velocity. Both configurations therefore yield consistent and compatible constraint maps across the full $(\alpha, \beta)$ parameter space.

\input{output.bbl}
\end{document}

%% file: output.bbl
%

%% file: paper_data.bbl
\begin{thebibliography}{121}%
\makeatletter
\providecommand \@ifxundefined [1]{%
 \@ifx{#1\undefined}
}%
\providecommand \@ifnum [1]{%
 \ifnum #1\expandafter \@firstoftwo
 \else \expandafter \@secondoftwo
 \fi
}%
\providecommand \@ifx [1]{%
 \ifx #1\expandafter \@firstoftwo
 \else \expandafter \@secondoftwo
 \fi
}%
\providecommand \natexlab [1]{#1}%
\providecommand \enquote  [1]{``#1''}%
\providecommand \bibnamefont  [1]{#1}%
\providecommand \bibfnamefont [1]{#1}%
\providecommand \citenamefont [1]{#1}%
\providecommand \href@noop [0]{\@secondoftwo}%
\providecommand \href [0]{\begingroup \@sanitize@url \@href}%
\providecommand \@href[1]{\@@startlink{#1}\@@href}%
\providecommand \@@href[1]{\endgroup#1\@@endlink}%
\providecommand \@sanitize@url [0]{\catcode `\\12\catcode `\$12\catcode `\&12\catcode `\#12\catcode `\^12\catcode `\_12\catcode `\%12\relax}%
\providecommand \@@startlink[1]{}%
\providecommand \@@endlink[0]{}%
\providecommand \url  [0]{\begingroup\@sanitize@url \@url }%
\providecommand \@url [1]{\endgroup\@href {#1}{\urlprefix }}%
\providecommand \urlprefix  [0]{URL }%
\providecommand \Eprint [0]{\href }%
\providecommand \doibase [0]{https://doi.org/}%
\providecommand \selectlanguage [0]{\@gobble}%
\providecommand \bibinfo  [0]{\@secondoftwo}%
\providecommand \bibfield  [0]{\@secondoftwo}%
\providecommand \translation [1]{[#1]}%
\providecommand \BibitemOpen [0]{}%
\providecommand \bibitemStop [0]{}%
\providecommand \bibitemNoStop [0]{.\EOS\space}%
\providecommand \EOS [0]{\spacefactor3000\relax}%
\providecommand \BibitemShut  [1]{\csname bibitem#1\endcsname}%
\let\auto@bib@innerbib\@empty
\bibitem [{\citenamefont {Einstein}(1916)}]{Einstein:1916cc}%
  \BibitemOpen
  \bibfield  {author} {\bibinfo {author} {\bibfnamefont {A.}~\bibnamefont {Einstein}},\ }\bibfield  {title} {\bibinfo {title} {{Approximative Integration of the Field Equations of Gravitation}},\ }\href@noop {} {\bibfield  {journal} {\bibinfo  {journal} {Sitzungsber. Preuss. Akad. Wiss. Berlin (Math. Phys. )}\ }\textbf {\bibinfo {volume} {1916}},\ \bibinfo {pages} {688} (\bibinfo {year} {1916})}\BibitemShut {NoStop}%
\bibitem [{\citenamefont {Peters}(1964)}]{peters1964}%
  \BibitemOpen
  \bibfield  {author} {\bibinfo {author} {\bibfnamefont {P.~C.}\ \bibnamefont {Peters}},\ }\bibfield  {title} {\bibinfo {title} {Gravitational radiation and the motion of two point masses},\ }\href {https://doi.org/10.1103/PhysRev.136.B1224} {\bibfield  {journal} {\bibinfo  {journal} {Phys. Rev.}\ }\textbf {\bibinfo {volume} {136}},\ \bibinfo {pages} {B1224} (\bibinfo {year} {1964})}\BibitemShut {NoStop}%
\bibitem [{\citenamefont {Cutler}\ and\ \citenamefont {Flanagan}(1994)}]{CutlerFlanagan1994}%
  \BibitemOpen
  \bibfield  {author} {\bibinfo {author} {\bibfnamefont {C.}~\bibnamefont {Cutler}}\ and\ \bibinfo {author} {\bibfnamefont {E.~E.}\ \bibnamefont {Flanagan}},\ }\bibfield  {title} {\bibinfo {title} {Gravitational waves from merging compact binaries: How accurately can one extract the binary's parameters from the inspiral waveform?},\ }\href {https://doi.org/10.1103/PhysRevD.49.2658} {\bibfield  {journal} {\bibinfo  {journal} {Phys. Rev. D}\ }\textbf {\bibinfo {volume} {49}},\ \bibinfo {pages} {2658} (\bibinfo {year} {1994})}\BibitemShut {NoStop}%
\bibitem [{\citenamefont {Abbott}\ \emph {et~al.}(2016)\citenamefont {Abbott} \emph {et~al.}}]{lvk2016feb}%
  \BibitemOpen
  \bibfield  {author} {\bibinfo {author} {\bibfnamefont {B.~P.}\ \bibnamefont {Abbott}} \emph {et~al.} (\bibinfo {collaboration} {LIGO Scientific, Virgo}),\ }\bibfield  {title} {\bibinfo {title} {{Observation of Gravitational Waves from a Binary Black Hole Merger}},\ }\href {https://doi.org/10.1103/PhysRevLett.116.061102} {\bibfield  {journal} {\bibinfo  {journal} {Phys. Rev. Lett.}\ }\textbf {\bibinfo {volume} {116}},\ \bibinfo {pages} {061102} (\bibinfo {year} {2016})},\ \Eprint {https://arxiv.org/abs/1602.03837} {arXiv:1602.03837 [gr-qc]} \BibitemShut {NoStop}%
\bibitem [{\citenamefont {Aasi}\ \emph {et~al.}(2015)\citenamefont {Aasi} \emph {et~al.}}]{LIGO}%
  \BibitemOpen
  \bibfield  {author} {\bibinfo {author} {\bibfnamefont {J.}~\bibnamefont {Aasi}} \emph {et~al.} (\bibinfo {collaboration} {LIGO Scientific}),\ }\bibfield  {title} {\bibinfo {title} {{Advanced LIGO}},\ }\href {https://doi.org/10.1088/0264-9381/32/7/074001} {\bibfield  {journal} {\bibinfo  {journal} {Class. Quant. Grav.}\ }\textbf {\bibinfo {volume} {32}},\ \bibinfo {pages} {074001} (\bibinfo {year} {2015})},\ \Eprint {https://arxiv.org/abs/1411.4547} {arXiv:1411.4547 [gr-qc]} \BibitemShut {NoStop}%
\bibitem [{\citenamefont {Acernese}\ \emph {et~al.}(2015)\citenamefont {Acernese} \emph {et~al.}}]{VIRGO}%
  \BibitemOpen
  \bibfield  {author} {\bibinfo {author} {\bibfnamefont {F.}~\bibnamefont {Acernese}} \emph {et~al.} (\bibinfo {collaboration} {VIRGO}),\ }\bibfield  {title} {\bibinfo {title} {{Advanced Virgo: a second-generation interferometric gravitational wave detector}},\ }\href {https://doi.org/10.1088/0264-9381/32/2/024001} {\bibfield  {journal} {\bibinfo  {journal} {Class. Quant. Grav.}\ }\textbf {\bibinfo {volume} {32}},\ \bibinfo {pages} {024001} (\bibinfo {year} {2015})},\ \Eprint {https://arxiv.org/abs/1408.3978} {arXiv:1408.3978 [gr-qc]} \BibitemShut {NoStop}%
\bibitem [{\citenamefont {Akutsu}\ \emph {et~al.}(2021)\citenamefont {Akutsu} \emph {et~al.}}]{KAGRA}%
  \BibitemOpen
  \bibfield  {author} {\bibinfo {author} {\bibfnamefont {T.}~\bibnamefont {Akutsu}} \emph {et~al.} (\bibinfo {collaboration} {KAGRA}),\ }\bibfield  {title} {\bibinfo {title} {{Overview of KAGRA: Detector design and construction history}},\ }\href {https://doi.org/10.1093/ptep/ptaa125} {\bibfield  {journal} {\bibinfo  {journal} {PTEP}\ }\textbf {\bibinfo {volume} {2021}},\ \bibinfo {pages} {05A101} (\bibinfo {year} {2021})},\ \Eprint {https://arxiv.org/abs/2005.05574} {arXiv:2005.05574 [physics.ins-det]} \BibitemShut {NoStop}%
\bibitem [{\citenamefont {Abac}\ \emph {et~al.}(2025{\natexlab{a}})\citenamefont {Abac} \emph {et~al.}}]{gwtc4Catalog}%
  \BibitemOpen
  \bibfield  {author} {\bibinfo {author} {\bibfnamefont {A.~G.}\ \bibnamefont {Abac}} \emph {et~al.} (\bibinfo {collaboration} {LIGO Scientific, VIRGO, KAGRA}),\ }\bibfield  {title} {\bibinfo {title} {{GWTC-4.0: Updating the Gravitational-Wave Transient Catalog with Observations from the First Part of the Fourth LIGO-Virgo-KAGRA Observing Run}},\ }\href@noop {} {\  (\bibinfo {year} {2025}{\natexlab{a}})},\ \Eprint {https://arxiv.org/abs/2508.18082} {arXiv:2508.18082 [gr-qc]} \BibitemShut {NoStop}%
\bibitem [{\citenamefont {Abac}\ \emph {et~al.}(2025{\natexlab{b}})\citenamefont {Abac} \emph {et~al.}}]{GW250114}%
  \BibitemOpen
  \bibfield  {author} {\bibinfo {author} {\bibfnamefont {A.~G.}\ \bibnamefont {Abac}} \emph {et~al.} (\bibinfo {collaboration} {LIGO Scientific, Virgo, KAGRA}),\ }\bibfield  {title} {\bibinfo {title} {{GW250114: Testing Hawking{\textquoteright}s Area Law and the Kerr Nature of Black Holes}},\ }\href {https://doi.org/10.1103/kw5g-d732} {\bibfield  {journal} {\bibinfo  {journal} {Phys. Rev. Lett.}\ }\textbf {\bibinfo {volume} {135}},\ \bibinfo {pages} {111403} (\bibinfo {year} {2025}{\natexlab{b}})},\ \Eprint {https://arxiv.org/abs/2509.08054} {arXiv:2509.08054 [gr-qc]} \BibitemShut {NoStop}%
\bibitem [{\citenamefont {Abac}\ \emph {et~al.}(2026)\citenamefont {Abac} \emph {et~al.}}]{gwtc4Tgr}%
  \BibitemOpen
  \bibfield  {author} {\bibinfo {author} {\bibfnamefont {A.~G.}\ \bibnamefont {Abac}} \emph {et~al.} (\bibinfo {collaboration} {LIGO Scientific, Virgo, KAGRA}),\ }\bibfield  {title} {\bibinfo {title} {{Black Hole Spectroscopy and Tests of General Relativity with GW250114}},\ }\href {https://doi.org/10.1103/6c61-fm1n} {\bibfield  {journal} {\bibinfo  {journal} {Phys. Rev. Lett.}\ }\textbf {\bibinfo {volume} {136}},\ \bibinfo {pages} {041403} (\bibinfo {year} {2026})},\ \Eprint {https://arxiv.org/abs/2509.08099} {arXiv:2509.08099 [gr-qc]} \BibitemShut {NoStop}%
\bibitem [{\citenamefont {Abac}\ \emph {et~al.}(2025{\natexlab{c}})\citenamefont {Abac} \emph {et~al.}}]{gwtc4Rnp}%
  \BibitemOpen
  \bibfield  {author} {\bibinfo {author} {\bibfnamefont {A.~G.}\ \bibnamefont {Abac}} \emph {et~al.} (\bibinfo {collaboration} {LIGO Scientific, VIRGO, KAGRA}),\ }\bibfield  {title} {\bibinfo {title} {{GWTC-4.0: Population Properties of Merging Compact Binaries}},\ }\href@noop {} {\  (\bibinfo {year} {2025}{\natexlab{c}})},\ \Eprint {https://arxiv.org/abs/2508.18083} {arXiv:2508.18083 [astro-ph.HE]} \BibitemShut {NoStop}%
\bibitem [{\citenamefont {Abac}\ \emph {et~al.}(2025{\natexlab{d}})\citenamefont {Abac} \emph {et~al.}}]{gwtc4Cosmo}%
  \BibitemOpen
  \bibfield  {author} {\bibinfo {author} {\bibfnamefont {A.~G.}\ \bibnamefont {Abac}} \emph {et~al.} (\bibinfo {collaboration} {LIGO Scientific, VIRGO, KAGRA}),\ }\bibfield  {title} {\bibinfo {title} {{GWTC-4.0: Constraints on the Cosmic Expansion Rate and Modified Gravitational-wave Propagation}},\ }\href@noop {} {\  (\bibinfo {year} {2025}{\natexlab{d}})},\ \Eprint {https://arxiv.org/abs/2509.04348} {arXiv:2509.04348 [astro-ph.CO]} \BibitemShut {NoStop}%
\bibitem [{\citenamefont {{Wainstein}}\ and\ \citenamefont {{Zubakov}}(1970)}]{signalextraction}%
  \BibitemOpen
  \bibfield  {author} {\bibinfo {author} {\bibfnamefont {L.~A.}\ \bibnamefont {{Wainstein}}}\ and\ \bibinfo {author} {\bibfnamefont {V.~D.}\ \bibnamefont {{Zubakov}}},\ }\href@noop {} {\emph {\bibinfo {title} {{Extraction of Signals from Noise}}}}\ (\bibinfo {year} {1970})\BibitemShut {NoStop}%
\bibitem [{\citenamefont {Sathyaprakash}\ and\ \citenamefont {Dhurandhar}(1991)}]{Sathyaprakash:1991mt}%
  \BibitemOpen
  \bibfield  {author} {\bibinfo {author} {\bibfnamefont {B.~S.}\ \bibnamefont {Sathyaprakash}}\ and\ \bibinfo {author} {\bibfnamefont {S.~V.}\ \bibnamefont {Dhurandhar}},\ }\bibfield  {title} {\bibinfo {title} {{Choice of filters for the detection of gravitational waves from coalescing binaries}},\ }\href {https://doi.org/10.1103/PhysRevD.44.3819} {\bibfield  {journal} {\bibinfo  {journal} {Phys. Rev. D}\ }\textbf {\bibinfo {volume} {44}},\ \bibinfo {pages} {3819} (\bibinfo {year} {1991})}\BibitemShut {NoStop}%
\bibitem [{\citenamefont {Dhurandhar}\ and\ \citenamefont {Sathyaprakash}(1994)}]{Dhurandhar:1992mw}%
  \BibitemOpen
  \bibfield  {author} {\bibinfo {author} {\bibfnamefont {S.~V.}\ \bibnamefont {Dhurandhar}}\ and\ \bibinfo {author} {\bibfnamefont {B.~S.}\ \bibnamefont {Sathyaprakash}},\ }\bibfield  {title} {\bibinfo {title} {{Choice of filters for the detection of gravitational waves from coalescing binaries. 2. Detection in colored noise}},\ }\href {https://doi.org/10.1103/PhysRevD.49.1707} {\bibfield  {journal} {\bibinfo  {journal} {Phys. Rev. D}\ }\textbf {\bibinfo {volume} {49}},\ \bibinfo {pages} {1707} (\bibinfo {year} {1994})}\BibitemShut {NoStop}%
\bibitem [{\citenamefont {Babak}\ \emph {et~al.}(2006)\citenamefont {Babak}, \citenamefont {Balasubramanian}, \citenamefont {Churches}, \citenamefont {Cokelaer},\ and\ \citenamefont {Sathyaprakash}}]{Babak:templates}%
  \BibitemOpen
  \bibfield  {author} {\bibinfo {author} {\bibfnamefont {S.}~\bibnamefont {Babak}}, \bibinfo {author} {\bibfnamefont {R.}~\bibnamefont {Balasubramanian}}, \bibinfo {author} {\bibfnamefont {D.}~\bibnamefont {Churches}}, \bibinfo {author} {\bibfnamefont {T.}~\bibnamefont {Cokelaer}},\ and\ \bibinfo {author} {\bibfnamefont {B.~S.}\ \bibnamefont {Sathyaprakash}},\ }\bibfield  {title} {\bibinfo {title} {{A Template bank to search for gravitational waves from inspiralling compact binaries. I. Physical models}},\ }\href {https://doi.org/10.1088/0264-9381/23/18/002} {\bibfield  {journal} {\bibinfo  {journal} {Class. Quant. Grav.}\ }\textbf {\bibinfo {volume} {23}},\ \bibinfo {pages} {5477} (\bibinfo {year} {2006})},\ \Eprint {https://arxiv.org/abs/gr-qc/0604037} {arXiv:gr-qc/0604037} \BibitemShut {NoStop}%
\bibitem [{\citenamefont {Gultekin}\ \emph {et~al.}(2006)\citenamefont {Gultekin}, \citenamefont {Coleman~Miller},\ and\ \citenamefont {Hamilton}}]{Gultekin:2005fd}%
  \BibitemOpen
  \bibfield  {author} {\bibinfo {author} {\bibfnamefont {K.}~\bibnamefont {Gultekin}}, \bibinfo {author} {\bibfnamefont {M.}~\bibnamefont {Coleman~Miller}},\ and\ \bibinfo {author} {\bibfnamefont {D.~P.}\ \bibnamefont {Hamilton}},\ }\bibfield  {title} {\bibinfo {title} {{Three-body dynamics with gravitational wave emission}},\ }\href {https://doi.org/10.1086/499917} {\bibfield  {journal} {\bibinfo  {journal} {Astrophys. J.}\ }\textbf {\bibinfo {volume} {640}},\ \bibinfo {pages} {156} (\bibinfo {year} {2006})},\ \Eprint {https://arxiv.org/abs/astro-ph/0509885} {arXiv:astro-ph/0509885} \BibitemShut {NoStop}%
\bibitem [{\citenamefont {Barausse}\ \emph {et~al.}(2015)\citenamefont {Barausse}, \citenamefont {Cardoso},\ and\ \citenamefont {Pani}}]{Barausse:2014pra}%
  \BibitemOpen
  \bibfield  {author} {\bibinfo {author} {\bibfnamefont {E.}~\bibnamefont {Barausse}}, \bibinfo {author} {\bibfnamefont {V.}~\bibnamefont {Cardoso}},\ and\ \bibinfo {author} {\bibfnamefont {P.}~\bibnamefont {Pani}},\ }\bibfield  {title} {\bibinfo {title} {{Environmental Effects for Gravitational-wave Astrophysics}},\ }\href {https://doi.org/10.1088/1742-6596/610/1/012044} {\bibfield  {journal} {\bibinfo  {journal} {J. Phys. Conf. Ser.}\ }\textbf {\bibinfo {volume} {610}},\ \bibinfo {pages} {012044} (\bibinfo {year} {2015})},\ \Eprint {https://arxiv.org/abs/1404.7140} {arXiv:1404.7140 [astro-ph.CO]} \BibitemShut {NoStop}%
\bibitem [{\citenamefont {Cardoso}\ and\ \citenamefont {Maselli}(2020)}]{Cardoso:2019rou}%
  \BibitemOpen
  \bibfield  {author} {\bibinfo {author} {\bibfnamefont {V.}~\bibnamefont {Cardoso}}\ and\ \bibinfo {author} {\bibfnamefont {A.}~\bibnamefont {Maselli}},\ }\bibfield  {title} {\bibinfo {title} {{Constraints on the astrophysical environment of binaries with gravitational-wave observations}},\ }\href {https://doi.org/10.1051/0004-6361/202037654} {\bibfield  {journal} {\bibinfo  {journal} {Astron. Astrophys.}\ }\textbf {\bibinfo {volume} {644}},\ \bibinfo {pages} {A147} (\bibinfo {year} {2020})},\ \Eprint {https://arxiv.org/abs/1909.05870} {arXiv:1909.05870 [astro-ph.HE]} \BibitemShut {NoStop}%
\bibitem [{\citenamefont {Tak{\'a}tsy}\ \emph {et~al.}(2025)\citenamefont {Tak{\'a}tsy}, \citenamefont {Zwick}, \citenamefont {Hendriks}, \citenamefont {Saini}, \citenamefont {Fabj},\ and\ \citenamefont {Samsing}}]{SamsingDephasing}%
  \BibitemOpen
  \bibfield  {author} {\bibinfo {author} {\bibfnamefont {J.}~\bibnamefont {Tak{\'a}tsy}}, \bibinfo {author} {\bibfnamefont {L.}~\bibnamefont {Zwick}}, \bibinfo {author} {\bibfnamefont {K.}~\bibnamefont {Hendriks}}, \bibinfo {author} {\bibfnamefont {P.}~\bibnamefont {Saini}}, \bibinfo {author} {\bibfnamefont {G.}~\bibnamefont {Fabj}},\ and\ \bibinfo {author} {\bibfnamefont {J.}~\bibnamefont {Samsing}},\ }\bibfield  {title} {\bibinfo {title} {{The construction and use of dephasing prescriptions for environmental effects in gravitational wave astronomy}},\ }\href {https://doi.org/10.1088/1361-6382/ae0fd4} {\bibfield  {journal} {\bibinfo  {journal} {Class. Quant. Grav.}\ }\textbf {\bibinfo {volume} {42}},\ \bibinfo {pages} {215006} (\bibinfo {year} {2025})},\ \Eprint {https://arxiv.org/abs/2505.09513} {arXiv:2505.09513 [astro-ph.HE]} \BibitemShut {NoStop}%
\bibitem [{\citenamefont {Li}\ and\ \citenamefont {Lai}(2022)}]{Li:2022pnc}%
  \BibitemOpen
  \bibfield  {author} {\bibinfo {author} {\bibfnamefont {R.}~\bibnamefont {Li}}\ and\ \bibinfo {author} {\bibfnamefont {D.}~\bibnamefont {Lai}},\ }\bibfield  {title} {\bibinfo {title} {{Hydrodynamical evolution of black-hole binaries embedded in AGN discs}},\ }\href {https://doi.org/10.1093/mnras/stac2577} {\bibfield  {journal} {\bibinfo  {journal} {Mon. Not. Roy. Astron. Soc.}\ }\textbf {\bibinfo {volume} {517}},\ \bibinfo {pages} {1602} (\bibinfo {year} {2022})},\ \Eprint {https://arxiv.org/abs/2202.07633} {arXiv:2202.07633 [astro-ph.HE]} \BibitemShut {NoStop}%
\bibitem [{\citenamefont {Codazzo}\ \emph {et~al.}(2023)\citenamefont {Codazzo}, \citenamefont {Di~Giovanni}, \citenamefont {Harms}, \citenamefont {Dall'Amico},\ and\ \citenamefont {Mapelli}}]{PhysRevD.107.023023}%
  \BibitemOpen
  \bibfield  {author} {\bibinfo {author} {\bibfnamefont {E.}~\bibnamefont {Codazzo}}, \bibinfo {author} {\bibfnamefont {M.}~\bibnamefont {Di~Giovanni}}, \bibinfo {author} {\bibfnamefont {J.}~\bibnamefont {Harms}}, \bibinfo {author} {\bibfnamefont {M.}~\bibnamefont {Dall'Amico}},\ and\ \bibinfo {author} {\bibfnamefont {M.}~\bibnamefont {Mapelli}},\ }\bibfield  {title} {\bibinfo {title} {Study on the detectability of gravitational radiation from single-binary encounters between black holes in nuclear star clusters: The case of hyperbolic flybys},\ }\href {https://doi.org/10.1103/PhysRevD.107.023023} {\bibfield  {journal} {\bibinfo  {journal} {Phys. Rev. D}\ }\textbf {\bibinfo {volume} {107}},\ \bibinfo {pages} {023023} (\bibinfo {year} {2023})}\BibitemShut {NoStop}%
\bibitem [{\citenamefont {Hendriks}\ \emph {et~al.}(2024)\citenamefont {Hendriks}, \citenamefont {Atallah}, \citenamefont {Martinez}, \citenamefont {Zevin}, \citenamefont {Zwick}, \citenamefont {Trani}, \citenamefont {Saini}, \citenamefont {Tak{\'a}tsy},\ and\ \citenamefont {Samsing}}]{Hendriks:2024gpp}%
  \BibitemOpen
  \bibfield  {author} {\bibinfo {author} {\bibfnamefont {K.}~\bibnamefont {Hendriks}}, \bibinfo {author} {\bibfnamefont {D.}~\bibnamefont {Atallah}}, \bibinfo {author} {\bibfnamefont {M.}~\bibnamefont {Martinez}}, \bibinfo {author} {\bibfnamefont {M.}~\bibnamefont {Zevin}}, \bibinfo {author} {\bibfnamefont {L.}~\bibnamefont {Zwick}}, \bibinfo {author} {\bibfnamefont {A.~A.}\ \bibnamefont {Trani}}, \bibinfo {author} {\bibfnamefont {P.}~\bibnamefont {Saini}}, \bibinfo {author} {\bibfnamefont {J.}~\bibnamefont {Tak{\'a}tsy}},\ and\ \bibinfo {author} {\bibfnamefont {J.}~\bibnamefont {Samsing}},\ }\bibfield  {title} {\bibinfo {title} {{Large Gravitational Wave Phase Shifts from Strong 3-body Interactions in Dense Stellar Clusters}},\ }\href@noop {} {\  (\bibinfo {year} {2024})},\ \Eprint {https://arxiv.org/abs/2411.08572} {arXiv:2411.08572 [astro-ph.HE]} \BibitemShut {NoStop}%
\bibitem [{\citenamefont {Maccarone}\ \emph {et~al.}(2007)\citenamefont {Maccarone}, \citenamefont {Kundu}, \citenamefont {Zepf},\ and\ \citenamefont {Rhode}}]{Maccarone:2007dd}%
  \BibitemOpen
  \bibfield  {author} {\bibinfo {author} {\bibfnamefont {T.~J.}\ \bibnamefont {Maccarone}}, \bibinfo {author} {\bibfnamefont {A.}~\bibnamefont {Kundu}}, \bibinfo {author} {\bibfnamefont {S.~E.}\ \bibnamefont {Zepf}},\ and\ \bibinfo {author} {\bibfnamefont {K.~L.}\ \bibnamefont {Rhode}},\ }\bibfield  {title} {\bibinfo {title} {{A black hole in a globular cluster}},\ }\href {https://doi.org/10.1038/nature05434} {\bibfield  {journal} {\bibinfo  {journal} {Nature}\ }\textbf {\bibinfo {volume} {445}},\ \bibinfo {pages} {183} (\bibinfo {year} {2007})},\ \Eprint {https://arxiv.org/abs/astro-ph/0701310} {arXiv:astro-ph/0701310} \BibitemShut {NoStop}%
\bibitem [{\citenamefont {{Downing}}\ \emph {et~al.}(2010)\citenamefont {{Downing}}, \citenamefont {{Benacquista}}, \citenamefont {{Giersz}},\ and\ \citenamefont {{Spurzem}}}]{2010MNRAS.407.1946D}%
  \BibitemOpen
  \bibfield  {author} {\bibinfo {author} {\bibfnamefont {J.~M.~B.}\ \bibnamefont {{Downing}}}, \bibinfo {author} {\bibfnamefont {M.~J.}\ \bibnamefont {{Benacquista}}}, \bibinfo {author} {\bibfnamefont {M.}~\bibnamefont {{Giersz}}},\ and\ \bibinfo {author} {\bibfnamefont {R.}~\bibnamefont {{Spurzem}}},\ }\bibfield  {title} {\bibinfo {title} {{Compact binaries in star clusters - I. Black hole binaries inside globular clusters}},\ }\href {https://doi.org/10.1111/j.1365-2966.2010.17040.x} {\bibfield  {journal} {\bibinfo  {journal} {\mnras}\ }\textbf {\bibinfo {volume} {407}},\ \bibinfo {pages} {1946} (\bibinfo {year} {2010})},\ \Eprint {https://arxiv.org/abs/0910.0546} {arXiv:0910.0546 [astro-ph.SR]} \BibitemShut {NoStop}%
\bibitem [{\citenamefont {Tiwari}\ \emph {et~al.}(2023)\citenamefont {Tiwari}, \citenamefont {Vijaykumar}, \citenamefont {Kapadia}, \citenamefont {Fragione},\ and\ \citenamefont {Chatterjee}}]{Tiwari:2023cpa}%
  \BibitemOpen
  \bibfield  {author} {\bibinfo {author} {\bibfnamefont {A.}~\bibnamefont {Tiwari}}, \bibinfo {author} {\bibfnamefont {A.}~\bibnamefont {Vijaykumar}}, \bibinfo {author} {\bibfnamefont {S.~J.}\ \bibnamefont {Kapadia}}, \bibinfo {author} {\bibfnamefont {G.}~\bibnamefont {Fragione}},\ and\ \bibinfo {author} {\bibfnamefont {S.}~\bibnamefont {Chatterjee}},\ }\bibfield  {title} {\bibinfo {title} {{Accelerated binary black holes in globular clusters: forecasts and detectability in the era of space-based gravitational-wave detectors}},\ }\href {https://doi.org/10.1093/mnras/stad3749} {\bibfield  {journal} {\bibinfo  {journal} {Mon. Not. Roy. Astron. Soc.}\ }\textbf {\bibinfo {volume} {527}},\ \bibinfo {pages} {8586} (\bibinfo {year} {2023})},\ \Eprint {https://arxiv.org/abs/2307.00930} {arXiv:2307.00930 [astro-ph.HE]} \BibitemShut {NoStop}%
\bibitem [{\citenamefont {{Torniamenti}}\ \emph {et~al.}(2022)\citenamefont {{Torniamenti}}, \citenamefont {{Rastello}}, \citenamefont {{Mapelli}}, \citenamefont {{Di Carlo}}, \citenamefont {{Ballone}},\ and\ \citenamefont {{Pasquato}}}]{2022MNRAS.517.2953T}%
  \BibitemOpen
  \bibfield  {author} {\bibinfo {author} {\bibfnamefont {S.}~\bibnamefont {{Torniamenti}}}, \bibinfo {author} {\bibfnamefont {S.}~\bibnamefont {{Rastello}}}, \bibinfo {author} {\bibfnamefont {M.}~\bibnamefont {{Mapelli}}}, \bibinfo {author} {\bibfnamefont {U.~N.}\ \bibnamefont {{Di Carlo}}}, \bibinfo {author} {\bibfnamefont {A.}~\bibnamefont {{Ballone}}},\ and\ \bibinfo {author} {\bibfnamefont {M.}~\bibnamefont {{Pasquato}}},\ }\bibfield  {title} {\bibinfo {title} {{Dynamics of binary black holes in young star clusters: the impact of cluster mass and long-term evolution}},\ }\href {https://doi.org/10.1093/mnras/stac2841} {\bibfield  {journal} {\bibinfo  {journal} {\mnras}\ }\textbf {\bibinfo {volume} {517}},\ \bibinfo {pages} {2953} (\bibinfo {year} {2022})},\ \Eprint {https://arxiv.org/abs/2203.08163} {arXiv:2203.08163 [astro-ph.GA]} \BibitemShut {NoStop}%
\bibitem [{\citenamefont {Mapelli}\ \emph {et~al.}(2022)\citenamefont {Mapelli}, \citenamefont {Bouffanais}, \citenamefont {Santoliquido}, \citenamefont {Sedda},\ and\ \citenamefont {Artale}}]{Mapelli:2021gyv}%
  \BibitemOpen
  \bibfield  {author} {\bibinfo {author} {\bibfnamefont {M.}~\bibnamefont {Mapelli}}, \bibinfo {author} {\bibfnamefont {Y.}~\bibnamefont {Bouffanais}}, \bibinfo {author} {\bibfnamefont {F.}~\bibnamefont {Santoliquido}}, \bibinfo {author} {\bibfnamefont {M.~A.}\ \bibnamefont {Sedda}},\ and\ \bibinfo {author} {\bibfnamefont {M.~C.}\ \bibnamefont {Artale}},\ }\bibfield  {title} {\bibinfo {title} {{The cosmic evolution of binary black holes in young, globular, and nuclear star clusters: rates, masses, spins, and mixing fractions}},\ }\href {https://doi.org/10.1093/mnras/stac422} {\bibfield  {journal} {\bibinfo  {journal} {Mon. Not. Roy. Astron. Soc.}\ }\textbf {\bibinfo {volume} {511}},\ \bibinfo {pages} {5797} (\bibinfo {year} {2022})},\ \Eprint {https://arxiv.org/abs/2109.06222} {arXiv:2109.06222 [astro-ph.HE]} \BibitemShut {NoStop}%
\bibitem [{\citenamefont {Joshi}\ \emph {et~al.}(2025)\citenamefont {Joshi}, \citenamefont {Bhake}, \citenamefont {Banerjee}, \citenamefont {Vaidya}, \citenamefont {Ruiz}, \citenamefont {Tsokaros}, \citenamefont {Mignone}, \citenamefont {Branchesi}, \citenamefont {Shukla},\ and\ \citenamefont {{\v{C}}emelji{\'c}}}]{Joshi:2025wcu}%
  \BibitemOpen
  \bibfield  {author} {\bibinfo {author} {\bibfnamefont {R.~K.}\ \bibnamefont {Joshi}}, \bibinfo {author} {\bibfnamefont {A.}~\bibnamefont {Bhake}}, \bibinfo {author} {\bibfnamefont {B.}~\bibnamefont {Banerjee}}, \bibinfo {author} {\bibfnamefont {B.}~\bibnamefont {Vaidya}}, \bibinfo {author} {\bibfnamefont {M.}~\bibnamefont {Ruiz}}, \bibinfo {author} {\bibfnamefont {A.}~\bibnamefont {Tsokaros}}, \bibinfo {author} {\bibfnamefont {A.}~\bibnamefont {Mignone}}, \bibinfo {author} {\bibfnamefont {M.}~\bibnamefont {Branchesi}}, \bibinfo {author} {\bibfnamefont {A.}~\bibnamefont {Shukla}},\ and\ \bibinfo {author} {\bibfnamefont {M.}~\bibnamefont {{\v{C}}emelji{\'c}}},\ }\bibfield  {title} {\bibinfo {title} {{Binary black holes in magnetized disks of active galactic nuclei}},\ }\href {https://doi.org/10.1051/0004-6361/202555874} {\bibfield  {journal} {\bibinfo  {journal} {Astron. Astrophys.}\ }\textbf {\bibinfo {volume} {703}},\ \bibinfo {pages} {A304} (\bibinfo {year} {2025})},\ \Eprint
  {https://arxiv.org/abs/2509.16796} {arXiv:2509.16796 [astro-ph.HE]} \BibitemShut {NoStop}%
\bibitem [{\citenamefont {Vijaykumar}\ \emph {et~al.}(2023)\citenamefont {Vijaykumar}, \citenamefont {Tiwari}, \citenamefont {Kapadia}, \citenamefont {Arun},\ and\ \citenamefont {Ajith}}]{Vijaykumar:2023tjg}%
  \BibitemOpen
  \bibfield  {author} {\bibinfo {author} {\bibfnamefont {A.}~\bibnamefont {Vijaykumar}}, \bibinfo {author} {\bibfnamefont {A.}~\bibnamefont {Tiwari}}, \bibinfo {author} {\bibfnamefont {S.~J.}\ \bibnamefont {Kapadia}}, \bibinfo {author} {\bibfnamefont {K.~G.}\ \bibnamefont {Arun}},\ and\ \bibinfo {author} {\bibfnamefont {P.}~\bibnamefont {Ajith}},\ }\bibfield  {title} {\bibinfo {title} {{Waltzing Binaries: Probing the Line-of-sight Acceleration of Merging Compact Objects with Gravitational Waves}},\ }\href {https://doi.org/10.3847/1538-4357/acd77d} {\bibfield  {journal} {\bibinfo  {journal} {Astrophys. J.}\ }\textbf {\bibinfo {volume} {954}},\ \bibinfo {pages} {105} (\bibinfo {year} {2023})},\ \Eprint {https://arxiv.org/abs/2302.09651} {arXiv:2302.09651 [astro-ph.HE]} \BibitemShut {NoStop}%
\bibitem [{\citenamefont {Britt}\ \emph {et~al.}(2021)\citenamefont {Britt}, \citenamefont {Johanson}, \citenamefont {Wood}, \citenamefont {Miller},\ and\ \citenamefont {Michaely}}]{Britt:2021dtg}%
  \BibitemOpen
  \bibfield  {author} {\bibinfo {author} {\bibfnamefont {D.}~\bibnamefont {Britt}}, \bibinfo {author} {\bibfnamefont {B.}~\bibnamefont {Johanson}}, \bibinfo {author} {\bibfnamefont {L.}~\bibnamefont {Wood}}, \bibinfo {author} {\bibfnamefont {M.~C.}\ \bibnamefont {Miller}},\ and\ \bibinfo {author} {\bibfnamefont {E.}~\bibnamefont {Michaely}},\ }\bibfield  {title} {\bibinfo {title} {{Binary black hole mergers from hierarchical triples in open clusters}},\ }\href {https://doi.org/10.1093/mnras/stab1570} {\bibfield  {journal} {\bibinfo  {journal} {Mon. Not. Roy. Astron. Soc.}\ }\textbf {\bibinfo {volume} {505}},\ \bibinfo {pages} {3844} (\bibinfo {year} {2021})},\ \Eprint {https://arxiv.org/abs/2103.14706} {arXiv:2103.14706 [astro-ph.HE]} \BibitemShut {NoStop}%
\bibitem [{\citenamefont {Bhalla}\ \emph {et~al.}(2025{\natexlab{a}})\citenamefont {Bhalla}, \citenamefont {Lehmann}, \citenamefont {Sinha},\ and\ \citenamefont {Xu}}]{Bhalla:2024jbu}%
  \BibitemOpen
  \bibfield  {author} {\bibinfo {author} {\bibfnamefont {B.}~\bibnamefont {Bhalla}}, \bibinfo {author} {\bibfnamefont {B.~V.}\ \bibnamefont {Lehmann}}, \bibinfo {author} {\bibfnamefont {K.}~\bibnamefont {Sinha}},\ and\ \bibinfo {author} {\bibfnamefont {T.}~\bibnamefont {Xu}},\ }\bibfield  {title} {\bibinfo {title} {{Three-body exchanges with primordial black holes}},\ }\href {https://doi.org/10.1103/PhysRevD.111.043029} {\bibfield  {journal} {\bibinfo  {journal} {Phys. Rev. D}\ }\textbf {\bibinfo {volume} {111}},\ \bibinfo {pages} {043029} (\bibinfo {year} {2025}{\natexlab{a}})},\ \Eprint {https://arxiv.org/abs/2408.04697} {arXiv:2408.04697 [hep-ph]} \BibitemShut {NoStop}%
\bibitem [{\citenamefont {Biermann}\ \emph {et~al.}(2002)\citenamefont {Biermann}, \citenamefont {Chirvasa}, \citenamefont {Falcke}, \citenamefont {Markoff},\ and\ \citenamefont {Zier}}]{Biermann:2002ky}%
  \BibitemOpen
  \bibfield  {author} {\bibinfo {author} {\bibfnamefont {P.~L.}\ \bibnamefont {Biermann}}, \bibinfo {author} {\bibfnamefont {M.}~\bibnamefont {Chirvasa}}, \bibinfo {author} {\bibfnamefont {H.}~\bibnamefont {Falcke}}, \bibinfo {author} {\bibfnamefont {S.}~\bibnamefont {Markoff}},\ and\ \bibinfo {author} {\bibfnamefont {C.}~\bibnamefont {Zier}},\ }\bibfield  {title} {\bibinfo {title} {{Single and binary black holes and their active environment}},\ }\href@noop {} {\  (\bibinfo {year} {2002})},\ \Eprint {https://arxiv.org/abs/astro-ph/0211503} {arXiv:astro-ph/0211503} \BibitemShut {NoStop}%
\bibitem [{\citenamefont {Bhalla}\ \emph {et~al.}(2025{\natexlab{b}})\citenamefont {Bhalla}, \citenamefont {Lehmann}, \citenamefont {Sinha},\ and\ \citenamefont {Xu}}]{Bhalla:2025xce}%
  \BibitemOpen
  \bibfield  {author} {\bibinfo {author} {\bibfnamefont {B.}~\bibnamefont {Bhalla}}, \bibinfo {author} {\bibfnamefont {B.~V.}\ \bibnamefont {Lehmann}}, \bibinfo {author} {\bibfnamefont {K.}~\bibnamefont {Sinha}},\ and\ \bibinfo {author} {\bibfnamefont {T.}~\bibnamefont {Xu}},\ }\bibfield  {title} {\bibinfo {title} {{Fast and Fewrious: Stochastic binary perturbations from fast compact objects}},\ }\href@noop {} {\  (\bibinfo {year} {2025}{\natexlab{b}})},\ \Eprint {https://arxiv.org/abs/2507.16894} {arXiv:2507.16894 [hep-ph]} \BibitemShut {NoStop}%
\bibitem [{\citenamefont {De~Lorenci}\ \emph {et~al.}(2025)\citenamefont {De~Lorenci}, \citenamefont {Kaiser}, \citenamefont {Peter}, \citenamefont {Ruiz},\ and\ \citenamefont {Wolfe}}]{DeLorenci:2025wbn}%
  \BibitemOpen
  \bibfield  {author} {\bibinfo {author} {\bibfnamefont {V.~A.}\ \bibnamefont {De~Lorenci}}, \bibinfo {author} {\bibfnamefont {D.~I.}\ \bibnamefont {Kaiser}}, \bibinfo {author} {\bibfnamefont {P.}~\bibnamefont {Peter}}, \bibinfo {author} {\bibfnamefont {L.~S.}\ \bibnamefont {Ruiz}},\ and\ \bibinfo {author} {\bibfnamefont {N.~E.}\ \bibnamefont {Wolfe}},\ }\bibfield  {title} {\bibinfo {title} {{Gravitational wave signals from primordial black holes orbiting solar-type stars}},\ }\href {https://doi.org/10.1103/294z-nfj4} {\bibfield  {journal} {\bibinfo  {journal} {Phys. Rev. D}\ }\textbf {\bibinfo {volume} {112}},\ \bibinfo {pages} {063063} (\bibinfo {year} {2025})},\ \Eprint {https://arxiv.org/abs/2504.07517} {arXiv:2504.07517 [gr-qc]} \BibitemShut {NoStop}%
\bibitem [{\citenamefont {van Die}\ \emph {et~al.}(2025)\citenamefont {van Die}, \citenamefont {Rapoport}, \citenamefont {Ginat},\ and\ \citenamefont {Desjacques}}]{vanDie:2024htf}%
  \BibitemOpen
  \bibfield  {author} {\bibinfo {author} {\bibfnamefont {F.}~\bibnamefont {van Die}}, \bibinfo {author} {\bibfnamefont {I.}~\bibnamefont {Rapoport}}, \bibinfo {author} {\bibfnamefont {Y.~B.}\ \bibnamefont {Ginat}},\ and\ \bibinfo {author} {\bibfnamefont {V.}~\bibnamefont {Desjacques}},\ }\bibfield  {title} {\bibinfo {title} {{Detection prospects for the GW background of galactic (sub)solar mass primordial black holes}},\ }\href {https://doi.org/10.1088/1475-7516/2025/05/036} {\bibfield  {journal} {\bibinfo  {journal} {JCAP}\ }\textbf {\bibinfo {volume} {05}},\ \bibinfo {pages} {036}},\ \Eprint {https://arxiv.org/abs/2410.04522} {arXiv:2410.04522 [astro-ph.CO]} \BibitemShut {NoStop}%
\bibitem [{\citenamefont {G{\`o}mez-Aguilar}\ \emph {et~al.}(2025)\citenamefont {G{\`o}mez-Aguilar}, \citenamefont {Erfani},\ and\ \citenamefont {Jim{\`e}nez~Cruz}}]{Gomez-Aguilar:2025wss}%
  \BibitemOpen
  \bibfield  {author} {\bibinfo {author} {\bibfnamefont {T.~D.}\ \bibnamefont {G{\`o}mez-Aguilar}}, \bibinfo {author} {\bibfnamefont {E.}~\bibnamefont {Erfani}},\ and\ \bibinfo {author} {\bibfnamefont {N.~M.}\ \bibnamefont {Jim{\`e}nez~Cruz}},\ }\bibfield  {title} {\bibinfo {title} {{Gravitational Waves from Hyperbolic Encounters of Primordial Black Holes in Dwarf Galaxies}},\ }\href@noop {} {\  (\bibinfo {year} {2025})},\ \Eprint {https://arxiv.org/abs/2509.19462} {arXiv:2509.19462 [astro-ph.CO]} \BibitemShut {NoStop}%
\bibitem [{\citenamefont {Garc{\'\i}a-Bellido}\ \emph {et~al.}(2022)\citenamefont {Garc{\'\i}a-Bellido}, \citenamefont {Jaraba},\ and\ \citenamefont {Kuroyanagi}}]{Garcia-Bellido:2021jlq}%
  \BibitemOpen
  \bibfield  {author} {\bibinfo {author} {\bibfnamefont {J.}~\bibnamefont {Garc{\'\i}a-Bellido}}, \bibinfo {author} {\bibfnamefont {S.}~\bibnamefont {Jaraba}},\ and\ \bibinfo {author} {\bibfnamefont {S.}~\bibnamefont {Kuroyanagi}},\ }\bibfield  {title} {\bibinfo {title} {{The stochastic gravitational wave background from close hyperbolic encounters of primordial black holes in dense clusters}},\ }\href {https://doi.org/10.1016/j.dark.2022.101009} {\bibfield  {journal} {\bibinfo  {journal} {Phys. Dark Univ.}\ }\textbf {\bibinfo {volume} {36}},\ \bibinfo {pages} {101009} (\bibinfo {year} {2022})},\ \Eprint {https://arxiv.org/abs/2109.11376} {arXiv:2109.11376 [gr-qc]} \BibitemShut {NoStop}%
\bibitem [{\citenamefont {Afroz}\ and\ \citenamefont {Mukherjee}(2025)}]{Afroz:2025urb}%
  \BibitemOpen
  \bibfield  {author} {\bibinfo {author} {\bibfnamefont {S.}~\bibnamefont {Afroz}}\ and\ \bibinfo {author} {\bibfnamefont {S.}~\bibnamefont {Mukherjee}},\ }\bibfield  {title} {\bibinfo {title} {{Gravitational Wave Burst from Bremsstrahlung in Milky Way Can Discover Sub-Solar Dark Matter in Near Future}},\ }\href@noop {} {\  (\bibinfo {year} {2025})},\ \Eprint {https://arxiv.org/abs/2507.22126} {arXiv:2507.22126 [astro-ph.CO]} \BibitemShut {NoStop}%
\bibitem [{\citenamefont {{Chandrasekhar}}(1931)}]{chandra1931}%
  \BibitemOpen
  \bibfield  {author} {\bibinfo {author} {\bibfnamefont {S.}~\bibnamefont {{Chandrasekhar}}},\ }\bibfield  {title} {\bibinfo {title} {{The Maximum Mass of Ideal White Dwarfs}},\ }\href {https://doi.org/10.1086/143324} {\bibfield  {journal} {\bibinfo  {journal} {\apj}\ }\textbf {\bibinfo {volume} {74}},\ \bibinfo {pages} {81} (\bibinfo {year} {1931})}\BibitemShut {NoStop}%
\bibitem [{\citenamefont {Tolman}(1939)}]{Tolman:1939jz}%
  \BibitemOpen
  \bibfield  {author} {\bibinfo {author} {\bibfnamefont {R.~C.}\ \bibnamefont {Tolman}},\ }\bibfield  {title} {\bibinfo {title} {{Static solutions of Einstein's field equations for spheres of fluid}},\ }\href {https://doi.org/10.1103/PhysRev.55.364} {\bibfield  {journal} {\bibinfo  {journal} {Phys. Rev.}\ }\textbf {\bibinfo {volume} {55}},\ \bibinfo {pages} {364} (\bibinfo {year} {1939})}\BibitemShut {NoStop}%
\bibitem [{\citenamefont {Oppenheimer}\ and\ \citenamefont {Volkoff}(1939)}]{Oppenheimer:1939ne}%
  \BibitemOpen
  \bibfield  {author} {\bibinfo {author} {\bibfnamefont {J.~R.}\ \bibnamefont {Oppenheimer}}\ and\ \bibinfo {author} {\bibfnamefont {G.~M.}\ \bibnamefont {Volkoff}},\ }\bibfield  {title} {\bibinfo {title} {{On massive neutron cores}},\ }\href {https://doi.org/10.1103/PhysRev.55.374} {\bibfield  {journal} {\bibinfo  {journal} {Phys. Rev.}\ }\textbf {\bibinfo {volume} {55}},\ \bibinfo {pages} {374} (\bibinfo {year} {1939})}\BibitemShut {NoStop}%
\bibitem [{\citenamefont {Hawking}(1971)}]{Hawking:1971ei}%
  \BibitemOpen
  \bibfield  {author} {\bibinfo {author} {\bibfnamefont {S.}~\bibnamefont {Hawking}},\ }\bibfield  {title} {\bibinfo {title} {{Gravitationally collapsed objects of very low mass}},\ }\href {https://doi.org/10.1093/mnras/152.1.75} {\bibfield  {journal} {\bibinfo  {journal} {Mon. Not. Roy. Astron. Soc.}\ }\textbf {\bibinfo {volume} {152}},\ \bibinfo {pages} {75} (\bibinfo {year} {1971})}\BibitemShut {NoStop}%
\bibitem [{\citenamefont {Kouvaris}\ \emph {et~al.}(2018)\citenamefont {Kouvaris}, \citenamefont {Tinyakov},\ and\ \citenamefont {Tytgat}}]{Kouvaris:2018wnh}%
  \BibitemOpen
  \bibfield  {author} {\bibinfo {author} {\bibfnamefont {C.}~\bibnamefont {Kouvaris}}, \bibinfo {author} {\bibfnamefont {P.}~\bibnamefont {Tinyakov}},\ and\ \bibinfo {author} {\bibfnamefont {M.~H.~G.}\ \bibnamefont {Tytgat}},\ }\bibfield  {title} {\bibinfo {title} {{NonPrimordial Solar Mass Black Holes}},\ }\href {https://doi.org/10.1103/PhysRevLett.121.221102} {\bibfield  {journal} {\bibinfo  {journal} {Phys. Rev. Lett.}\ }\textbf {\bibinfo {volume} {121}},\ \bibinfo {pages} {221102} (\bibinfo {year} {2018})},\ \Eprint {https://arxiv.org/abs/1804.06740} {arXiv:1804.06740 [astro-ph.HE]} \BibitemShut {NoStop}%
\bibitem [{\citenamefont {Samsing}\ \emph {et~al.}(2025)\citenamefont {Samsing}, \citenamefont {Hendriks}, \citenamefont {Zwick}, \citenamefont {D'Orazio},\ and\ \citenamefont {Liu}}]{Samsing:2024syt}%
  \BibitemOpen
  \bibfield  {author} {\bibinfo {author} {\bibfnamefont {J.}~\bibnamefont {Samsing}}, \bibinfo {author} {\bibfnamefont {K.}~\bibnamefont {Hendriks}}, \bibinfo {author} {\bibfnamefont {L.}~\bibnamefont {Zwick}}, \bibinfo {author} {\bibfnamefont {D.~J.}\ \bibnamefont {D'Orazio}},\ and\ \bibinfo {author} {\bibfnamefont {B.}~\bibnamefont {Liu}},\ }\bibfield  {title} {\bibinfo {title} {{Gravitational-wave Phase Shifts in Eccentric Black Hole Mergers as a Probe of Dynamical Formation Environments}},\ }\href {https://doi.org/10.3847/1538-4357/ad9f3d} {\bibfield  {journal} {\bibinfo  {journal} {Astrophys. J.}\ }\textbf {\bibinfo {volume} {990}},\ \bibinfo {pages} {211} (\bibinfo {year} {2025})},\ \Eprint {https://arxiv.org/abs/2403.05625} {arXiv:2403.05625 [astro-ph.HE]} \BibitemShut {NoStop}%
\bibitem [{\citenamefont {Hendriks}\ \emph {et~al.}(2025)\citenamefont {Hendriks}, \citenamefont {Zwick},\ and\ \citenamefont {Samsing}}]{Hendriks:2024zbu}%
  \BibitemOpen
  \bibfield  {author} {\bibinfo {author} {\bibfnamefont {K.}~\bibnamefont {Hendriks}}, \bibinfo {author} {\bibfnamefont {L.}~\bibnamefont {Zwick}},\ and\ \bibinfo {author} {\bibfnamefont {J.}~\bibnamefont {Samsing}},\ }\bibfield  {title} {\bibinfo {title} {{Eccentric Features in the Gravitational-wave Phase of Dynamically Formed Black Hole Binaries}},\ }\href {https://doi.org/10.3847/1538-4357/adcb35} {\bibfield  {journal} {\bibinfo  {journal} {Astrophys. J.}\ }\textbf {\bibinfo {volume} {985}},\ \bibinfo {pages} {252} (\bibinfo {year} {2025})},\ \Eprint {https://arxiv.org/abs/2408.04603} {arXiv:2408.04603 [gr-qc]} \BibitemShut {NoStop}%
\bibitem [{\citenamefont {Zwick}\ \emph {et~al.}(2025)\citenamefont {Zwick}, \citenamefont {Tak{\'a}tsy}, \citenamefont {Saini}, \citenamefont {Hendriks}, \citenamefont {Samsing}, \citenamefont {Tiede}, \citenamefont {Rowan},\ and\ \citenamefont {Trani}}]{Zwick:2025wkt}%
  \BibitemOpen
  \bibfield  {author} {\bibinfo {author} {\bibfnamefont {L.}~\bibnamefont {Zwick}}, \bibinfo {author} {\bibfnamefont {J.}~\bibnamefont {Tak{\'a}tsy}}, \bibinfo {author} {\bibfnamefont {P.}~\bibnamefont {Saini}}, \bibinfo {author} {\bibfnamefont {K.}~\bibnamefont {Hendriks}}, \bibinfo {author} {\bibfnamefont {J.}~\bibnamefont {Samsing}}, \bibinfo {author} {\bibfnamefont {C.}~\bibnamefont {Tiede}}, \bibinfo {author} {\bibfnamefont {C.}~\bibnamefont {Rowan}},\ and\ \bibinfo {author} {\bibfnamefont {A.~A.}\ \bibnamefont {Trani}},\ }\bibfield  {title} {\bibinfo {title} {{Environmental Effects in Stellar Mass Gravitational-wave Sources. I. Expected Fraction of Signals with Significant Dephasing in the Dynamical and Active Galactic Nucleus Channels}},\ }\href {https://doi.org/10.3847/1538-4357/adf6b8} {\bibfield  {journal} {\bibinfo  {journal} {Astrophys. J.}\ }\textbf {\bibinfo {volume} {991}},\ \bibinfo {pages} {131} (\bibinfo {year} {2025})},\ \Eprint {https://arxiv.org/abs/2503.24084} {arXiv:2503.24084
  [astro-ph.HE]} \BibitemShut {NoStop}%
\bibitem [{\citenamefont {Yunes}\ \emph {et~al.}(2011)\citenamefont {Yunes}, \citenamefont {Coleman~Miller},\ and\ \citenamefont {Thornburg}}]{Yunes:2010sm}%
  \BibitemOpen
  \bibfield  {author} {\bibinfo {author} {\bibfnamefont {N.}~\bibnamefont {Yunes}}, \bibinfo {author} {\bibfnamefont {M.}~\bibnamefont {Coleman~Miller}},\ and\ \bibinfo {author} {\bibfnamefont {J.}~\bibnamefont {Thornburg}},\ }\bibfield  {title} {\bibinfo {title} {{The Effect of Massive Perturbers on Extreme Mass-Ratio Inspiral Waveforms}},\ }\href {https://doi.org/10.1103/PhysRevD.83.044030} {\bibfield  {journal} {\bibinfo  {journal} {Phys. Rev. D}\ }\textbf {\bibinfo {volume} {83}},\ \bibinfo {pages} {044030} (\bibinfo {year} {2011})},\ \Eprint {https://arxiv.org/abs/1010.1721} {arXiv:1010.1721 [astro-ph.GA]} \BibitemShut {NoStop}%
\bibitem [{\citenamefont {Bahramian}\ \emph {et~al.}(2013)\citenamefont {Bahramian}, \citenamefont {Heinke}, \citenamefont {Sivakoff},\ and\ \citenamefont {Gladstone}}]{Bahramian:2013ihw}%
  \BibitemOpen
  \bibfield  {author} {\bibinfo {author} {\bibfnamefont {A.}~\bibnamefont {Bahramian}}, \bibinfo {author} {\bibfnamefont {C.~O.}\ \bibnamefont {Heinke}}, \bibinfo {author} {\bibfnamefont {G.~R.}\ \bibnamefont {Sivakoff}},\ and\ \bibinfo {author} {\bibfnamefont {J.~C.}\ \bibnamefont {Gladstone}},\ }\bibfield  {title} {\bibinfo {title} {{Stellar Encounter Rate in Galactic Globular Clusters}},\ }\href {https://doi.org/10.1088/0004-637X/766/2/136} {\bibfield  {journal} {\bibinfo  {journal} {Astrophys. J.}\ }\textbf {\bibinfo {volume} {766}},\ \bibinfo {pages} {136} (\bibinfo {year} {2013})},\ \Eprint {https://arxiv.org/abs/1302.2549} {arXiv:1302.2549 [astro-ph.HE]} \BibitemShut {NoStop}%
\bibitem [{\citenamefont {{Neumayer}}\ \emph {et~al.}(2020)\citenamefont {{Neumayer}}, \citenamefont {{Seth}},\ and\ \citenamefont {{B{\"o}ker}}}]{NSC2020}%
  \BibitemOpen
  \bibfield  {author} {\bibinfo {author} {\bibfnamefont {N.}~\bibnamefont {{Neumayer}}}, \bibinfo {author} {\bibfnamefont {A.}~\bibnamefont {{Seth}}},\ and\ \bibinfo {author} {\bibfnamefont {T.}~\bibnamefont {{B{\"o}ker}}},\ }\bibfield  {title} {\bibinfo {title} {{Nuclear star clusters}},\ }\href {https://doi.org/10.1007/s00159-020-00125-0} {\bibfield  {journal} {\bibinfo  {journal} {\aapr}\ }\textbf {\bibinfo {volume} {28}},\ \bibinfo {eid} {4} (\bibinfo {year} {2020})},\ \Eprint {https://arxiv.org/abs/2001.03626} {arXiv:2001.03626 [astro-ph.GA]} \BibitemShut {NoStop}%
\bibitem [{\citenamefont {{B{\"o}ker}}(2010)}]{NSC2010}%
  \BibitemOpen
  \bibfield  {author} {\bibinfo {author} {\bibfnamefont {T.}~\bibnamefont {{B{\"o}ker}}},\ }\bibfield  {title} {\bibinfo {title} {{Nuclear star clusters}},\ }in\ \href {https://doi.org/10.1017/S1743921309990871} {\emph {\bibinfo {booktitle} {Star Clusters: Basic Galactic Building Blocks Throughout Time and Space}}},\ \bibinfo {series} {IAU Symposium}, Vol.\ \bibinfo {volume} {266},\ \bibinfo {editor} {edited by\ \bibinfo {editor} {\bibfnamefont {R.}~\bibnamefont {{de Grijs}}}\ and\ \bibinfo {editor} {\bibfnamefont {J.~R.~D.}\ \bibnamefont {{L{\'e}pine}}}}\ (\bibinfo {year} {2010})\ pp.\ \bibinfo {pages} {58--63},\ \Eprint {https://arxiv.org/abs/0910.4863} {arXiv:0910.4863 [astro-ph.CO]} \BibitemShut {NoStop}%
\bibitem [{\citenamefont {Gnedin}\ \emph {et~al.}(2014)\citenamefont {Gnedin}, \citenamefont {Ostriker},\ and\ \citenamefont {Tremaine}}]{Gnedin:2013cda}%
  \BibitemOpen
  \bibfield  {author} {\bibinfo {author} {\bibfnamefont {O.~Y.}\ \bibnamefont {Gnedin}}, \bibinfo {author} {\bibfnamefont {J.~P.}\ \bibnamefont {Ostriker}},\ and\ \bibinfo {author} {\bibfnamefont {S.}~\bibnamefont {Tremaine}},\ }\bibfield  {title} {\bibinfo {title} {{Co-Evolution of Galactic Nuclei and Globular Cluster Systems}},\ }\href {https://doi.org/10.1088/0004-637X/785/1/71} {\bibfield  {journal} {\bibinfo  {journal} {Astrophys. J.}\ }\textbf {\bibinfo {volume} {785}},\ \bibinfo {pages} {71} (\bibinfo {year} {2014})},\ \Eprint {https://arxiv.org/abs/1308.0021} {arXiv:1308.0021 [astro-ph.CO]} \BibitemShut {NoStop}%
\bibitem [{\citenamefont {Bahcall}\ and\ \citenamefont {Wolf}(1976)}]{Bahcall:1976aa}%
  \BibitemOpen
  \bibfield  {author} {\bibinfo {author} {\bibfnamefont {J.~N.}\ \bibnamefont {Bahcall}}\ and\ \bibinfo {author} {\bibfnamefont {R.~A.}\ \bibnamefont {Wolf}},\ }\bibfield  {title} {\bibinfo {title} {{Star distribution around a massive black hole in a globular cluster}},\ }\href {https://doi.org/10.1086/154711} {\bibfield  {journal} {\bibinfo  {journal} {Astrophys. J.}\ }\textbf {\bibinfo {volume} {209}},\ \bibinfo {pages} {214} (\bibinfo {year} {1976})}\BibitemShut {NoStop}%
\bibitem [{\citenamefont {Campanelli}\ \emph {et~al.}(2006)\citenamefont {Campanelli}, \citenamefont {Dettwyler}, \citenamefont {Hannam},\ and\ \citenamefont {Lousto}}]{Campanelli:2005kr}%
  \BibitemOpen
  \bibfield  {author} {\bibinfo {author} {\bibfnamefont {M.}~\bibnamefont {Campanelli}}, \bibinfo {author} {\bibfnamefont {M.}~\bibnamefont {Dettwyler}}, \bibinfo {author} {\bibfnamefont {M.}~\bibnamefont {Hannam}},\ and\ \bibinfo {author} {\bibfnamefont {C.~O.}\ \bibnamefont {Lousto}},\ }\bibfield  {title} {\bibinfo {title} {{Relativistic three-body effects in black hole coalescence}},\ }\href {https://doi.org/10.1103/PhysRevD.74.087503} {\bibfield  {journal} {\bibinfo  {journal} {Phys. Rev. D}\ }\textbf {\bibinfo {volume} {74}},\ \bibinfo {pages} {087503} (\bibinfo {year} {2006})},\ \Eprint {https://arxiv.org/abs/astro-ph/0509814} {arXiv:astro-ph/0509814} \BibitemShut {NoStop}%
\bibitem [{\citenamefont {{Miller}}(2002)}]{MillerIMBHs}%
  \BibitemOpen
  \bibfield  {author} {\bibinfo {author} {\bibfnamefont {M.~C.}\ \bibnamefont {{Miller}}},\ }\bibfield  {title} {\bibinfo {title} {{Gravitational Radiation from Intermediate-Mass Black Holes}},\ }\href {https://doi.org/10.1086/344156} {\bibfield  {journal} {\bibinfo  {journal} {\apj}\ }\textbf {\bibinfo {volume} {581}},\ \bibinfo {pages} {438} (\bibinfo {year} {2002})},\ \Eprint {https://arxiv.org/abs/astro-ph/0206404} {arXiv:astro-ph/0206404 [astro-ph]} \BibitemShut {NoStop}%
\bibitem [{\citenamefont {Gultekin}\ \emph {et~al.}(2003)\citenamefont {Gultekin}, \citenamefont {Miller},\ and\ \citenamefont {Hamilton}}]{Gultekin:2003xd}%
  \BibitemOpen
  \bibfield  {author} {\bibinfo {author} {\bibfnamefont {K.}~\bibnamefont {Gultekin}}, \bibinfo {author} {\bibfnamefont {M.~C.}\ \bibnamefont {Miller}},\ and\ \bibinfo {author} {\bibfnamefont {D.~P.}\ \bibnamefont {Hamilton}},\ }\bibfield  {title} {\bibinfo {title} {{Three-body encounters of black holes in globular clusters}},\ }\href {https://doi.org/10.1063/1.1629425} {\bibfield  {journal} {\bibinfo  {journal} {AIP Conf. Proc.}\ }\textbf {\bibinfo {volume} {686}},\ \bibinfo {pages} {135} (\bibinfo {year} {2003})},\ \Eprint {https://arxiv.org/abs/astro-ph/0306204} {arXiv:astro-ph/0306204} \BibitemShut {NoStop}%
\bibitem [{\citenamefont {Trani}\ \emph {et~al.}(2024)\citenamefont {Trani}, \citenamefont {Quaini},\ and\ \citenamefont {Colpi}}]{Trani:2023oqa}%
  \BibitemOpen
  \bibfield  {author} {\bibinfo {author} {\bibfnamefont {A.~A.}\ \bibnamefont {Trani}}, \bibinfo {author} {\bibfnamefont {S.}~\bibnamefont {Quaini}},\ and\ \bibinfo {author} {\bibfnamefont {M.}~\bibnamefont {Colpi}},\ }\bibfield  {title} {\bibinfo {title} {{Three-body encounters in black hole discs around a supermassive black hole - The disc velocity dispersion and the Keplerian tidal field determine the eccentricity and spin-orbit alignment of gravitational wave mergers}},\ }\href {https://doi.org/10.1051/0004-6361/202347920} {\bibfield  {journal} {\bibinfo  {journal} {Astron. Astrophys.}\ }\textbf {\bibinfo {volume} {683}},\ \bibinfo {pages} {A135} (\bibinfo {year} {2024})},\ \Eprint {https://arxiv.org/abs/2312.13281} {arXiv:2312.13281 [astro-ph.HE]} \BibitemShut {NoStop}%
\bibitem [{\citenamefont {Laeuger}\ \emph {et~al.}(2024)\citenamefont {Laeuger}, \citenamefont {Seymour}, \citenamefont {Chen},\ and\ \citenamefont {Yu}}]{Laeuger:2023qyz}%
  \BibitemOpen
  \bibfield  {author} {\bibinfo {author} {\bibfnamefont {A.}~\bibnamefont {Laeuger}}, \bibinfo {author} {\bibfnamefont {B.}~\bibnamefont {Seymour}}, \bibinfo {author} {\bibfnamefont {Y.}~\bibnamefont {Chen}},\ and\ \bibinfo {author} {\bibfnamefont {H.}~\bibnamefont {Yu}},\ }\bibfield  {title} {\bibinfo {title} {{Measuring supermassive black hole properties via gravitational radiation from eccentrically orbiting stellar mass black hole binaries}},\ }\href {https://doi.org/10.1103/PhysRevD.109.064086} {\bibfield  {journal} {\bibinfo  {journal} {Phys. Rev. D}\ }\textbf {\bibinfo {volume} {109}},\ \bibinfo {pages} {064086} (\bibinfo {year} {2024})},\ \Eprint {https://arxiv.org/abs/2310.16799} {arXiv:2310.16799 [gr-qc]} \BibitemShut {NoStop}%
\bibitem [{\citenamefont {Amaro-Seoane}(2018)}]{Amaro-Seoane:2012lgq}%
  \BibitemOpen
  \bibfield  {author} {\bibinfo {author} {\bibfnamefont {P.}~\bibnamefont {Amaro-Seoane}},\ }\bibfield  {title} {\bibinfo {title} {{Relativistic dynamics and extreme mass ratio inspirals}},\ }\href {https://doi.org/10.1007/s41114-018-0013-8} {\bibfield  {journal} {\bibinfo  {journal} {Living Rev. Rel.}\ }\textbf {\bibinfo {volume} {21}},\ \bibinfo {pages} {4} (\bibinfo {year} {2018})},\ \Eprint {https://arxiv.org/abs/1205.5240} {arXiv:1205.5240 [astro-ph.CO]} \BibitemShut {NoStop}%
\bibitem [{\citenamefont {Ozernoy}\ \emph {et~al.}(1997)\citenamefont {Ozernoy}, \citenamefont {Genzel},\ and\ \citenamefont {Usov}}]{Ozernoy:1997pa}%
  \BibitemOpen
  \bibfield  {author} {\bibinfo {author} {\bibfnamefont {L.~M.}\ \bibnamefont {Ozernoy}}, \bibinfo {author} {\bibfnamefont {R.}~\bibnamefont {Genzel}},\ and\ \bibinfo {author} {\bibfnamefont {V.~V.}\ \bibnamefont {Usov}},\ }\bibfield  {title} {\bibinfo {title} {{Colliding winds in the stellar core at the galactic center: some implications}},\ }\href@noop {} {\  (\bibinfo {year} {1997})},\ \Eprint {https://arxiv.org/abs/astro-ph/9706196} {arXiv:astro-ph/9706196} \BibitemShut {NoStop}%
\bibitem [{\citenamefont {Ott}\ \emph {et~al.}(2003)\citenamefont {Ott} \emph {et~al.}}]{Ott:2003gr}%
  \BibitemOpen
  \bibfield  {author} {\bibinfo {author} {\bibfnamefont {T.}~\bibnamefont {Ott}} \emph {et~al.},\ }\bibfield  {title} {\bibinfo {title} {{Inward bound: Studying the Galactic center with NAOS / CONICA}},\ }\href@noop {} {\bibfield  {journal} {\bibinfo  {journal} {The Messenger}\ }\textbf {\bibinfo {volume} {111}},\ \bibinfo {pages} {1} (\bibinfo {year} {2003})},\ \Eprint {https://arxiv.org/abs/astro-ph/0303408} {arXiv:astro-ph/0303408} \BibitemShut {NoStop}%
\bibitem [{\citenamefont {Willems}\ \emph {et~al.}(2007)\citenamefont {Willems}, \citenamefont {Kalogera}, \citenamefont {Vecchio}, \citenamefont {Ivanova}, \citenamefont {Rasio}, \citenamefont {Fregeau},\ and\ \citenamefont {Belczynski}}]{Willems:2007xe}%
  \BibitemOpen
  \bibfield  {author} {\bibinfo {author} {\bibfnamefont {B.}~\bibnamefont {Willems}}, \bibinfo {author} {\bibfnamefont {V.}~\bibnamefont {Kalogera}}, \bibinfo {author} {\bibfnamefont {A.}~\bibnamefont {Vecchio}}, \bibinfo {author} {\bibfnamefont {N.}~\bibnamefont {Ivanova}}, \bibinfo {author} {\bibfnamefont {F.~A.}\ \bibnamefont {Rasio}}, \bibinfo {author} {\bibfnamefont {J.~M.}\ \bibnamefont {Fregeau}},\ and\ \bibinfo {author} {\bibfnamefont {K.}~\bibnamefont {Belczynski}},\ }\bibfield  {title} {\bibinfo {title} {{Eccentric double white dwarfs as LISA sources in globular clusters}},\ }\href {https://doi.org/10.1086/521049} {\bibfield  {journal} {\bibinfo  {journal} {Astrophys. J. Lett.}\ }\textbf {\bibinfo {volume} {665}},\ \bibinfo {pages} {L59} (\bibinfo {year} {2007})},\ \Eprint {https://arxiv.org/abs/0705.4287} {arXiv:0705.4287 [astro-ph]} \BibitemShut {NoStop}%
\bibitem [{\citenamefont {Ivanova}\ \emph {et~al.}(2003)\citenamefont {Ivanova}, \citenamefont {Belczynski}, \citenamefont {Fregeau},\ and\ \citenamefont {Rasio}}]{Ivanova:2003by}%
  \BibitemOpen
  \bibfield  {author} {\bibinfo {author} {\bibfnamefont {N.}~\bibnamefont {Ivanova}}, \bibinfo {author} {\bibfnamefont {K.}~\bibnamefont {Belczynski}}, \bibinfo {author} {\bibfnamefont {J.~M.}\ \bibnamefont {Fregeau}},\ and\ \bibinfo {author} {\bibfnamefont {F.~A.}\ \bibnamefont {Rasio}},\ }\bibfield  {title} {\bibinfo {title} {{On the maximum binary fraction in globular cluster cores}},\ }\href@noop {} {\  (\bibinfo {year} {2003})},\ \Eprint {https://arxiv.org/abs/astro-ph/0312497} {arXiv:astro-ph/0312497} \BibitemShut {NoStop}%
\bibitem [{\citenamefont {Miller}\ and\ \citenamefont {Hamilton}(2002)}]{Miller:2002pg}%
  \BibitemOpen
  \bibfield  {author} {\bibinfo {author} {\bibfnamefont {M.~C.}\ \bibnamefont {Miller}}\ and\ \bibinfo {author} {\bibfnamefont {D.~P.}\ \bibnamefont {Hamilton}},\ }\bibfield  {title} {\bibinfo {title} {{Four-body effects in globular cluster black hole coalescence}},\ }\href {https://doi.org/10.1086/341788} {\bibfield  {journal} {\bibinfo  {journal} {Astrophys. J.}\ }\textbf {\bibinfo {volume} {576}},\ \bibinfo {pages} {894} (\bibinfo {year} {2002})},\ \Eprint {https://arxiv.org/abs/astro-ph/0202298} {arXiv:astro-ph/0202298} \BibitemShut {NoStop}%
\bibitem [{\citenamefont {{Baumgardt}}\ and\ \citenamefont {{Hilker}}(2018)}]{Baumgardt2018}%
  \BibitemOpen
  \bibfield  {author} {\bibinfo {author} {\bibfnamefont {H.}~\bibnamefont {{Baumgardt}}}\ and\ \bibinfo {author} {\bibfnamefont {M.}~\bibnamefont {{Hilker}}},\ }\bibfield  {title} {\bibinfo {title} {{A catalogue of masses, structural parameters, and velocity dispersion profiles of 112 Milky Way globular clusters}},\ }\href {https://doi.org/10.1093/mnras/sty1057} {\bibfield  {journal} {\bibinfo  {journal} {\mnras}\ }\textbf {\bibinfo {volume} {478}},\ \bibinfo {pages} {1520} (\bibinfo {year} {2018})},\ \Eprint {https://arxiv.org/abs/1804.08359} {arXiv:1804.08359 [astro-ph.GA]} \BibitemShut {NoStop}%
\bibitem [{\citenamefont {Navarro}\ \emph {et~al.}(1996)\citenamefont {Navarro}, \citenamefont {Frenk},\ and\ \citenamefont {White}}]{Navarro:1995iw}%
  \BibitemOpen
  \bibfield  {author} {\bibinfo {author} {\bibfnamefont {J.~F.}\ \bibnamefont {Navarro}}, \bibinfo {author} {\bibfnamefont {C.~S.}\ \bibnamefont {Frenk}},\ and\ \bibinfo {author} {\bibfnamefont {S.~D.~M.}\ \bibnamefont {White}},\ }\bibfield  {title} {\bibinfo {title} {{The Structure of cold dark matter halos}},\ }\href {https://doi.org/10.1086/177173} {\bibfield  {journal} {\bibinfo  {journal} {Astrophys. J.}\ }\textbf {\bibinfo {volume} {462}},\ \bibinfo {pages} {563} (\bibinfo {year} {1996})},\ \Eprint {https://arxiv.org/abs/astro-ph/9508025} {arXiv:astro-ph/9508025} \BibitemShut {NoStop}%
\bibitem [{\citenamefont {Bertone}\ \emph {et~al.}(2005)\citenamefont {Bertone}, \citenamefont {Hooper},\ and\ \citenamefont {Silk}}]{Bertone:2004pz}%
  \BibitemOpen
  \bibfield  {author} {\bibinfo {author} {\bibfnamefont {G.}~\bibnamefont {Bertone}}, \bibinfo {author} {\bibfnamefont {D.}~\bibnamefont {Hooper}},\ and\ \bibinfo {author} {\bibfnamefont {J.}~\bibnamefont {Silk}},\ }\bibfield  {title} {\bibinfo {title} {{Particle dark matter: Evidence, candidates and constraints}},\ }\href {https://doi.org/10.1016/j.physrep.2004.08.031} {\bibfield  {journal} {\bibinfo  {journal} {Phys. Rept.}\ }\textbf {\bibinfo {volume} {405}},\ \bibinfo {pages} {279} (\bibinfo {year} {2005})},\ \Eprint {https://arxiv.org/abs/hep-ph/0404175} {arXiv:hep-ph/0404175} \BibitemShut {NoStop}%
\bibitem [{\citenamefont {{Cautun}}\ \emph {et~al.}(2020)\citenamefont {{Cautun}}, \citenamefont {{Ben{\'\i}tez-Llambay}}, \citenamefont {{Deason}}, \citenamefont {{Frenk}}, \citenamefont {{Fattahi}}, \citenamefont {{G{\'o}mez}}, \citenamefont {{Grand}}, \citenamefont {{Oman}}, \citenamefont {{Navarro}},\ and\ \citenamefont {{Simpson}}}]{Cautun:2019eaf}%
  \BibitemOpen
  \bibfield  {author} {\bibinfo {author} {\bibfnamefont {M.}~\bibnamefont {{Cautun}}}, \bibinfo {author} {\bibfnamefont {A.}~\bibnamefont {{Ben{\'\i}tez-Llambay}}}, \bibinfo {author} {\bibfnamefont {A.~J.}\ \bibnamefont {{Deason}}}, \bibinfo {author} {\bibfnamefont {C.~S.}\ \bibnamefont {{Frenk}}}, \bibinfo {author} {\bibfnamefont {A.}~\bibnamefont {{Fattahi}}}, \bibinfo {author} {\bibfnamefont {F.~A.}\ \bibnamefont {{G{\'o}mez}}}, \bibinfo {author} {\bibfnamefont {R.~J.~J.}\ \bibnamefont {{Grand}}}, \bibinfo {author} {\bibfnamefont {K.~A.}\ \bibnamefont {{Oman}}}, \bibinfo {author} {\bibfnamefont {J.~F.}\ \bibnamefont {{Navarro}}},\ and\ \bibinfo {author} {\bibfnamefont {C.~M.}\ \bibnamefont {{Simpson}}},\ }\bibfield  {title} {\bibinfo {title} {{The milky way total mass profile as inferred from Gaia DR2}},\ }\href {https://doi.org/10.1093/mnras/staa1017} {\bibfield  {journal} {\bibinfo  {journal} {\mnras}\ }\textbf {\bibinfo {volume} {494}},\ \bibinfo {pages} {4291} (\bibinfo {year} {2020})},\ \Eprint
  {https://arxiv.org/abs/1911.04557} {arXiv:1911.04557 [astro-ph.GA]} \BibitemShut {NoStop}%
\bibitem [{\citenamefont {{Klypin}}\ \emph {et~al.}(2002)\citenamefont {{Klypin}}, \citenamefont {{Zhao}},\ and\ \citenamefont {{Somerville}}}]{Klypin2002}%
  \BibitemOpen
  \bibfield  {author} {\bibinfo {author} {\bibfnamefont {A.}~\bibnamefont {{Klypin}}}, \bibinfo {author} {\bibfnamefont {H.}~\bibnamefont {{Zhao}}},\ and\ \bibinfo {author} {\bibfnamefont {R.~S.}\ \bibnamefont {{Somerville}}},\ }\bibfield  {title} {\bibinfo {title} {{{\ensuremath{\Lambda}}CDM-based Models for the Milky Way and M31. I. Dynamical Models}},\ }\href {https://doi.org/10.1086/340656} {\bibfield  {journal} {\bibinfo  {journal} {\apj}\ }\textbf {\bibinfo {volume} {573}},\ \bibinfo {pages} {597} (\bibinfo {year} {2002})},\ \Eprint {https://arxiv.org/abs/astro-ph/0110390} {arXiv:astro-ph/0110390 [astro-ph]} \BibitemShut {NoStop}%
\bibitem [{\citenamefont {{Sofue}}(2012)}]{Sofue2012}%
  \BibitemOpen
  \bibfield  {author} {\bibinfo {author} {\bibfnamefont {Y.}~\bibnamefont {{Sofue}}},\ }\bibfield  {title} {\bibinfo {title} {{Grand Rotation Curve and Dark Matter Halo in the Milky Way Galaxy}},\ }\href {https://doi.org/10.1093/pasj/64.4.75} {\bibfield  {journal} {\bibinfo  {journal} {\pasj}\ }\textbf {\bibinfo {volume} {64}},\ \bibinfo {eid} {75} (\bibinfo {year} {2012})},\ \Eprint {https://arxiv.org/abs/1110.4431} {arXiv:1110.4431 [astro-ph.GA]} \BibitemShut {NoStop}%
\bibitem [{\citenamefont {Pato}\ \emph {et~al.}(2015)\citenamefont {Pato}, \citenamefont {Iocco},\ and\ \citenamefont {Bertone}}]{Pato:2015dua}%
  \BibitemOpen
  \bibfield  {author} {\bibinfo {author} {\bibfnamefont {M.}~\bibnamefont {Pato}}, \bibinfo {author} {\bibfnamefont {F.}~\bibnamefont {Iocco}},\ and\ \bibinfo {author} {\bibfnamefont {G.}~\bibnamefont {Bertone}},\ }\bibfield  {title} {\bibinfo {title} {{Dynamical constraints on the dark matter distribution in the Milky Way}},\ }\href {https://doi.org/10.1088/1475-7516/2015/12/001} {\bibfield  {journal} {\bibinfo  {journal} {JCAP}\ }\textbf {\bibinfo {volume} {12}},\ \bibinfo {pages} {001}},\ \Eprint {https://arxiv.org/abs/1504.06324} {arXiv:1504.06324 [astro-ph.GA]} \BibitemShut {NoStop}%
\bibitem [{\citenamefont {Branchesi}\ \emph {et~al.}(2023)\citenamefont {Branchesi} \emph {et~al.}}]{EinsteinTelescope}%
  \BibitemOpen
  \bibfield  {author} {\bibinfo {author} {\bibfnamefont {M.}~\bibnamefont {Branchesi}} \emph {et~al.},\ }\bibfield  {title} {\bibinfo {title} {{Science with the Einstein Telescope: a comparison of different designs}},\ }\href {https://doi.org/10.1088/1475-7516/2023/07/068} {\bibfield  {journal} {\bibinfo  {journal} {JCAP}\ }\textbf {\bibinfo {volume} {07}},\ \bibinfo {pages} {068}},\ \Eprint {https://arxiv.org/abs/2303.15923} {arXiv:2303.15923 [gr-qc]} \BibitemShut {NoStop}%
\bibitem [{\citenamefont {Evans}\ \emph {et~al.}(2021)\citenamefont {Evans} \emph {et~al.}}]{CosmicExplorer}%
  \BibitemOpen
  \bibfield  {author} {\bibinfo {author} {\bibfnamefont {M.}~\bibnamefont {Evans}} \emph {et~al.},\ }\bibfield  {title} {\bibinfo {title} {{A Horizon Study for Cosmic Explorer: Science, Observatories, and Community}},\ }\href@noop {} {\  (\bibinfo {year} {2021})},\ \Eprint {https://arxiv.org/abs/2109.09882} {arXiv:2109.09882 [astro-ph.IM]} \BibitemShut {NoStop}%
\bibitem [{\citenamefont {Kawamura}\ \emph {et~al.}(2021)\citenamefont {Kawamura} \emph {et~al.}}]{DECIGO}%
  \BibitemOpen
  \bibfield  {author} {\bibinfo {author} {\bibfnamefont {S.}~\bibnamefont {Kawamura}} \emph {et~al.},\ }\bibfield  {title} {\bibinfo {title} {{Current status of space gravitational wave antenna DECIGO and B-DECIGO}},\ }\href {https://doi.org/10.1093/ptep/ptab019} {\bibfield  {journal} {\bibinfo  {journal} {PTEP}\ }\textbf {\bibinfo {volume} {2021}},\ \bibinfo {pages} {05A105} (\bibinfo {year} {2021})},\ \Eprint {https://arxiv.org/abs/2006.13545} {arXiv:2006.13545 [gr-qc]} \BibitemShut {NoStop}%
\bibitem [{\citenamefont {Ajith}\ \emph {et~al.}(2025{\natexlab{a}})\citenamefont {Ajith} \emph {et~al.}}]{LGWA}%
  \BibitemOpen
  \bibfield  {author} {\bibinfo {author} {\bibfnamefont {P.}~\bibnamefont {Ajith}} \emph {et~al.},\ }\bibfield  {title} {\bibinfo {title} {{The Lunar Gravitational-wave Antenna: mission studies and science case}},\ }\href {https://doi.org/10.1088/1475-7516/2025/01/108} {\bibfield  {journal} {\bibinfo  {journal} {JCAP}\ }\textbf {\bibinfo {volume} {01}},\ \bibinfo {pages} {108}},\ \Eprint {https://arxiv.org/abs/2404.09181} {arXiv:2404.09181 [gr-qc]} \BibitemShut {NoStop}%
\bibitem [{\citenamefont {Jani}\ \emph {et~al.}(2025)\citenamefont {Jani} \emph {et~al.}}]{LILA}%
  \BibitemOpen
  \bibfield  {author} {\bibinfo {author} {\bibfnamefont {K.}~\bibnamefont {Jani}} \emph {et~al.},\ }\bibfield  {title} {\bibinfo {title} {{Laser Interferometer Lunar Antenna (LILA): Advancing the U.S. Priorities in Gravitational-wave and Lunar Science}},\ }\href@noop {} {\  (\bibinfo {year} {2025})},\ \Eprint {https://arxiv.org/abs/2508.11631} {arXiv:2508.11631 [gr-qc]} \BibitemShut {NoStop}%
\bibitem [{\citenamefont {Berti}\ \emph {et~al.}(2026)\citenamefont {Berti} \emph {et~al.}}]{deciHz_status}%
  \BibitemOpen
  \bibfield  {author} {\bibinfo {author} {\bibfnamefont {E.}~\bibnamefont {Berti}} \emph {et~al.},\ }\bibfield  {title} {\bibinfo {title} {{deci-Hz Gravitational Wave Observations on the Moon and Beyond}},\ }\href@noop {} {\  (\bibinfo {year} {2026})},\ \Eprint {https://arxiv.org/abs/2602.05923} {arXiv:2602.05923 [gr-qc]} \BibitemShut {NoStop}%
\bibitem [{\citenamefont {Colpi}\ \emph {et~al.}(2024)\citenamefont {Colpi} \emph {et~al.}}]{LISA:2024hlh}%
  \BibitemOpen
  \bibfield  {author} {\bibinfo {author} {\bibfnamefont {M.}~\bibnamefont {Colpi}} \emph {et~al.} (\bibinfo {collaboration} {LISA}),\ }\bibfield  {title} {\bibinfo {title} {{LISA Definition Study Report}},\ }\href@noop {} {\  (\bibinfo {year} {2024})},\ \Eprint {https://arxiv.org/abs/2402.07571} {arXiv:2402.07571 [astro-ph.CO]} \BibitemShut {NoStop}%
\bibitem [{\citenamefont {{de Blok}}(2010)}]{2010AdAst2010E...5D}%
  \BibitemOpen
  \bibfield  {author} {\bibinfo {author} {\bibfnamefont {W.~J.~G.}\ \bibnamefont {{de Blok}}},\ }\bibfield  {title} {\bibinfo {title} {{The Core-Cusp Problem}},\ }\href {https://doi.org/10.1155/2010/789293} {\bibfield  {journal} {\bibinfo  {journal} {Advances in Astronomy}\ }\textbf {\bibinfo {volume} {2010}},\ \bibinfo {eid} {789293} (\bibinfo {year} {2010})},\ \Eprint {https://arxiv.org/abs/0910.3538} {arXiv:0910.3538 [astro-ph.CO]} \BibitemShut {NoStop}%
\bibitem [{\citenamefont {Burke}\ and\ \citenamefont {Thorne}(1970)}]{Burke:1970dnm}%
  \BibitemOpen
  \bibfield  {author} {\bibinfo {author} {\bibfnamefont {W.~L.}\ \bibnamefont {Burke}}\ and\ \bibinfo {author} {\bibfnamefont {K.~S.}\ \bibnamefont {Thorne}},\ }\bibfield  {title} {\bibinfo {title} {{Gravitational Radiation Damping}},\ }in\ \href {https://doi.org/10.1007/978-1-4684-0721-1_12} {\emph {\bibinfo {booktitle} {{Relativity Conference in the Midwest}}}}\ (\bibinfo {year} {1970})\ pp.\ \bibinfo {pages} {209--228}\BibitemShut {NoStop}%
\bibitem [{\citenamefont {Blanchet}(2002)}]{Blanchet:2002av}%
  \BibitemOpen
  \bibfield  {author} {\bibinfo {author} {\bibfnamefont {L.}~\bibnamefont {Blanchet}},\ }\bibfield  {title} {\bibinfo {title} {{Gravitational radiation from postNewtonian sources and inspiraling compact binaries}},\ }\href {https://doi.org/10.12942/lrr-2002-3} {\bibfield  {journal} {\bibinfo  {journal} {Living Rev. Rel.}\ }\textbf {\bibinfo {volume} {5}},\ \bibinfo {pages} {3} (\bibinfo {year} {2002})},\ \Eprint {https://arxiv.org/abs/gr-qc/0202016} {arXiv:gr-qc/0202016} \BibitemShut {NoStop}%
\bibitem [{\citenamefont {Blanchet}(2014)}]{Blanchet:2013haa}%
  \BibitemOpen
  \bibfield  {author} {\bibinfo {author} {\bibfnamefont {L.}~\bibnamefont {Blanchet}},\ }\bibfield  {title} {\bibinfo {title} {{Post-Newtonian Theory for Gravitational Waves}},\ }\href {https://doi.org/10.12942/lrr-2014-2} {\bibfield  {journal} {\bibinfo  {journal} {Living Rev. Rel.}\ }\textbf {\bibinfo {volume} {17}},\ \bibinfo {pages} {2} (\bibinfo {year} {2014})},\ \Eprint {https://arxiv.org/abs/1310.1528} {arXiv:1310.1528 [gr-qc]} \BibitemShut {NoStop}%
\bibitem [{\citenamefont {Hairer}\ \emph {et~al.}(1993)\citenamefont {Hairer}, \citenamefont {N{\o}rsett},\ and\ \citenamefont {Wanner}}]{HairerNorsettWanner1993}%
  \BibitemOpen
  \bibfield  {author} {\bibinfo {author} {\bibfnamefont {E.}~\bibnamefont {Hairer}}, \bibinfo {author} {\bibfnamefont {S.~P.}\ \bibnamefont {N{\o}rsett}},\ and\ \bibinfo {author} {\bibfnamefont {G.}~\bibnamefont {Wanner}},\ }\href {https://doi.org/10.1007/978-3-540-78862-1} {\emph {\bibinfo {title} {Solving Ordinary Differential Equations I: Nonstiff Problems}}},\ \bibinfo {edition} {2nd}\ ed.,\ \bibinfo {series} {Springer Series in Computational Mathematics}, Vol.~\bibinfo {volume} {8}\ (\bibinfo  {publisher} {Springer},\ \bibinfo {address} {Berlin, Heidelberg},\ \bibinfo {year} {1993})\BibitemShut {NoStop}%
\bibitem [{\citenamefont {Hairer}(nd)}]{hairer_software}%
  \BibitemOpen
  \bibfield  {author} {\bibinfo {author} {\bibfnamefont {E.}~\bibnamefont {Hairer}},\ }\href {https://www.unige.ch/~hairer/software.html} {\bibinfo {title} {Software for ordinary differential equations}} (\bibinfo {year} {n.d.})\BibitemShut {NoStop}%
\bibitem [{\citenamefont {Virtanen}\ \emph {et~al.}(2020)\citenamefont {Virtanen}, \citenamefont {Gommers}, \citenamefont {Oliphant}, \citenamefont {Haberland}, \citenamefont {Reddy}, \citenamefont {Cournapeau}, \citenamefont {Burovski}, \citenamefont {Peterson}, \citenamefont {Weckesser}, \citenamefont {Bright}, \citenamefont {{van der Walt}}, \citenamefont {Brett}, \citenamefont {Wilson}, \citenamefont {Millman}, \citenamefont {Mayorov}, \citenamefont {Nelson}, \citenamefont {Jones}, \citenamefont {Kern}, \citenamefont {Larson}, \citenamefont {Carey}, \citenamefont {Polat}, \citenamefont {Feng}, \citenamefont {Moore}, \citenamefont {{VanderPlas}}, \citenamefont {Laxalde}, \citenamefont {Perktold}, \citenamefont {Cimrman}, \citenamefont {Henriksen}, \citenamefont {Quintero}, \citenamefont {Harris}, \citenamefont {Archibald}, \citenamefont {Ribeiro}, \citenamefont {Pedregosa}, \citenamefont {{van Mulbregt}},\ and\ \citenamefont {{SciPy 1.0 Contributors}}}]{2020SciPy-NMeth}%
  \BibitemOpen
  \bibfield  {author} {\bibinfo {author} {\bibfnamefont {P.}~\bibnamefont {Virtanen}}, \bibinfo {author} {\bibfnamefont {R.}~\bibnamefont {Gommers}}, \bibinfo {author} {\bibfnamefont {T.~E.}\ \bibnamefont {Oliphant}}, \bibinfo {author} {\bibfnamefont {M.}~\bibnamefont {Haberland}}, \bibinfo {author} {\bibfnamefont {T.}~\bibnamefont {Reddy}}, \bibinfo {author} {\bibfnamefont {D.}~\bibnamefont {Cournapeau}}, \bibinfo {author} {\bibfnamefont {E.}~\bibnamefont {Burovski}}, \bibinfo {author} {\bibfnamefont {P.}~\bibnamefont {Peterson}}, \bibinfo {author} {\bibfnamefont {W.}~\bibnamefont {Weckesser}}, \bibinfo {author} {\bibfnamefont {J.}~\bibnamefont {Bright}}, \bibinfo {author} {\bibfnamefont {S.~J.}\ \bibnamefont {{van der Walt}}}, \bibinfo {author} {\bibfnamefont {M.}~\bibnamefont {Brett}}, \bibinfo {author} {\bibfnamefont {J.}~\bibnamefont {Wilson}}, \bibinfo {author} {\bibfnamefont {K.~J.}\ \bibnamefont {Millman}}, \bibinfo {author} {\bibfnamefont {N.}~\bibnamefont {Mayorov}}, \bibinfo {author} {\bibfnamefont
  {A.~R.~J.}\ \bibnamefont {Nelson}}, \bibinfo {author} {\bibfnamefont {E.}~\bibnamefont {Jones}}, \bibinfo {author} {\bibfnamefont {R.}~\bibnamefont {Kern}}, \bibinfo {author} {\bibfnamefont {E.}~\bibnamefont {Larson}}, \bibinfo {author} {\bibfnamefont {C.~J.}\ \bibnamefont {Carey}}, \bibinfo {author} {\bibfnamefont {{\.I}.}~\bibnamefont {Polat}}, \bibinfo {author} {\bibfnamefont {Y.}~\bibnamefont {Feng}}, \bibinfo {author} {\bibfnamefont {E.~W.}\ \bibnamefont {Moore}}, \bibinfo {author} {\bibfnamefont {J.}~\bibnamefont {{VanderPlas}}}, \bibinfo {author} {\bibfnamefont {D.}~\bibnamefont {Laxalde}}, \bibinfo {author} {\bibfnamefont {J.}~\bibnamefont {Perktold}}, \bibinfo {author} {\bibfnamefont {R.}~\bibnamefont {Cimrman}}, \bibinfo {author} {\bibfnamefont {I.}~\bibnamefont {Henriksen}}, \bibinfo {author} {\bibfnamefont {E.~A.}\ \bibnamefont {Quintero}}, \bibinfo {author} {\bibfnamefont {C.~R.}\ \bibnamefont {Harris}}, \bibinfo {author} {\bibfnamefont {A.~M.}\ \bibnamefont {Archibald}}, \bibinfo {author}
  {\bibfnamefont {A.~H.}\ \bibnamefont {Ribeiro}}, \bibinfo {author} {\bibfnamefont {F.}~\bibnamefont {Pedregosa}}, \bibinfo {author} {\bibfnamefont {P.}~\bibnamefont {{van Mulbregt}}},\ and\ \bibinfo {author} {\bibnamefont {{SciPy 1.0 Contributors}}},\ }\bibfield  {title} {\bibinfo {title} {{{SciPy} 1.0: Fundamental Algorithms for Scientific Computing in Python}},\ }\href {https://doi.org/10.1038/s41592-019-0686-2} {\bibfield  {journal} {\bibinfo  {journal} {Nature Methods}\ }\textbf {\bibinfo {volume} {17}},\ \bibinfo {pages} {261} (\bibinfo {year} {2020})}\BibitemShut {NoStop}%
\bibitem [{\citenamefont {de~Boor}(1978)}]{deBoor1978PracticalGuideSplines}%
  \BibitemOpen
  \bibfield  {author} {\bibinfo {author} {\bibfnamefont {C.}~\bibnamefont {de~Boor}},\ }\href {https://link.springer.com/book/9780387953663} {\emph {\bibinfo {title} {A Practical Guide to Splines}}}\ (\bibinfo  {publisher} {Springer-Verlag},\ \bibinfo {address} {New York},\ \bibinfo {year} {1978})\BibitemShut {NoStop}%
\bibitem [{\citenamefont {Maggiore}(2007)}]{Maggiore:2007ulw}%
  \BibitemOpen
  \bibfield  {author} {\bibinfo {author} {\bibfnamefont {M.}~\bibnamefont {Maggiore}},\ }\href {https://doi.org/10.1093/acprof:oso/9780198570745.001.0001} {\emph {\bibinfo {title} {{Gravitational Waves. Vol. 1: Theory and Experiments}}}}\ (\bibinfo  {publisher} {Oxford University Press},\ \bibinfo {year} {2007})\BibitemShut {NoStop}%
\bibitem [{\citenamefont {Khan}\ \emph {et~al.}(2019)\citenamefont {Khan}, \citenamefont {Chatziioannou}, \citenamefont {Hannam},\ and\ \citenamefont {Ohme}}]{IMRPhenomPv}%
  \BibitemOpen
  \bibfield  {author} {\bibinfo {author} {\bibfnamefont {S.}~\bibnamefont {Khan}}, \bibinfo {author} {\bibfnamefont {K.}~\bibnamefont {Chatziioannou}}, \bibinfo {author} {\bibfnamefont {M.}~\bibnamefont {Hannam}},\ and\ \bibinfo {author} {\bibfnamefont {F.}~\bibnamefont {Ohme}},\ }\bibfield  {title} {\bibinfo {title} {{Phenomenological model for the gravitational-wave signal from precessing binary black holes with two-spin effects}},\ }\href {https://doi.org/10.1103/PhysRevD.100.024059} {\bibfield  {journal} {\bibinfo  {journal} {Phys. Rev. D}\ }\textbf {\bibinfo {volume} {100}},\ \bibinfo {pages} {024059} (\bibinfo {year} {2019})},\ \Eprint {https://arxiv.org/abs/1809.10113} {arXiv:1809.10113 [gr-qc]} \BibitemShut {NoStop}%
\bibitem [{\citenamefont {Pratten}\ \emph {et~al.}(2021)\citenamefont {Pratten} \emph {et~al.}}]{IMRPhenomXPHM}%
  \BibitemOpen
  \bibfield  {author} {\bibinfo {author} {\bibfnamefont {G.}~\bibnamefont {Pratten}} \emph {et~al.},\ }\bibfield  {title} {\bibinfo {title} {{Computationally efficient models for the dominant and subdominant harmonic modes of precessing binary black holes}},\ }\href {https://doi.org/10.1103/PhysRevD.103.104056} {\bibfield  {journal} {\bibinfo  {journal} {Phys. Rev. D}\ }\textbf {\bibinfo {volume} {103}},\ \bibinfo {pages} {104056} (\bibinfo {year} {2021})},\ \Eprint {https://arxiv.org/abs/2004.06503} {arXiv:2004.06503 [gr-qc]} \BibitemShut {NoStop}%
\bibitem [{\citenamefont {Ramos-Buades}\ \emph {et~al.}(2023)\citenamefont {Ramos-Buades}, \citenamefont {Buonanno}, \citenamefont {Estell{\'e}s}, \citenamefont {Khalil}, \citenamefont {Mihaylov}, \citenamefont {Ossokine}, \citenamefont {Pompili},\ and\ \citenamefont {Shiferaw}}]{SEOBNRv5PHM}%
  \BibitemOpen
  \bibfield  {author} {\bibinfo {author} {\bibfnamefont {A.}~\bibnamefont {Ramos-Buades}}, \bibinfo {author} {\bibfnamefont {A.}~\bibnamefont {Buonanno}}, \bibinfo {author} {\bibfnamefont {H.}~\bibnamefont {Estell{\'e}s}}, \bibinfo {author} {\bibfnamefont {M.}~\bibnamefont {Khalil}}, \bibinfo {author} {\bibfnamefont {D.~P.}\ \bibnamefont {Mihaylov}}, \bibinfo {author} {\bibfnamefont {S.}~\bibnamefont {Ossokine}}, \bibinfo {author} {\bibfnamefont {L.}~\bibnamefont {Pompili}},\ and\ \bibinfo {author} {\bibfnamefont {M.}~\bibnamefont {Shiferaw}},\ }\bibfield  {title} {\bibinfo {title} {{Next generation of accurate and efficient multipolar precessing-spin effective-one-body waveforms for binary black holes}},\ }\href {https://doi.org/10.1103/PhysRevD.108.124037} {\bibfield  {journal} {\bibinfo  {journal} {Phys. Rev. D}\ }\textbf {\bibinfo {volume} {108}},\ \bibinfo {pages} {124037} (\bibinfo {year} {2023})},\ \Eprint {https://arxiv.org/abs/2303.18046} {arXiv:2303.18046 [gr-qc]} \BibitemShut {NoStop}%
\bibitem [{\citenamefont {{LIGO Scientific Collaboration}}\ \emph {et~al.}(2024)\citenamefont {{LIGO Scientific Collaboration}}, \citenamefont {{Virgo Collaboration}},\ and\ \citenamefont {{KAGRA Collaboration}}}]{GWOSC_GWTC4_EventAPI}%
  \BibitemOpen
  \bibfield  {author} {\bibinfo {author} {\bibnamefont {{LIGO Scientific Collaboration}}}, \bibinfo {author} {\bibnamefont {{Virgo Collaboration}}},\ and\ \bibinfo {author} {\bibnamefont {{KAGRA Collaboration}}},\ }\href@noop {} {\bibinfo {title} {{GWTC-4.0 Event API}}},\ \bibinfo {howpublished} {\url{https://gwosc.org/eventapi/html/GWTC-4.0/}} (\bibinfo {year} {2024}),\ \bibinfo {note} {gravitational Wave Open Science Center}\BibitemShut {NoStop}%
\bibitem [{\citenamefont {Abbott}\ \emph {et~al.}(2020)\citenamefont {Abbott} \emph {et~al.}}]{LIGOScientific:2020aai}%
  \BibitemOpen
  \bibfield  {author} {\bibinfo {author} {\bibfnamefont {B.~P.}\ \bibnamefont {Abbott}} \emph {et~al.} (\bibinfo {collaboration} {LIGO Scientific, Virgo}),\ }\bibfield  {title} {\bibinfo {title} {{GW190425: Observation of a Compact Binary Coalescence with Total Mass $\sim 3.4 M_{\odot}$}},\ }\href {https://doi.org/10.3847/2041-8213/ab75f5} {\bibfield  {journal} {\bibinfo  {journal} {Astrophys. J. Lett.}\ }\textbf {\bibinfo {volume} {892}},\ \bibinfo {pages} {L3} (\bibinfo {year} {2020})},\ \Eprint {https://arxiv.org/abs/2001.01761} {arXiv:2001.01761 [astro-ph.HE]} \BibitemShut {NoStop}%
\bibitem [{\citenamefont {{LIGO Scientific Collaboration}}\ \emph {et~al.}(2025)\citenamefont {{LIGO Scientific Collaboration}}, \citenamefont {{Virgo Collaboration}},\ and\ \citenamefont {{KAGRA Collaboration}}}]{LVK_GWTC4_PE_2025}%
  \BibitemOpen
  \bibfield  {author} {\bibinfo {author} {\bibnamefont {{LIGO Scientific Collaboration}}}, \bibinfo {author} {\bibnamefont {{Virgo Collaboration}}},\ and\ \bibinfo {author} {\bibnamefont {{KAGRA Collaboration}}},\ }\bibfield  {title} {\bibinfo {title} {Gwtc-4.0: Parameter estimation data release},\ }\href {https://doi.org/10.5281/zenodo.17014085} {10.5281/zenodo.17014085} (\bibinfo {year} {2025})\BibitemShut {NoStop}%
\bibitem [{\citenamefont {Finn}\ and\ \citenamefont {Chernoff}(1993)}]{Finn:1992xs}%
  \BibitemOpen
  \bibfield  {author} {\bibinfo {author} {\bibfnamefont {L.~S.}\ \bibnamefont {Finn}}\ and\ \bibinfo {author} {\bibfnamefont {D.~F.}\ \bibnamefont {Chernoff}},\ }\bibfield  {title} {\bibinfo {title} {{Observing binary inspiral in gravitational radiation: One interferometer}},\ }\href {https://doi.org/10.1103/PhysRevD.47.2198} {\bibfield  {journal} {\bibinfo  {journal} {Phys. Rev. D}\ }\textbf {\bibinfo {volume} {47}},\ \bibinfo {pages} {2198} (\bibinfo {year} {1993})},\ \Eprint {https://arxiv.org/abs/gr-qc/9301003} {arXiv:gr-qc/9301003} \BibitemShut {NoStop}%
\bibitem [{\citenamefont {Allen}\ and\ \citenamefont {Romano}(1999)}]{Allen:1997ad}%
  \BibitemOpen
  \bibfield  {author} {\bibinfo {author} {\bibfnamefont {B.}~\bibnamefont {Allen}}\ and\ \bibinfo {author} {\bibfnamefont {J.~D.}\ \bibnamefont {Romano}},\ }\bibfield  {title} {\bibinfo {title} {{Detecting a stochastic background of gravitational radiation: Signal processing strategies and sensitivities}},\ }\href {https://doi.org/10.1103/PhysRevD.59.102001} {\bibfield  {journal} {\bibinfo  {journal} {Phys. Rev. D}\ }\textbf {\bibinfo {volume} {59}},\ \bibinfo {pages} {102001} (\bibinfo {year} {1999})},\ \Eprint {https://arxiv.org/abs/gr-qc/9710117} {arXiv:gr-qc/9710117} \BibitemShut {NoStop}%
\bibitem [{\citenamefont {Abbott}\ \emph {et~al.}(2007)\citenamefont {Abbott} \emph {et~al.}}]{LIGOScientific:2007gwp}%
  \BibitemOpen
  \bibfield  {author} {\bibinfo {author} {\bibfnamefont {B.}~\bibnamefont {Abbott}} \emph {et~al.} (\bibinfo {collaboration} {LIGO Scientific}),\ }\bibfield  {title} {\bibinfo {title} {{Upper limit map of a background of gravitational waves}},\ }\href {https://doi.org/10.1103/PhysRevD.76.082003} {\bibfield  {journal} {\bibinfo  {journal} {Phys. Rev. D}\ }\textbf {\bibinfo {volume} {76}},\ \bibinfo {pages} {082003} (\bibinfo {year} {2007})},\ \Eprint {https://arxiv.org/abs/astro-ph/0703234} {arXiv:astro-ph/0703234} \BibitemShut {NoStop}%
\bibitem [{\citenamefont {Abbott}\ \emph {et~al.}(2017)\citenamefont {Abbott} \emph {et~al.}}]{LIGOScientific:2016jlg}%
  \BibitemOpen
  \bibfield  {author} {\bibinfo {author} {\bibfnamefont {B.~P.}\ \bibnamefont {Abbott}} \emph {et~al.} (\bibinfo {collaboration} {LIGO Scientific, Virgo}),\ }\bibfield  {title} {\bibinfo {title} {{Upper Limits on the Stochastic Gravitational-Wave Background from Advanced LIGO{\textquoteright}s First Observing Run}},\ }\href {https://doi.org/10.1103/PhysRevLett.118.121101} {\bibfield  {journal} {\bibinfo  {journal} {Phys. Rev. Lett.}\ }\textbf {\bibinfo {volume} {118}},\ \bibinfo {pages} {121101} (\bibinfo {year} {2017})},\ \bibinfo {note} {[Erratum: Phys.Rev.Lett. 119, 029901 (2017)]},\ \Eprint {https://arxiv.org/abs/1612.02029} {arXiv:1612.02029 [gr-qc]} \BibitemShut {NoStop}%
\bibitem [{\citenamefont {Klimenko}\ \emph {et~al.}(2008)\citenamefont {Klimenko}, \citenamefont {Yakushin}, \citenamefont {Mercer},\ and\ \citenamefont {Mitselmakher}}]{Klimenko:2008fu}%
  \BibitemOpen
  \bibfield  {author} {\bibinfo {author} {\bibfnamefont {S.}~\bibnamefont {Klimenko}}, \bibinfo {author} {\bibfnamefont {I.}~\bibnamefont {Yakushin}}, \bibinfo {author} {\bibfnamefont {A.}~\bibnamefont {Mercer}},\ and\ \bibinfo {author} {\bibfnamefont {G.}~\bibnamefont {Mitselmakher}},\ }\bibfield  {title} {\bibinfo {title} {{Coherent method for detection of gravitational wave bursts}},\ }\href {https://doi.org/10.1088/0264-9381/25/11/114029} {\bibfield  {journal} {\bibinfo  {journal} {Class. Quant. Grav.}\ }\textbf {\bibinfo {volume} {25}},\ \bibinfo {pages} {114029} (\bibinfo {year} {2008})},\ \Eprint {https://arxiv.org/abs/0802.3232} {arXiv:0802.3232 [gr-qc]} \BibitemShut {NoStop}%
\bibitem [{\citenamefont {Klimenko}\ \emph {et~al.}(2016)\citenamefont {Klimenko} \emph {et~al.}}]{Klimenko:2015ypf}%
  \BibitemOpen
  \bibfield  {author} {\bibinfo {author} {\bibfnamefont {S.}~\bibnamefont {Klimenko}} \emph {et~al.},\ }\bibfield  {title} {\bibinfo {title} {{Method for detection and reconstruction of gravitational wave transients with networks of advanced detectors}},\ }\href {https://doi.org/10.1103/PhysRevD.93.042004} {\bibfield  {journal} {\bibinfo  {journal} {Phys. Rev. D}\ }\textbf {\bibinfo {volume} {93}},\ \bibinfo {pages} {042004} (\bibinfo {year} {2016})},\ \Eprint {https://arxiv.org/abs/1511.05999} {arXiv:1511.05999 [gr-qc]} \BibitemShut {NoStop}%
\bibitem [{\citenamefont {Abbott}\ \emph {et~al.}(2019)\citenamefont {Abbott} \emph {et~al.}}]{LIGOScientific:2019gaw}%
  \BibitemOpen
  \bibfield  {author} {\bibinfo {author} {\bibfnamefont {B.~P.}\ \bibnamefont {Abbott}} \emph {et~al.} (\bibinfo {collaboration} {LIGO Scientific, Virgo}),\ }\bibfield  {title} {\bibinfo {title} {{Directional limits on persistent gravitational waves using data from Advanced LIGO's first two observing runs}},\ }\href {https://doi.org/10.1103/PhysRevD.100.062001} {\bibfield  {journal} {\bibinfo  {journal} {Phys. Rev. D}\ }\textbf {\bibinfo {volume} {100}},\ \bibinfo {pages} {062001} (\bibinfo {year} {2019})},\ \Eprint {https://arxiv.org/abs/1903.08844} {arXiv:1903.08844 [gr-qc]} \BibitemShut {NoStop}%
\bibitem [{\citenamefont {Dideron}\ \emph {et~al.}(2023)\citenamefont {Dideron}, \citenamefont {Mukherjee},\ and\ \citenamefont {Lehner}}]{Dideron:2022tap}%
  \BibitemOpen
  \bibfield  {author} {\bibinfo {author} {\bibfnamefont {G.}~\bibnamefont {Dideron}}, \bibinfo {author} {\bibfnamefont {S.}~\bibnamefont {Mukherjee}},\ and\ \bibinfo {author} {\bibfnamefont {L.}~\bibnamefont {Lehner}},\ }\bibfield  {title} {\bibinfo {title} {{New framework to study unmodeled physics from gravitational wave data}},\ }\href {https://doi.org/10.1103/PhysRevD.107.104023} {\bibfield  {journal} {\bibinfo  {journal} {Phys. Rev. D}\ }\textbf {\bibinfo {volume} {107}},\ \bibinfo {pages} {104023} (\bibinfo {year} {2023})},\ \Eprint {https://arxiv.org/abs/2209.14321} {arXiv:2209.14321 [gr-qc]} \BibitemShut {NoStop}%
\bibitem [{\citenamefont {Chakraborty}\ and\ \citenamefont {Mukherjee}(2025{\natexlab{a}})}]{Chakraborty:2024mbr}%
  \BibitemOpen
  \bibfield  {author} {\bibinfo {author} {\bibfnamefont {A.}~\bibnamefont {Chakraborty}}\ and\ \bibinfo {author} {\bibfnamefont {S.}~\bibnamefont {Mukherjee}},\ }\bibfield  {title} {\bibinfo {title} {{{\ensuremath{\mu}}-GLANCE: A Novel Technique to Detect Chromatically and Achromatically Lensed Gravitational-wave Signals}},\ }\href {https://doi.org/10.3847/1538-4357/adc578} {\bibfield  {journal} {\bibinfo  {journal} {Astrophys. J.}\ }\textbf {\bibinfo {volume} {984}},\ \bibinfo {pages} {107} (\bibinfo {year} {2025}{\natexlab{a}})},\ \Eprint {https://arxiv.org/abs/2410.06995} {arXiv:2410.06995 [gr-qc]} \BibitemShut {NoStop}%
\bibitem [{\citenamefont {Chakraborty}\ and\ \citenamefont {Mukherjee}(2024)}]{Chakraborty:2024net}%
  \BibitemOpen
  \bibfield  {author} {\bibinfo {author} {\bibfnamefont {A.}~\bibnamefont {Chakraborty}}\ and\ \bibinfo {author} {\bibfnamefont {S.}~\bibnamefont {Mukherjee}},\ }\bibfield  {title} {\bibinfo {title} {{GLANCE {\textendash} Gravitational Lensing Authenticator using Non-modelled Cross-correlation Exploration of Gravitational Wave Signals}},\ }\href {https://doi.org/10.1093/mnras/stae1800} {\bibfield  {journal} {\bibinfo  {journal} {Mon. Not. Roy. Astron. Soc.}\ }\textbf {\bibinfo {volume} {532}},\ \bibinfo {pages} {4842} (\bibinfo {year} {2024})},\ \Eprint {https://arxiv.org/abs/2403.03982} {arXiv:2403.03982 [gr-qc]} \BibitemShut {NoStop}%
\bibitem [{\citenamefont {Dideron}\ \emph {et~al.}(2025)\citenamefont {Dideron}, \citenamefont {Mukherjee},\ and\ \citenamefont {Lehner}}]{Dideron:2024xwm}%
  \BibitemOpen
  \bibfield  {author} {\bibinfo {author} {\bibfnamefont {G.}~\bibnamefont {Dideron}}, \bibinfo {author} {\bibfnamefont {S.}~\bibnamefont {Mukherjee}},\ and\ \bibinfo {author} {\bibfnamefont {L.}~\bibnamefont {Lehner}},\ }\bibfield  {title} {\bibinfo {title} {{Detecting unmodeled, source-dependent signals in gravitational waves with SCoRe}},\ }\href {https://doi.org/10.1103/PhysRevD.111.064029} {\bibfield  {journal} {\bibinfo  {journal} {Phys. Rev. D}\ }\textbf {\bibinfo {volume} {111}},\ \bibinfo {pages} {064029} (\bibinfo {year} {2025})},\ \Eprint {https://arxiv.org/abs/2411.04198} {arXiv:2411.04198 [gr-qc]} \BibitemShut {NoStop}%
\bibitem [{\citenamefont {Chakraborty}\ and\ \citenamefont {Mukherjee}(2025{\natexlab{b}})}]{Chakraborty:2025maj}%
  \BibitemOpen
  \bibfield  {author} {\bibinfo {author} {\bibfnamefont {A.}~\bibnamefont {Chakraborty}}\ and\ \bibinfo {author} {\bibfnamefont {S.}~\bibnamefont {Mukherjee}},\ }\bibfield  {title} {\bibinfo {title} {{The First Model-independent Chromatic Microlensing Search: No Evidence in the Gravitational Wave Catalog of LIGO{\textendash}Virgo{\textendash}KAGRA}},\ }\href {https://doi.org/10.3847/1538-4357/adf330} {\bibfield  {journal} {\bibinfo  {journal} {Astrophys. J.}\ }\textbf {\bibinfo {volume} {990}},\ \bibinfo {pages} {68} (\bibinfo {year} {2025}{\natexlab{b}})},\ \Eprint {https://arxiv.org/abs/2503.16281} {arXiv:2503.16281 [gr-qc]} \BibitemShut {NoStop}%
\bibitem [{\citenamefont {Carr}\ and\ \citenamefont {Kuhnel}(2020)}]{Carr:2020xqk}%
  \BibitemOpen
  \bibfield  {author} {\bibinfo {author} {\bibfnamefont {B.}~\bibnamefont {Carr}}\ and\ \bibinfo {author} {\bibfnamefont {F.}~\bibnamefont {Kuhnel}},\ }\bibfield  {title} {\bibinfo {title} {{Primordial Black Holes as Dark Matter: Recent Developments}},\ }\href {https://doi.org/10.1146/annurev-nucl-050520-125911} {\bibfield  {journal} {\bibinfo  {journal} {Ann. Rev. Nucl. Part. Sci.}\ }\textbf {\bibinfo {volume} {70}},\ \bibinfo {pages} {355} (\bibinfo {year} {2020})},\ \Eprint {https://arxiv.org/abs/2006.02838} {arXiv:2006.02838 [astro-ph.CO]} \BibitemShut {NoStop}%
\bibitem [{\citenamefont {Weber}\ and\ \citenamefont {de~Boer}(2010)}]{Weber_2010}%
  \BibitemOpen
  \bibfield  {author} {\bibinfo {author} {\bibfnamefont {M.}~\bibnamefont {Weber}}\ and\ \bibinfo {author} {\bibfnamefont {W.}~\bibnamefont {de~Boer}},\ }\bibfield  {title} {\bibinfo {title} {Determination of the local dark matter density in our galaxy},\ }\href {https://doi.org/10.1051/0004-6361/200913381} {\bibfield  {journal} {\bibinfo  {journal} {Astronomy and Astrophysics}\ }\textbf {\bibinfo {volume} {509}},\ \bibinfo {pages} {A25} (\bibinfo {year} {2010})}\BibitemShut {NoStop}%
\bibitem [{\citenamefont {Iocco}\ \emph {et~al.}(2015)\citenamefont {Iocco}, \citenamefont {Pato},\ and\ \citenamefont {Bertone}}]{Iocco:2015xga}%
  \BibitemOpen
  \bibfield  {author} {\bibinfo {author} {\bibfnamefont {F.}~\bibnamefont {Iocco}}, \bibinfo {author} {\bibfnamefont {M.}~\bibnamefont {Pato}},\ and\ \bibinfo {author} {\bibfnamefont {G.}~\bibnamefont {Bertone}},\ }\bibfield  {title} {\bibinfo {title} {{Evidence for dark matter in the inner Milky Way}},\ }\href {https://doi.org/10.1038/nphys3237} {\bibfield  {journal} {\bibinfo  {journal} {Nature Phys.}\ }\textbf {\bibinfo {volume} {11}},\ \bibinfo {pages} {245} (\bibinfo {year} {2015})},\ \Eprint {https://arxiv.org/abs/1502.03821} {arXiv:1502.03821 [astro-ph.GA]} \BibitemShut {NoStop}%
\bibitem [{\citenamefont {Salucci}(2019)}]{Salucci:2018hqu}%
  \BibitemOpen
  \bibfield  {author} {\bibinfo {author} {\bibfnamefont {P.}~\bibnamefont {Salucci}},\ }\bibfield  {title} {\bibinfo {title} {{The distribution of dark matter in galaxies}},\ }\href {https://doi.org/10.1007/s00159-018-0113-1} {\bibfield  {journal} {\bibinfo  {journal} {Astron. Astrophys. Rev.}\ }\textbf {\bibinfo {volume} {27}},\ \bibinfo {pages} {2} (\bibinfo {year} {2019})},\ \Eprint {https://arxiv.org/abs/1811.08843} {arXiv:1811.08843 [astro-ph.GA]} \BibitemShut {NoStop}%
\bibitem [{\citenamefont {Mandel}\ \emph {et~al.}(2018)\citenamefont {Mandel}, \citenamefont {Sesana},\ and\ \citenamefont {Vecchio}}]{Mandel:2017pzd}%
  \BibitemOpen
  \bibfield  {author} {\bibinfo {author} {\bibfnamefont {I.}~\bibnamefont {Mandel}}, \bibinfo {author} {\bibfnamefont {A.}~\bibnamefont {Sesana}},\ and\ \bibinfo {author} {\bibfnamefont {A.}~\bibnamefont {Vecchio}},\ }\bibfield  {title} {\bibinfo {title} {{The astrophysical science case for a decihertz gravitational-wave detector}},\ }\href {https://doi.org/10.1088/1361-6382/aaa7e0} {\bibfield  {journal} {\bibinfo  {journal} {Class. Quant. Grav.}\ }\textbf {\bibinfo {volume} {35}},\ \bibinfo {pages} {054004} (\bibinfo {year} {2018})},\ \Eprint {https://arxiv.org/abs/1710.11187} {arXiv:1710.11187 [astro-ph.HE]} \BibitemShut {NoStop}%
\bibitem [{\citenamefont {Ajith}\ \emph {et~al.}(2025{\natexlab{b}})\citenamefont {Ajith} \emph {et~al.}}]{Ajith:2024mie}%
  \BibitemOpen
  \bibfield  {author} {\bibinfo {author} {\bibfnamefont {P.}~\bibnamefont {Ajith}} \emph {et~al.},\ }\bibfield  {title} {\bibinfo {title} {{The Lunar Gravitational-wave Antenna: mission studies and science case}},\ }\href {https://doi.org/10.1088/1475-7516/2025/01/108} {\bibfield  {journal} {\bibinfo  {journal} {JCAP}\ }\textbf {\bibinfo {volume} {01}},\ \bibinfo {pages} {108}},\ \Eprint {https://arxiv.org/abs/2404.09181} {arXiv:2404.09181 [gr-qc]} \BibitemShut {NoStop}%
\bibitem [{\citenamefont {Sesana}(2017)}]{Sesana:2017vsj}%
  \BibitemOpen
  \bibfield  {author} {\bibinfo {author} {\bibfnamefont {A.}~\bibnamefont {Sesana}},\ }\bibfield  {title} {\bibinfo {title} {{Multi-band gravitational wave astronomy: science with joint space- and ground-based observations of black hole binaries}},\ }\href {https://doi.org/10.1088/1742-6596/840/1/012018} {\bibfield  {journal} {\bibinfo  {journal} {J. Phys. Conf. Ser.}\ }\textbf {\bibinfo {volume} {840}},\ \bibinfo {pages} {012018} (\bibinfo {year} {2017})},\ \Eprint {https://arxiv.org/abs/1702.04356} {arXiv:1702.04356 [astro-ph.HE]} \BibitemShut {NoStop}%
\bibitem [{\citenamefont {Harris}\ \emph {et~al.}(2020)\citenamefont {Harris}, \citenamefont {Millman}, \citenamefont {van~der Walt}, \citenamefont {Gommers}, \citenamefont {Virtanen}, \citenamefont {Cournapeau}, \citenamefont {Wieser}, \citenamefont {Taylor}, \citenamefont {Berg}, \citenamefont {Smith}, \citenamefont {Kern}, \citenamefont {Picus}, \citenamefont {Hoyer}, \citenamefont {van Kerkwijk}, \citenamefont {Brett}, \citenamefont {Haldane}, \citenamefont {del R{\'{i}}o}, \citenamefont {Wiebe}, \citenamefont {Peterson}, \citenamefont {G{\'{e}}rard-Marchant}, \citenamefont {Sheppard}, \citenamefont {Reddy}, \citenamefont {Weckesser}, \citenamefont {Abbasi}, \citenamefont {Gohlke},\ and\ \citenamefont {Oliphant}}]{numpy}%
  \BibitemOpen
  \bibfield  {author} {\bibinfo {author} {\bibfnamefont {C.~R.}\ \bibnamefont {Harris}}, \bibinfo {author} {\bibfnamefont {K.~J.}\ \bibnamefont {Millman}}, \bibinfo {author} {\bibfnamefont {S.~J.}\ \bibnamefont {van~der Walt}}, \bibinfo {author} {\bibfnamefont {R.}~\bibnamefont {Gommers}}, \bibinfo {author} {\bibfnamefont {P.}~\bibnamefont {Virtanen}}, \bibinfo {author} {\bibfnamefont {D.}~\bibnamefont {Cournapeau}}, \bibinfo {author} {\bibfnamefont {E.}~\bibnamefont {Wieser}}, \bibinfo {author} {\bibfnamefont {J.}~\bibnamefont {Taylor}}, \bibinfo {author} {\bibfnamefont {S.}~\bibnamefont {Berg}}, \bibinfo {author} {\bibfnamefont {N.~J.}\ \bibnamefont {Smith}}, \bibinfo {author} {\bibfnamefont {R.}~\bibnamefont {Kern}}, \bibinfo {author} {\bibfnamefont {M.}~\bibnamefont {Picus}}, \bibinfo {author} {\bibfnamefont {S.}~\bibnamefont {Hoyer}}, \bibinfo {author} {\bibfnamefont {M.~H.}\ \bibnamefont {van Kerkwijk}}, \bibinfo {author} {\bibfnamefont {M.}~\bibnamefont {Brett}}, \bibinfo {author} {\bibfnamefont
  {A.}~\bibnamefont {Haldane}}, \bibinfo {author} {\bibfnamefont {J.~F.}\ \bibnamefont {del R{\'{i}}o}}, \bibinfo {author} {\bibfnamefont {M.}~\bibnamefont {Wiebe}}, \bibinfo {author} {\bibfnamefont {P.}~\bibnamefont {Peterson}}, \bibinfo {author} {\bibfnamefont {P.}~\bibnamefont {G{\'{e}}rard-Marchant}}, \bibinfo {author} {\bibfnamefont {K.}~\bibnamefont {Sheppard}}, \bibinfo {author} {\bibfnamefont {T.}~\bibnamefont {Reddy}}, \bibinfo {author} {\bibfnamefont {W.}~\bibnamefont {Weckesser}}, \bibinfo {author} {\bibfnamefont {H.}~\bibnamefont {Abbasi}}, \bibinfo {author} {\bibfnamefont {C.}~\bibnamefont {Gohlke}},\ and\ \bibinfo {author} {\bibfnamefont {T.~E.}\ \bibnamefont {Oliphant}},\ }\bibfield  {title} {\bibinfo {title} {Array programming with {NumPy}},\ }\href {https://doi.org/10.1038/s41586-020-2649-2} {\bibfield  {journal} {\bibinfo  {journal} {Nature}\ }\textbf {\bibinfo {volume} {585}},\ \bibinfo {pages} {357} (\bibinfo {year} {2020})}\BibitemShut {NoStop}%
\bibitem [{\citenamefont {Nitz}\ \emph {et~al.}(2024)\citenamefont {Nitz}, \citenamefont {Harry}, \citenamefont {Brown}, \citenamefont {Biwer}, \citenamefont {Willis}, \citenamefont {Canton}, \citenamefont {Capano}, \citenamefont {Dent}, \citenamefont {Pekowsky}, \citenamefont {Davies}, \citenamefont {De}, \citenamefont {Cabero}, \citenamefont {Wu}, \citenamefont {Williamson}, \citenamefont {Machenschalk}, \citenamefont {Macleod}, \citenamefont {Pannarale}, \citenamefont {Kumar}, \citenamefont {Reyes}, \citenamefont {dfinstad}, \citenamefont {Kumar}, \citenamefont {Tápai}, \citenamefont {Singer}, \citenamefont {Kumar}, \citenamefont {veronica villa}, \citenamefont {maxtrevor}, \citenamefont {Gadre}, \citenamefont {Khan}, \citenamefont {Fairhurst},\ and\ \citenamefont {Tolley}}]{pycbc}%
  \BibitemOpen
  \bibfield  {author} {\bibinfo {author} {\bibfnamefont {A.}~\bibnamefont {Nitz}}, \bibinfo {author} {\bibfnamefont {I.}~\bibnamefont {Harry}}, \bibinfo {author} {\bibfnamefont {D.}~\bibnamefont {Brown}}, \bibinfo {author} {\bibfnamefont {C.~M.}\ \bibnamefont {Biwer}}, \bibinfo {author} {\bibfnamefont {J.}~\bibnamefont {Willis}}, \bibinfo {author} {\bibfnamefont {T.~D.}\ \bibnamefont {Canton}}, \bibinfo {author} {\bibfnamefont {C.}~\bibnamefont {Capano}}, \bibinfo {author} {\bibfnamefont {T.}~\bibnamefont {Dent}}, \bibinfo {author} {\bibfnamefont {L.}~\bibnamefont {Pekowsky}}, \bibinfo {author} {\bibfnamefont {G.~S.~C.}\ \bibnamefont {Davies}}, \bibinfo {author} {\bibfnamefont {S.}~\bibnamefont {De}}, \bibinfo {author} {\bibfnamefont {M.}~\bibnamefont {Cabero}}, \bibinfo {author} {\bibfnamefont {S.}~\bibnamefont {Wu}}, \bibinfo {author} {\bibfnamefont {A.~R.}\ \bibnamefont {Williamson}}, \bibinfo {author} {\bibfnamefont {B.}~\bibnamefont {Machenschalk}}, \bibinfo {author} {\bibfnamefont {D.}~\bibnamefont
  {Macleod}}, \bibinfo {author} {\bibfnamefont {F.}~\bibnamefont {Pannarale}}, \bibinfo {author} {\bibfnamefont {P.}~\bibnamefont {Kumar}}, \bibinfo {author} {\bibfnamefont {S.}~\bibnamefont {Reyes}}, \bibinfo {author} {\bibnamefont {dfinstad}}, \bibinfo {author} {\bibfnamefont {S.}~\bibnamefont {Kumar}}, \bibinfo {author} {\bibfnamefont {M.}~\bibnamefont {Tápai}}, \bibinfo {author} {\bibfnamefont {L.}~\bibnamefont {Singer}}, \bibinfo {author} {\bibfnamefont {P.}~\bibnamefont {Kumar}}, \bibinfo {author} {\bibnamefont {veronica villa}}, \bibinfo {author} {\bibnamefont {maxtrevor}}, \bibinfo {author} {\bibfnamefont {B.~U.~V.}\ \bibnamefont {Gadre}}, \bibinfo {author} {\bibfnamefont {S.}~\bibnamefont {Khan}}, \bibinfo {author} {\bibfnamefont {S.}~\bibnamefont {Fairhurst}},\ and\ \bibinfo {author} {\bibfnamefont {A.}~\bibnamefont {Tolley}},\ }\href {https://doi.org/10.5281/zenodo.10473621} {\bibinfo {title} {gwastro/pycbc: v2.3.3 release of pycbc}} (\bibinfo {year} {2024})\BibitemShut {NoStop}%
\bibitem [{\citenamefont {{Numba Developers}}(2025)}]{numba}%
  \BibitemOpen
  \bibfield  {author} {\bibinfo {author} {\bibnamefont {{Numba Developers}}},\ }\href {https://numba.pydata.org/} {\bibinfo {title} {Numba}} (\bibinfo {year} {2025}),\ \bibinfo {note} {open-source JIT compiler for Python}\BibitemShut {NoStop}%
\bibitem [{\citenamefont {joblib developers}(2025)}]{joblib}%
  \BibitemOpen
  \bibfield  {author} {\bibinfo {author} {\bibfnamefont {T.}~\bibnamefont {joblib developers}},\ }\href {https://doi.org/10.5281/zenodo.15496554} {\bibinfo {title} {joblib}} (\bibinfo {year} {2025})\BibitemShut {NoStop}%
\bibitem [{\citenamefont {Hoy}\ and\ \citenamefont {Raymond}(2021)}]{pesummary}%
  \BibitemOpen
  \bibfield  {author} {\bibinfo {author} {\bibfnamefont {C.}~\bibnamefont {Hoy}}\ and\ \bibinfo {author} {\bibfnamefont {V.}~\bibnamefont {Raymond}},\ }\bibfield  {title} {\bibinfo {title} {{PESummary: the code agnostic Parameter Estimation Summary page builder}},\ }\href {https://doi.org/10.1016/j.softx.2021.100765} {\bibfield  {journal} {\bibinfo  {journal} {SoftwareX}\ }\textbf {\bibinfo {volume} {15}},\ \bibinfo {pages} {100765} (\bibinfo {year} {2021})},\ \Eprint {https://arxiv.org/abs/2006.06639} {arXiv:2006.06639 [astro-ph.IM]} \BibitemShut {NoStop}%
\bibitem [{\citenamefont {{LIGO Scientific Collaboration}}\ \emph {et~al.}(2018)\citenamefont {{LIGO Scientific Collaboration}}, \citenamefont {{Virgo Collaboration}},\ and\ \citenamefont {{KAGRA Collaboration}}}]{lalsuite}%
  \BibitemOpen
  \bibfield  {author} {\bibinfo {author} {\bibnamefont {{LIGO Scientific Collaboration}}}, \bibinfo {author} {\bibnamefont {{Virgo Collaboration}}},\ and\ \bibinfo {author} {\bibnamefont {{KAGRA Collaboration}}},\ }\href {https://doi.org/10.7935/GT1W-FZ16} {\bibinfo {title} {{LVK} {A}lgorithm {L}ibrary - {LALS}uite}},\ \bibinfo {howpublished} {Free software (GPL)} (\bibinfo {year} {2018})\BibitemShut {NoStop}%
\bibitem [{\citenamefont {pandas~development team}(2020)}]{pandas}%
  \BibitemOpen
  \bibfield  {author} {\bibinfo {author} {\bibfnamefont {T.}~\bibnamefont {pandas~development team}},\ }\href {https://doi.org/10.5281/zenodo.3509134} {\bibinfo {title} {pandas-dev/pandas: Pandas}} (\bibinfo {year} {2020})\BibitemShut {NoStop}%
\bibitem [{\citenamefont {Collette}\ \emph {et~al.}(2022)\citenamefont {Collette}, \citenamefont {Kluyver}, \citenamefont {Caswell}, \citenamefont {Tocknell}, \citenamefont {Kieffer}, \citenamefont {Jelenak}, \citenamefont {Scopatz}, \citenamefont {Dale}, \citenamefont {Chen}, \citenamefont {VINCENT}, \citenamefont {Einhorn}, \citenamefont {payno}, \citenamefont {juliagarriga}, \citenamefont {Sciarelli}, \citenamefont {Valls}, \citenamefont {Ghosh}, \citenamefont {Pedersen}, \citenamefont {jakirkham}, \citenamefont {Raspaud}, \citenamefont {Danilevski}, \citenamefont {Abbasi}, \citenamefont {Readey}, \citenamefont {Mühlbauer}, \citenamefont {Paramonov}, \citenamefont {Chan}, \citenamefont {Solé}, \citenamefont {jialin}, \citenamefont {Guest}, \citenamefont {Feng},\ and\ \citenamefont {Kittisopikul}}]{h5py}%
  \BibitemOpen
  \bibfield  {author} {\bibinfo {author} {\bibfnamefont {A.}~\bibnamefont {Collette}}, \bibinfo {author} {\bibfnamefont {T.}~\bibnamefont {Kluyver}}, \bibinfo {author} {\bibfnamefont {T.~A.}\ \bibnamefont {Caswell}}, \bibinfo {author} {\bibfnamefont {J.}~\bibnamefont {Tocknell}}, \bibinfo {author} {\bibfnamefont {J.}~\bibnamefont {Kieffer}}, \bibinfo {author} {\bibfnamefont {A.}~\bibnamefont {Jelenak}}, \bibinfo {author} {\bibfnamefont {A.}~\bibnamefont {Scopatz}}, \bibinfo {author} {\bibfnamefont {D.}~\bibnamefont {Dale}}, \bibinfo {author} {\bibnamefont {Chen}}, \bibinfo {author} {\bibfnamefont {T.}~\bibnamefont {VINCENT}}, \bibinfo {author} {\bibfnamefont {M.}~\bibnamefont {Einhorn}}, \bibinfo {author} {\bibnamefont {payno}}, \bibinfo {author} {\bibnamefont {juliagarriga}}, \bibinfo {author} {\bibfnamefont {P.}~\bibnamefont {Sciarelli}}, \bibinfo {author} {\bibfnamefont {V.}~\bibnamefont {Valls}}, \bibinfo {author} {\bibfnamefont {S.}~\bibnamefont {Ghosh}}, \bibinfo {author} {\bibfnamefont {U.~K.}\
  \bibnamefont {Pedersen}}, \bibinfo {author} {\bibnamefont {jakirkham}}, \bibinfo {author} {\bibfnamefont {M.}~\bibnamefont {Raspaud}}, \bibinfo {author} {\bibfnamefont {C.}~\bibnamefont {Danilevski}}, \bibinfo {author} {\bibfnamefont {H.}~\bibnamefont {Abbasi}}, \bibinfo {author} {\bibfnamefont {J.}~\bibnamefont {Readey}}, \bibinfo {author} {\bibfnamefont {K.}~\bibnamefont {Mühlbauer}}, \bibinfo {author} {\bibfnamefont {A.}~\bibnamefont {Paramonov}}, \bibinfo {author} {\bibfnamefont {L.}~\bibnamefont {Chan}}, \bibinfo {author} {\bibfnamefont {V.~A.}\ \bibnamefont {Solé}}, \bibinfo {author} {\bibnamefont {jialin}}, \bibinfo {author} {\bibfnamefont {D.~H.}\ \bibnamefont {Guest}}, \bibinfo {author} {\bibfnamefont {Y.}~\bibnamefont {Feng}},\ and\ \bibinfo {author} {\bibfnamefont {M.}~\bibnamefont {Kittisopikul}},\ }\href {https://doi.org/10.5281/zenodo.6575970} {\bibinfo {title} {h5py/h5py: 3.7.0}} (\bibinfo {year} {2022})\BibitemShut {NoStop}%
\bibitem [{\citenamefont {Hunter}(2007)}]{matplotlib}%
  \BibitemOpen
  \bibfield  {author} {\bibinfo {author} {\bibfnamefont {J.~D.}\ \bibnamefont {Hunter}},\ }\bibfield  {title} {\bibinfo {title} {Matplotlib: A 2d graphics environment},\ }\href {https://doi.org/10.1109/MCSE.2007.55} {\bibfield  {journal} {\bibinfo  {journal} {Computing in Science \& Engineering}\ }\textbf {\bibinfo {volume} {9}},\ \bibinfo {pages} {90} (\bibinfo {year} {2007})}\BibitemShut {NoStop}%
\end{thebibliography}
